\documentclass{article}
\usepackage[left=1.25in,top=1.25in,right=1.25in,bottom=1.25in,head=1.25in]{geometry}
\usepackage{amsfonts,amsmath,amssymb,amsthm}
\usepackage{verbatim,float,url,enumerate}
\usepackage{graphicx,subfigure,epsfig,psfrag}
\usepackage{dsfont,bm,color,appendix}
\usepackage{natbib}
\usepackage{pdfpages}
\usepackage{algorithm}
\graphicspath{{figures/}}
\def\by {{\bf y}}

\def\bB{{\bf B}}
\def\bF{{\bf F}}
\def\bX{{\bf X}}
\def\bZ{{\bf Z}}
\def\eps{\varepsilon}

\newcommand{\norm}[1]{\left\|{#1}\right\|} 
\newcommand{\ltwo}[1]{\norm{#1}_2} 

\title{Reluctant generalized additive modeling}
\author{ J. Kenneth Tay and Robert Tibshirani  \\
Department of Statistics, and Department of  Biomedical Data Science \\ Stanford University }

\begin{document}
\maketitle

\begin{abstract}
Sparse generalized additive models (GAMs) are an extension of sparse generalized linear models which allow a model's prediction to vary non-linearly with an input variable. This enables the data analyst build more accurate models, especially when the linearity assumption is known to be a poor approximation of reality. Motivated by reluctant interaction modeling \citep{Yu2019}, we propose a multi-stage algorithm, called \textit{reluctant generalized additive modeling (RGAM)}, that can fit sparse generalized additive models at scale. It is guided by the principle that, if all else is equal, one should prefer a linear feature over a non-linear feature. Unlike existing methods for sparse GAMs, RGAM can be extended easily to binary, count and survival data. We demonstrate the method's effectiveness on real and simulated examples.
\end{abstract}

\section{Introduction}\label{sec:intro}

Consider the supervised learning setting, where we have $n$ observations of $p$ features $\bX = \{ x_{ij} \}$ for $i = 1, 2, \dots, n$ and $j = 1, 2, \dots, p$, along with $n$ responses $y = (y_1, \dots, y_n)$. We assume that $\by$ and the columns of $\bX$ are mean-centered at zero so that we need not fit an intercept term. Letting $X_j \in \mathbb{R}^n$ denote the values of the $j$th feature, generalized linear models (GLMs) assume that the relationship between the response and the features is
\begin{equation}\label{eqn:ols}
\eta(y) = \sum_{j=1}^p \beta_j X_j + \eps,
\end{equation}
where $\eta$ is a link function and $\eps$ is a mean-zero error term. While GLMs are highly interpretable, they make the (possibly) unrealistic assumption that each feature influences the transformed response $\eta(y)$ in a linear fashion. Generalized additive models (GAMs), introduced by \cite{Hastie1986}, avoid this issue by modeling the relationship as
\begin{equation}\label{eqn:gam}
\eta(y) = \sum_{j=1}^p f_j(X_j) + \eps,
\end{equation}
where the $f_j(\cdot)$'s are unknown component functions, assumed to be smooth or to have low complexity. Even though the transformed response can vary with the individual features in a non-linear fashion, GAMs remain interpretable since the effect of $X_j$ on $\eta(y)$ (and hence, on $y$) does not depend on any $X_k$ with $k \neq j$.

One drawback of GAMs is that they assume that the response is influenced by every feature available to the data analyst. When $p$ is large, this seems to be an unreasonable assumption. In fact, once $p \geq n$, GAMs are unidentifiable: we can find two different fits $\hat{f}_1(\cdot), \dots, \hat{f}_p(\cdot)$ and $\hat{f}_1'(\cdot), \dots, \hat{f}_p'(\cdot)$ such that $\sum_j \hat{f}_j(x_j) = \sum_j \hat{f}_j'(x_j)$ for all possible $(x_1, \dots, x_p) \in \mathbb{R}^p$. This causes GAMs to lose their interpretability. An additional problem in this setting is that GAMs will tend to overfit to the noise in the data. As a result, there has been demand for additive models which are \textit{sparse}, i.e. consisting of just a handful of the features available to the data analyst. Previous methods for estimating sparse additive models, detailed in Section \ref{sec:litreview}, have cast model-fitting as an optimization problem

\begin{equation}\label{eqn:mingam}
\underset{f_1, \dots, f_p \in \mathcal{F}}{\text{minimize}} \quad \ell (y; f_1, \dots, f_p) + \sum_{j=1}^p J(f_j),
\end{equation}

where $\ell$ is the negative log-likelihood of the data, $J$ is some penalty function, and $\mathcal{F}$ is some space of allowable functions for the $f_j$'s.

When building a sparse additive model, the algorithm needs to make a choice: for some signal in the response, should we attribute it to a linear term in some feature $X_j$, or should we attribute it to a non-linear term in some (possibly other) feature $X_k$? Some of the earlier sparse additive methods ignore this choice: the $f_j$'s are all modeled as non-linear functions. This may result in needlessly complex models when having some of the $f_j$'s as linear functions would have sufficed. Later methods recognize this deficiency and have the flexibility to model each $f_j$ as either a linear or non-linear function through clever choices of penalty functions. However, the tradeoff between having a linear or non-linear function is often implicit and controlled via a tuning parameter.

Inspired by ``reluctant interaction modeling" \citep{Yu2019}, we propose a new algorithm for fitting sparse generalized additive models that has an explicit bias toward linear relationships over non-linear ones. As a guiding principle, we prefer a model to contain only effects that are linear in the original set of features: non-linearities are only included thereafter if they add to predictive performance. To operationalize this, we first construct a sparse model for $\eta(y)$ with just linear features. Next, we use the residual from the first step to construct new non-linear features. Finally, we fit another sparse model for $\eta(y)$ utilizing both the linear and non-linear features.

The rest of the paper is organized as follows: in the next section, we review previous methods which have sought to estimate sparse additive models from the given data. In Section \ref{sec:rgam}, we give a brief review of the ideas in \cite{Yu2019} and introduce our method, called ``reluctant generalized additive modeling" (RGAM), in greater detail. In Section \ref{sec:computation}, we point parameter choices that a practitioner should be cognizant of when using RGAM, as well as the computational advantages of the method. We demonstrate the method on synthetic and real data examples in Section \ref{sec:examples}, briefly discuss RGAM's effective degrees of freedom in Section \ref{sec:df} and end off with a discussion in Section \ref{sec:discussion}.

\section{Related work}\label{sec:litreview}

This review closely follows that in \cite{Chouldechova2015} and \cite{Petersen2019}. As mentioned in the introduction,  previous methods for fitting sparse additive models involve solving an optimization problem
\begin{equation}
\underset{f_1, \dots, f_p \in \mathcal{F}}{\text{minimize}} \quad \ell (y; f_1, \dots, f_p) + \sum_{j=1}^p J(f_j),
\end{equation}
with different methods choosing different penalty functions $J(\cdot)$ and different family of functions $\mathcal{F}$. The \textit{component selection and smoothing operator} (COSSO) \citep{Lin2006} models the $f_j$'s as belonging to a reproducing kernel Hilbert space (RKHS) and penalizes the sum of the RKHS norms of the component functions. \cite{Ravikumar2007} proposed the \textit{sparse additive model} (SpAM), which is essentially a functional version of the group lasso \citep{Yuan2006}. For each $j$, $f_j$ is modeled as a linear combination of $d$ basis functions $f_j = \beta_{j1} g_{j1} + \dots + \beta_{jd} g_{jd}$. Letting $\bB_j$ denote the $n \times d$ matrix with $(B_j)_{k\ell} = g_{j \ell} (x_{kj})$, SpAM penalizes the sum of $\ell_2$ norms of the $\bB_j \beta_j$, i.e. $J(f_j) = \lambda \ltwo{\bB_j \beta_j}$ for some hyperparameter $\lambda \geq 0$. \cite{Meier2009} parametrize each $f_j$ in a similar fashion, and propose a penalty which is the quadratic mean of the component function norm and a second derivative smoothness penalty, summed over the component functions. \cite{Sadhanala2017} proposed \textit{additive models with trend filtering}, where the penalty for $f_j$ is the (discrete) total variation of its $k$th (discrete) derivative, $k$ being an integer hyperparameter chosen by the user. The fit for each variable is restricted to piecewise polynomials of degreee $k$. Additive models with trend filtering are a generalization of the \textit{fused lasso additive model} (FLAM) \citep{Petersen2016}, where each $f_j$ is either all zero or piecewise constant with a small number of adaptively chosen knots.

While these earlier methods are able to capture non-linear fits between the features and the response, they will continue to do so even when a linear fit would have sufficed. (Additive models with trend filtering will only give a linear fit when $k = 1$.) In these cases, the methods above may overfit to the data, resulting in less interpretable models with possibly worse predictive performance. To address this issue, more recent methods have the ability to decide whether to model a feature linearly or non-linearly, given that it is included in the model. This ability is achieved with the use of more complex penalty functions. For example, the \textit{sparse partially linear additive model} (SPLAM) \citep{Lou2016} does so using a hierarchical group lasso penalty \citep{Yan2017}, while \textit{generalized additive model selection} (GAMSEL) \citep{Chouldechova2015} does so with an overlap group lasso penalty \citep{Jacob2009}. Most recently, \cite{Petersen2019} introduced \textit{sparse partially linear additive trend filtering} (SPLAT), which allows the knots for the non-linear fits to be adaptively chosen. It does so using a three-term penalty for each $f_j$ that is a combination of $\ell_1$ and $\ell_2$ norms of different quantities.

\section{Reluctant generalized additive modeling (``RGAM")}\label{sec:rgam}

Our method, which we call \textit{reluctant generalized additive modeling} (RGAM), was inspired by the ideas behind \textit{reluctant interaction modeling} \citep{Yu2019}. We give a brief overview of reluctant interaction modeling here.

\subsection{Reluctant interaction modeling}

\cite{Yu2019} considers the following interaction model:

\begin{equation}
y = \sum_{j=1}^p \beta_j X_j + \sum_{k=1}^{p^2} \gamma_k Z_k + \eps,
\end{equation}

where the $Z_k$'s index the $q = (p^2 + p) / 2$ two-way interaction terms $X_j * X_{j'}$, $1 \leq j \leq j' \leq p$. The key difficulty in fitting such a model is to pick a small but relevant subset of the $q$ interaction terms. Instead of the commonly used \textit{hierarchical principle}, where one includes an interaction only if the corresponding main effects are also included, \cite{Yu2019} propose a new guiding principle:

\setlength{\leftskip}{0.5in}
\setlength{\rightskip}{0.5in}

\textbf{The reluctant interaction selection principle:} \textit{One should prefer a main effect over an interaction if all else is equal.}

\setlength{\leftskip}{0in}
\setlength{\rightskip}{0in}

One way to interpret this principle is to fit the response as well as possible using only the main effects; only after that do we include interaction terms to capture signal in the response which could not be captured by the main effects. \cite{Yu2019}'s full algorithm, called \textit{sprinter}, is detailed in Algorithm \ref{alg:rim}. The authors note that the lasso \citep{Tibshirani1996} in Steps 1 and 3 could be substituted by other regression methods.

\begin{algorithm}
\caption{ \em Reluctant interaction model algorithm}
\label{alg:rim}

\textbf{Require:} Design matrix $\bX \in \mathbb{R}^{n \times p}$, response $y \in \mathbb{R}^n$, screening hyperparameter $\eta > 0$.

\begin{enumerate}
\item Fit the lasso of $y$ on $\bX$ to get coefficients $\hat{\beta}$. Compute the residuals $r = y  - \bX \hat{\beta}$, using the $\lambda$ hyperparameter selected by cross-validation.

\item For the hyperparameter $\eta > 0$, screen the interaction terms based on the residual:

\begin{equation}
\hat{\mathcal{I}}_\eta = \left\{ \ell \in \{ 1, \dots, q\}: \overline{\text{sd}}(r) |\overline{\text{cor}}(Z_\ell, r)| > \eta \right\}.
\end{equation}

\item Fit the lasso of $y$ on $\bX$ and $\bZ_{\hat{\mathcal{I}}_\eta}$.

\end{enumerate}

\end{algorithm}

\subsection{Reluctant generalized additive modeling}

We adapt the reluctant interaction selection principle for GAMs:

\setlength{\leftskip}{0.5in}
\setlength{\rightskip}{0.5in}

\textbf{The reluctant non-linear selection principle:} \textit{One should prefer a linear term over a non-linear term if all else is equal.}

\setlength{\leftskip}{0in}
\setlength{\rightskip}{0in}

To operationalize this principle, we mimic the three-step process of reluctant interaction modeling. In Step 1, we fit the response as well as we can using only the main effects, and in Step 3, we re-fit the response on all the main effects and the additional features which were constructed in Step 2. Where our proposal differs from reluctant interaction modeling is in the construction of the additional features in Step 2. Given a hyperparameter $d \in \mathbb{N}$, for each $j \in \{1, \dots, p\}$, we build a smoothing spline with $d$ degrees of freedom of $r$, the residual from Step 1, on $X_j$. This new feature, which we denote by $\hat{f}_j$, captures signal in the residual using a non-linear relationship with $X_j$. Full details of our proposal, which we call \textit{reluctant generalized additive modeling (RGAM)}, can be found in Algorithm \ref{alg:rgam}. While the lasso in Steps 1 and 3 could be substituted with a different regression method, we recommend it strongly in this context as it performs variable selection, giving us the sparsity we want for the final model.

\begin{algorithm}
\caption{ \em Reluctant generalized additive model algorithm}
\label{alg:rgam}

\textbf{Require:} Design matrix $\bX \in \mathbb{R}^{n \times p}$, response $y \in \mathbb{R}^n$, degrees of freedom hyperparameter $d \in \mathbb{N}$, scaling hyperparameter $\gamma \in [0, 1]$ and a path of lasso hyperparameters $\lambda_1 > \dots > \lambda_m \geq 0$. (Note that the lasso hyperparameters will only be used in Step 3.)

\begin{enumerate}
\item Fit the lasso of $y$ on $\bX$ to get coefficients $\hat{\beta}$. Compute the residuals $r = y  - \bX \hat{\beta}$, using the $\lambda$ hyperparameter selected by cross-validation.

\item For each $j \in \{ 1, \dots, p \}$, fit a smoothing spline with $d$ degrees of freedom of $r$ on $X_j$ which we denote by $\hat{f}_j$. Rescale $\hat{f}_j$ so that $\overline{\text{sd}}(\hat{f}_j) / \text{mean}(\overline{\text{sd}}(X_j)) = \gamma$. Let $\bF \in \mathbb{R}^{n \times p}$ denote the matrix whose columns are the $\hat{f}_j(X_j)$'s.

\item Fit the lasso of $y$ on $\bX$ and $\bF$ for the path of tuning parameters $\lambda_1 > \dots > \lambda_m \geq 0$.

\end{enumerate}

\end{algorithm}

Our proposal is ``reluctant" to include non-linearities in a few ways. First, as with reluctant interaction modeling, the non-linear features are only allowed to model signal which the main effects were unable to capture in Step 1. Second, by rescaling the non-linear features so that their sample standard deviation is just a fraction $\gamma$ compared to that of the main effects, it means that the non-linearity must be strong enough so that its associated coefficient is important enough to survive variable selection by the lasso in Step 3. Third, if we think of the non-linear feature for variable $j$ as a linear combination of spline basis functions for variable $j$, the construction in Step 2 forces this linear combination to be fixed up to a global scaling factor. As such, we expect RGAM to have smaller effective degrees of freedom than methods which allow these coefficients to vary independently of each other. We explore this last point in more detail in Section \ref{sec:df}.

We note that in Algorithm \ref{alg:rgam}, we construct non-linear counterparts for all $p$ features in Step 2. In some settings, we may wish to be conservative in allowing non-linear features into the model. One can tweak Step 2 of Algorithm \ref{alg:rgam} to achieve this outcome. For example, let $\mathcal{A} = \{j: \hat{\beta}_j \neq 0 \}$ be the active set of features after Step 1, i.e. the set of features which were selected by the lasso on the main effects. We could constrain Step 2 to compute non-linear features only for $j \in \mathcal{A}$. This version of RGAM, which we call RGAM\_SEL, weakly assumes a hierarchical principle where we expect a non-linear version of a variable to have an effect only if we expect the variable to have a linear effect in the first place. (The hierarchy is not strictly enforced as it is still possible for the non-linear version of variable $j$ to be selected without variable $j$ itself being selected in Step 3.) As a side benefit, RGAM\_SEL is more computationally efficient than RGAM since Step 3 involves computing the lasso solution for $p + |\mathcal{A}|$, rather than $2p$, features. In our simulations, RGAM\_SEL is 1.5 to 3 times as fast as RGAM, but does not appear to perform as well in terms of test error.

\section{Computation}\label{sec:computation}

We have developed an R package, \texttt{relgam}, which implements our proposal. Steps 1 and 3 of the RGAM algorithm are implemented using the \texttt{cv.glmnet()} and \texttt{glmnet()} functions from the \texttt{glmnet} package \citep{Friedman2010}, while Step 2 is implemented with the \texttt{smooth.spline()} function in the \texttt{stats} package.

The \texttt{rgam()} function (which fits our model) has an option \texttt{init\_nz} which admits a vector of indices. For a given feature index $j$, a non-linear feature is computed for $X_j$ if it appears in \texttt{init\_nz} or if it appears in the active set $\mathcal{A}$. The default behavior is to compute non-linear features for all $p$ variables, i.e. \texttt{init\_nz = 1:p}. To compute non-linear features for just the active set of Step 1, the user can set \texttt{init\_nz} to the empty vector: \texttt{init\_nz = c()}. Hence, this option allows \texttt{rgam()} to compute the solutions to both RGAM (as in Algorithm \ref{alg:rgam}) and RGAM\_SEL (defined in the previous section). The \texttt{init\_nz} option is also useful if the user has some prior information on the relevance of the variables to the response: variables with high relevance can always have non-linear features included in Step 3 by including them in the vector passed to \texttt{init\_nz}.

With respect to hyperparameters, the user can specify the $\lambda$, $\gamma$ and $d$ values using the \texttt{lambda}, \texttt{gamma} and \texttt{df} options respectively. \texttt{rgam()} selects a path of $\lambda$ values in the same manner as \text{glmnet()}; we recommend that the user stick with this choice of $\lambda$ values. The default value for \texttt{gamma} is $0.8$ if \texttt{init\_nz = c()} (i.e. RGAM\_SEL), and is $0.6$ otherwise. We recommend that the user perform cross-validation to pick an optimal value of \texttt{gamma}. In our simulations, we find that values of \texttt{gamma} below $0.5$ usually result in models without any non-linearities. The default value for \texttt{df} is set conservatively at $4$. We recommend using cross-validation to pick an optimal value of \texttt{df} but over just a handful of values as the model is not that sensitive to this choice.

\subsection{Extension to other likelihood functions}

In Section \ref{sec:litreview}, we noted that previous methods for fitting sparse GAMs solve an optimization problem of the form

\begin{equation}
\underset{f_1, \dots, f_p \in \mathcal{F}}{\text{minimize}} \quad \ell (y; f_1, \dots, f_p) + \sum_{j=1}^p J(f_j).
\end{equation}

The optimization is typically performed via an iterative algorithm such as coordinate descent or block coordinate descent. In theory, these methods work with a large class of likelihoods $\ell$. For example, in the case of GLMs, $\ell$ is repeatedly approximated by a quadratic term $\ell'$ and the sum $\ell' + \sum_{j=1}^p J(f_j)$ is minimized until convergence is attained. Implementing this procedure in practice, however, can be tedious. This is also the case for extending the methods to Cox regression models, where $\ell$ is the partial likelihood of the data.

Unlike previous methods, the RGAM algorithm (Algorithm \ref{alg:rgam}) can be extended easily to different likelihood functions. As long as the likelihood can be handled by the \texttt{glmnet()} function in the \texttt{glmnet} package through its \texttt{family} option, Steps 1 and 3 of the RGAM algorithm can be adapted immediately by passing that \texttt{family} option to \texttt{glmnet()}. The only remaining work is to compute the analog of the residual in Step 2, which is much easier than solving a modified optimization problem. At the time of writing, apart from the Gaussian likelihood for continuous responses, we have working software implementing the logistic, Poisson and Cox regression models for binary, count and survival data respectively.

\subsection{Timing comparison}

Since RGAM uses $k$-fold cross-validation of the lasso in Step 1 (our software sets $k = 5$ as a default) and the lasso on all the linear and non-linear features in Step 2, we expect RGAM to take at least $k+1$ times as much time as \texttt{glmnet()}, which implements the lasso. Nevertheless, we find that RGAM is very competitive with other sparse additive modeling techniques in terms of computational efficiency. 

\begin{figure}[!htbp]
\centering
\includegraphics[width=6in]{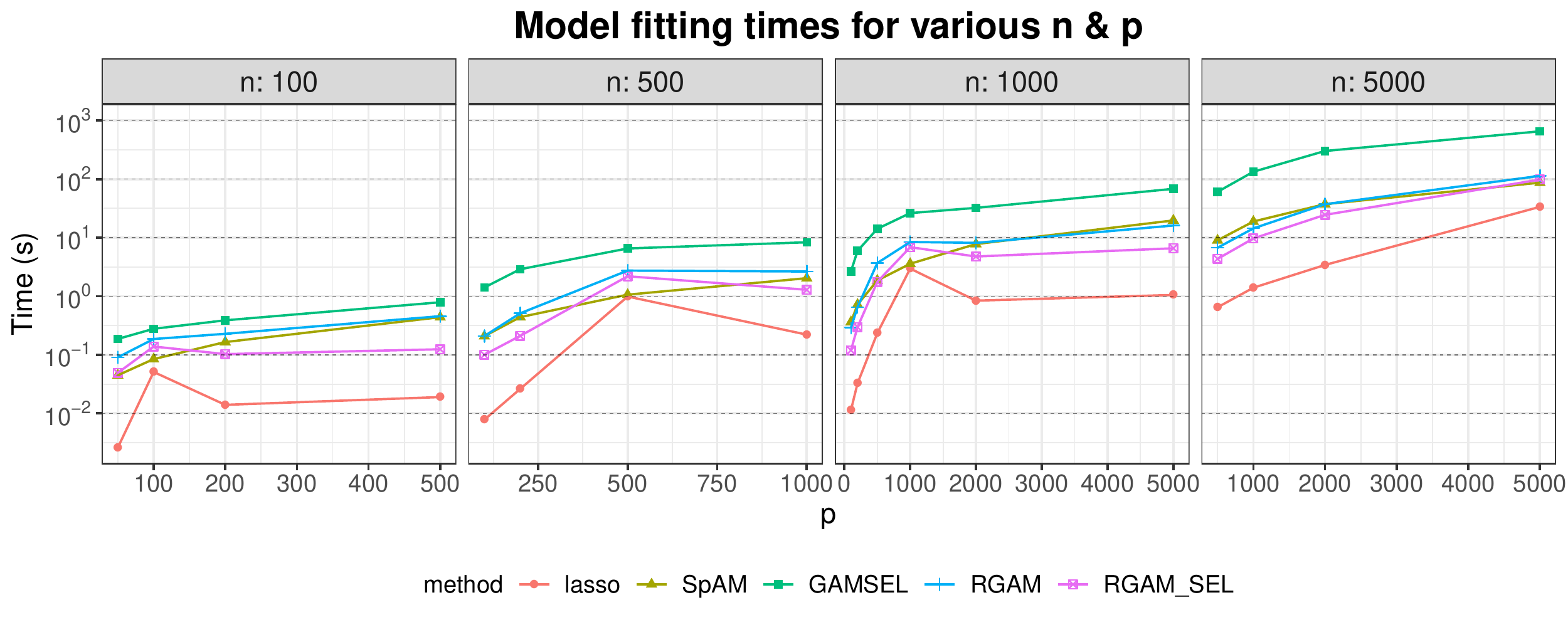}
\caption[fig:timing1]{\em Model fitting times (in seconds) for different combinations of $n$ (number of observations) and $p$ (number of features). Each point is the mean of 5 simulation runs. Note that the $y$-axis is on a logarithmic scale.}
\label{fig:timing1}
\end{figure}

\begin{figure}[!htbp]
\centering
\includegraphics[width=6in]{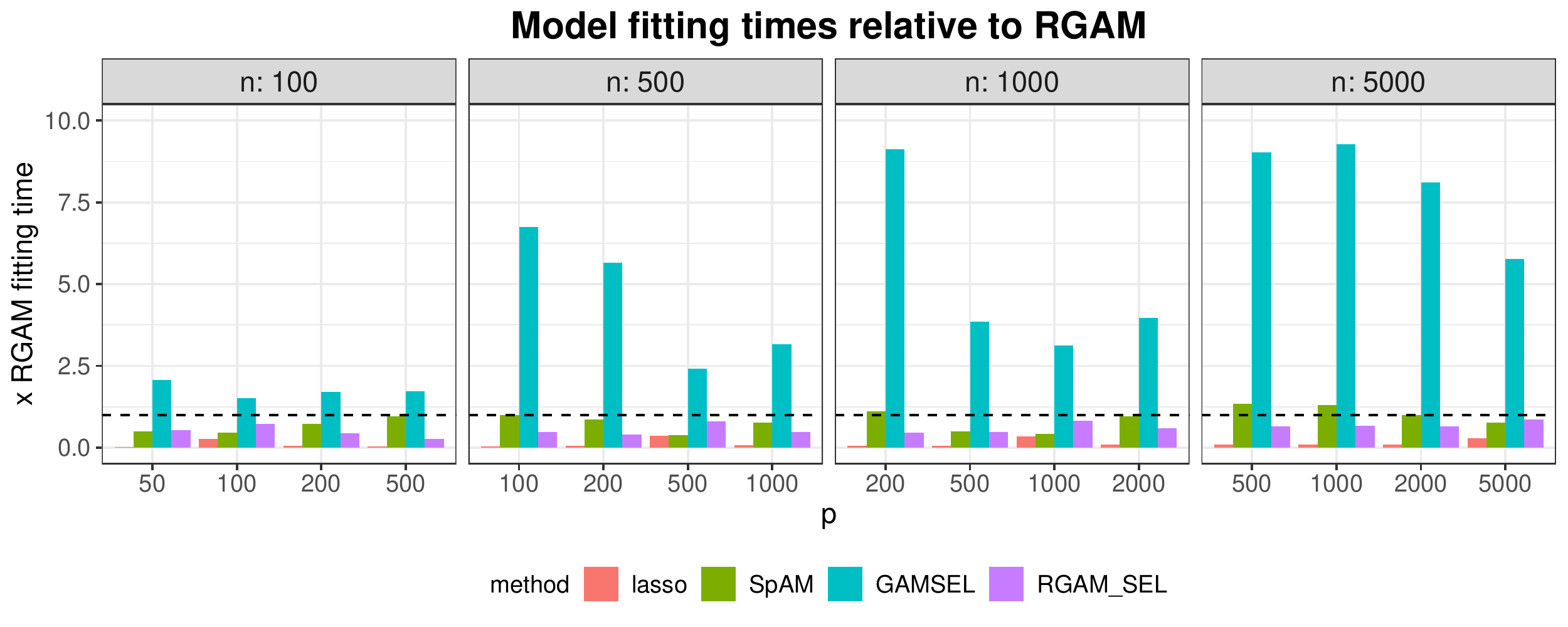}
\caption[fig:timing2]{\em Model fitting times for different combinations of $n$ (number of observations) and $p$ (number of features), expressed as a multiple of the model fitting time for RGAM. Each bar is the mean of 5 simulation runs. The dotted horizontal line indicates the fitting time for RGAM.}
\label{fig:timing2}
\end{figure}

Figure \ref{fig:timing1} presents the absolute time taken to fit the models for various values of $n$ (number of observations) and $p$ (number of features), while Figure \ref{fig:timing2} presents these times relative to that for RGAM. (Recall that RGAM refers to procedure in Algorithm \ref{alg:rgam} while RGAM\_SEL refers to the procedure where non-linear features are only constructed for features in the active set from Step 1.) Each point or bar is the mean of 5 simulation runs. We see that GAMSEL takes anywhere from 1.5 to 8 times as long as RGAM for model fitting, with the factors being bigger for larger simulation settings. The computation burden for SpAM and RGAM are comparable, while RGAM\_SEL can often be faster than these two methods.

\section{Simulated and real data examples}\label{sec:examples}

We conducted an extensive simulation study comparing our method with the lasso and GAMSEL\footnote{We did not compare RGAM with SPLAM \citep{Lou2016} and SPLAT \citep{Petersen2019} as we did not find R packages implementing these methods at the time of writing.}. The full simulation results can be found in Appendix \ref{app:simstudy}; we present some illustrative snippets here. For RGAM, we considered two versions: one where non-linear features were constructed for all $p$ main effects, and one where non-linear features were constructed for only the main effects in the active set from Step 1. 

In all the simulations that follow, the feature values $X_{ij}$ are independent draws from the $\text{Unif}[-1,1]$ distribution. The response is $y_i = \mu_i + \eps_i = f(X_{i1}, \dots, X_{ip}) + \eps_i$, where $f$ is a function that depends on the simulation and the $\eps_i$'s are independent $\mathcal{N}(0, \sigma^2)$ draws. The data generating process for the signal is such that the linear and non-linear components are orthogonal. $\sigma^2$ is set so that the data has the desired signal-to-noise ratio (SNR).

For all methods, 5-fold cross-validation was performed to select the hyperparameter $\lambda$ only: default values were used for all other hyperparameters. Each boxplot is the result of 30 simulation runs. The test error metric is mean-squared error where the target is the true signal value, i.e. $\mathbb{E}[(\hat{y}_{test} - \mu_{test})^2]$ instead of $\mathbb{E}[(\hat{y}_{test} - y_{test})^2]$. (With this test error metric, the oracle which knows the data generating model would have a test error of $0$.) Test error is estimated using 5,000 test points.

\subsection{Simulation 1: Hierarchical setting}

In this setting, we have 100 observations and 200 features with the signal being a function of the first five features. The setting is ``hierarchical" in the sense that all the features that make up the non-linear component of the signal also have a linear component. More explicitly, the signal is $f(X_1, \dots, X_p) = \sum_{j=1}^5 [X_j + \frac{2}{3} (3X_j^2 - 1)]$. The SNR of the overall response is set to 2, with roughly equal SNR in each of the non-linear and linear components.

\begin{figure}[!htbp]
\centering
\includegraphics[width=2.0in]{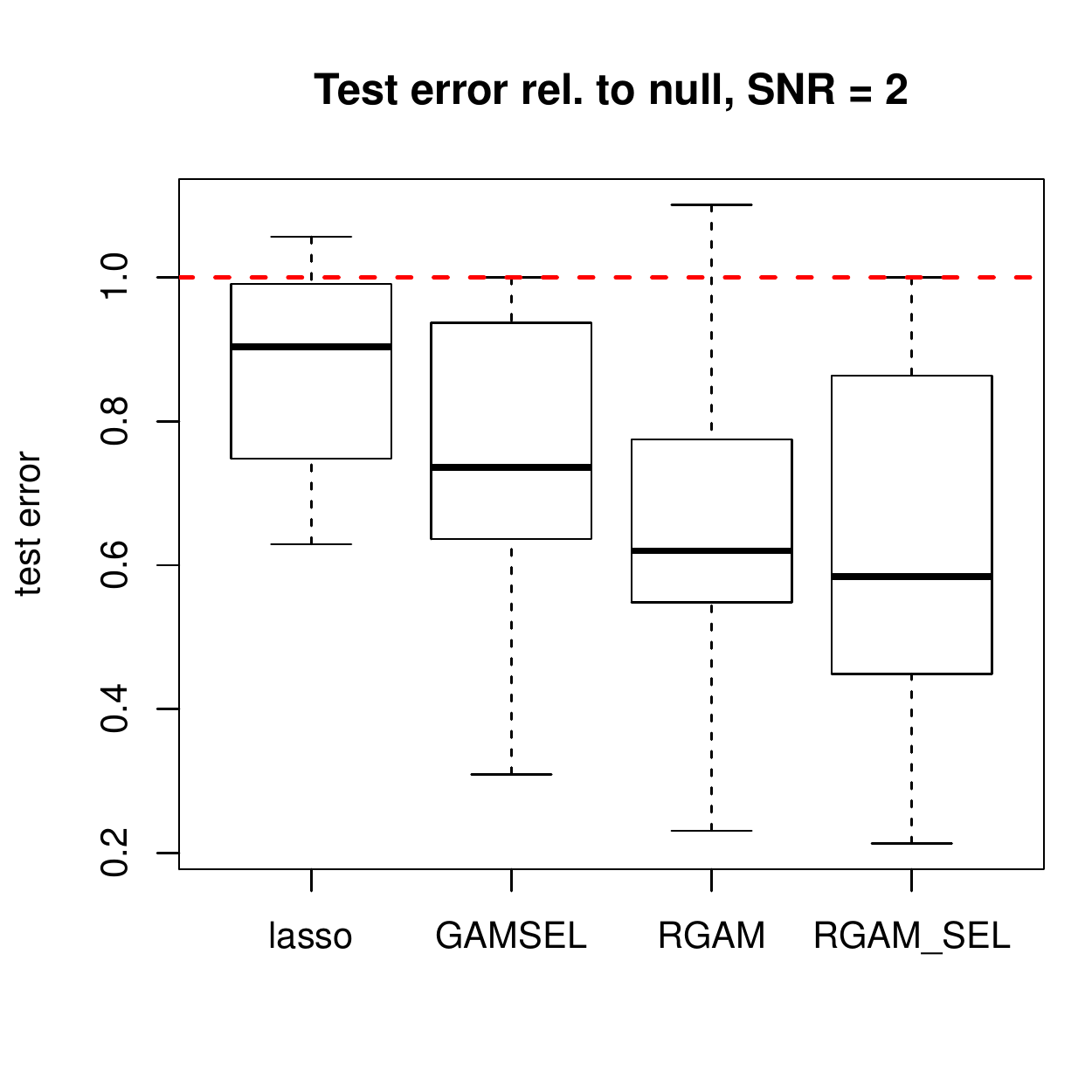}\includegraphics[width=2.0in]{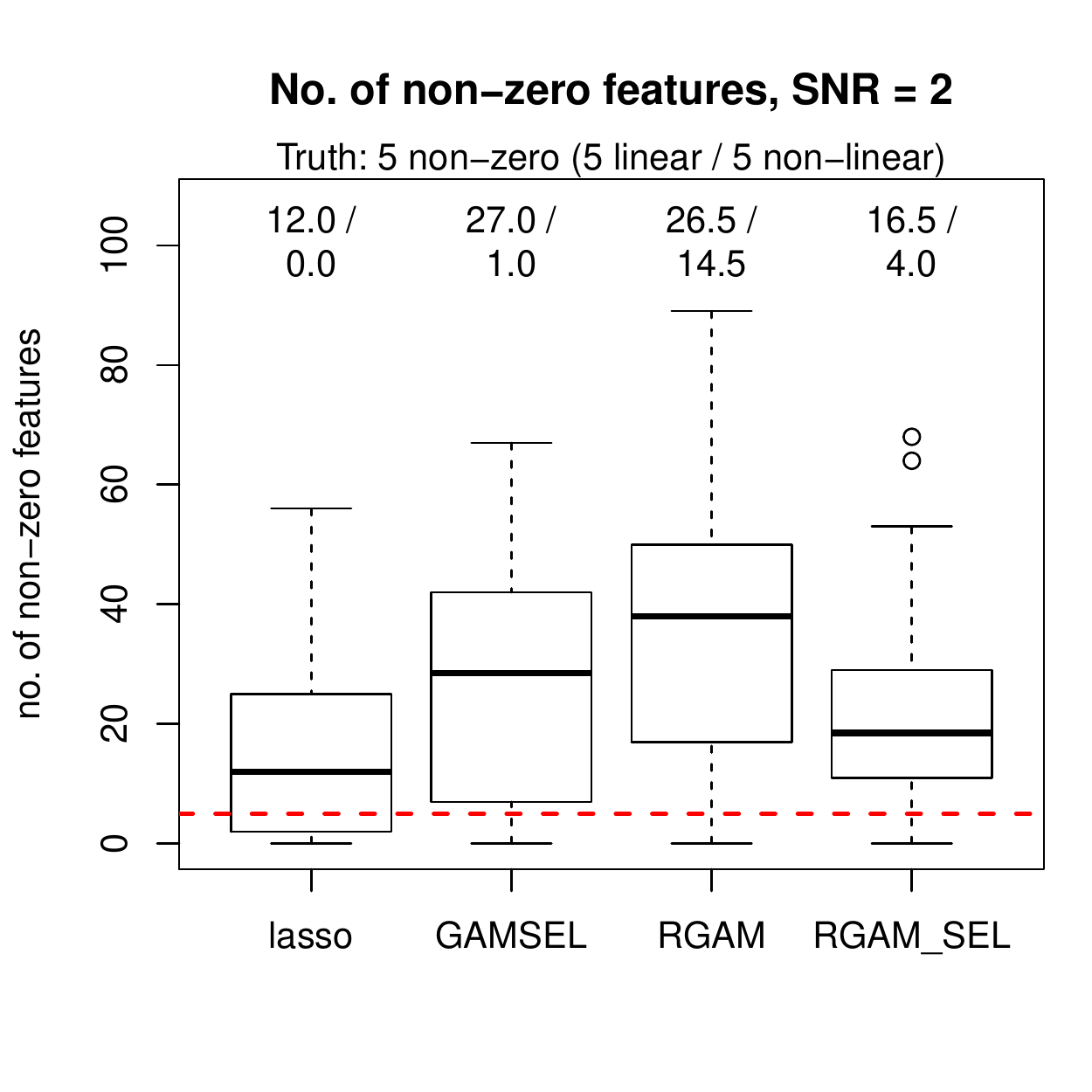}\includegraphics[width=2.0in]{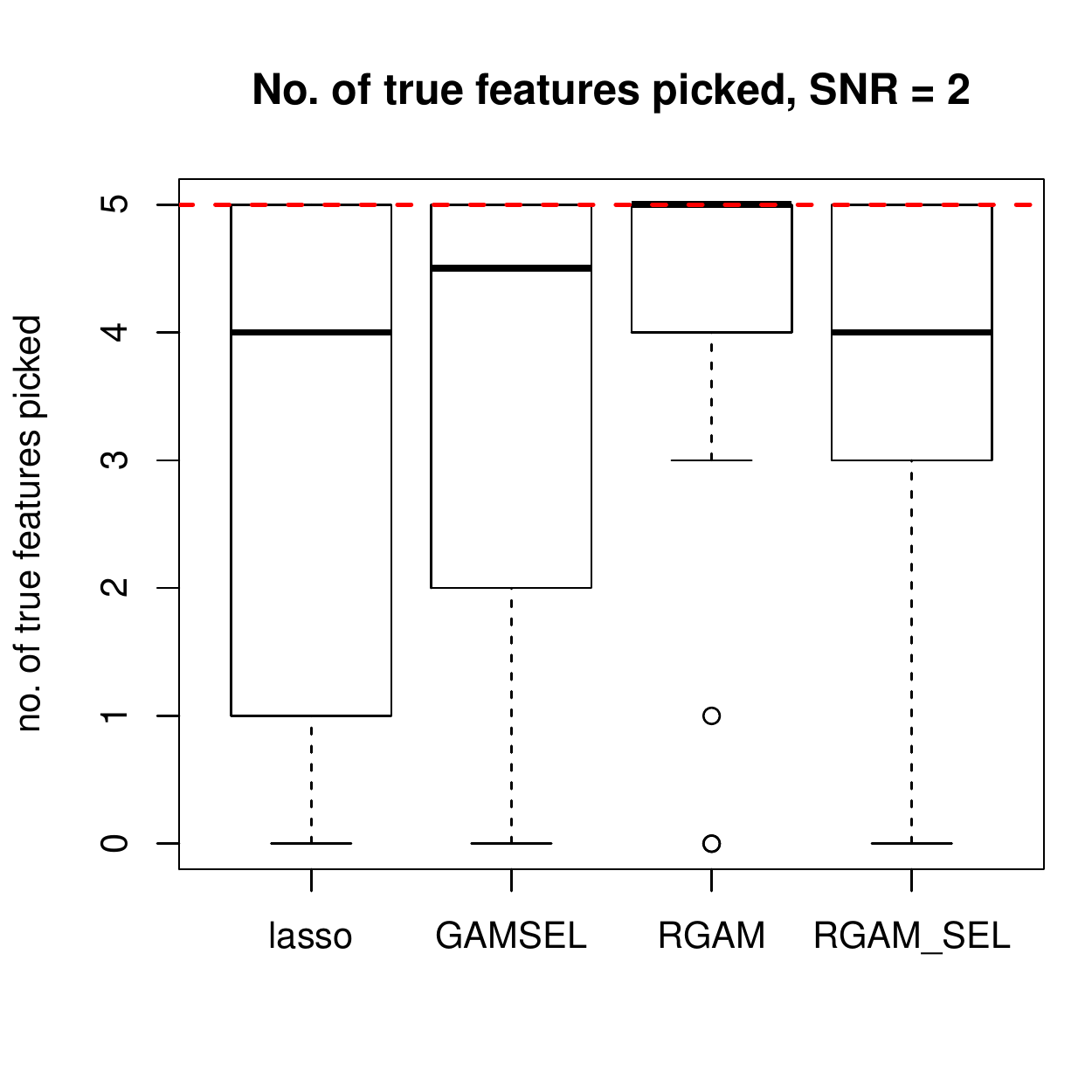}
\caption[fig:sim1]{\em Hierarchical setting: $n = 100$, $p = 200$. The left-most plot presents test error as a fraction of the null model's test error. The middle plot row presents the number of features each model selected, with the two numbers on top of the boxplots being the median number of linear components and non-linear components selected. The right-most plot presents the number of true features each method selected, with the total number of true features indicated by the dotted red line. RGAM\_SEL performs best in test error due to the hierarchical nature of the non-linearities.}
\label{fig:sim1}
\end{figure}

The results are shown in Figure \ref{fig:sim1}. Both versions of RGAM outperform the other methods, with RGAM\_SEL being the best. This makes intuitive sense: since the signal is hierarchical, the main effects selected by RGAM\_SEL for Step 2 will be smaller than $p$, yet will very likely include the true main effects. Thus, in Step 3, the true non-linear features only have to compete with a smaller set of features to enter the final model as opposed to RGAM, where they have to compete with all $p$ non-linear features.

\subsection{Simulation 2: Signal is purely non-linear}

In this setting, we again have 100 observations and 200 features, with the signal being a function of the first five features. However, the signal, $f(X_1, \dots, X_p) = \sum_{j=1}^5 2(5X_j^3 - 3X_j)$,  only depends on non-linear functions of the features which are orthogonal to the feature itself. Since the $X_j$'s are drawn from a $\text{Unif}[-1,1]$ distribution, we have $\text{Cov}(5X_j^3 - 3X_j, X_j) = 0$. The SNR of the response is set to 2.

\begin{figure}[!htbp]
\centering
\includegraphics[width=2.0in]{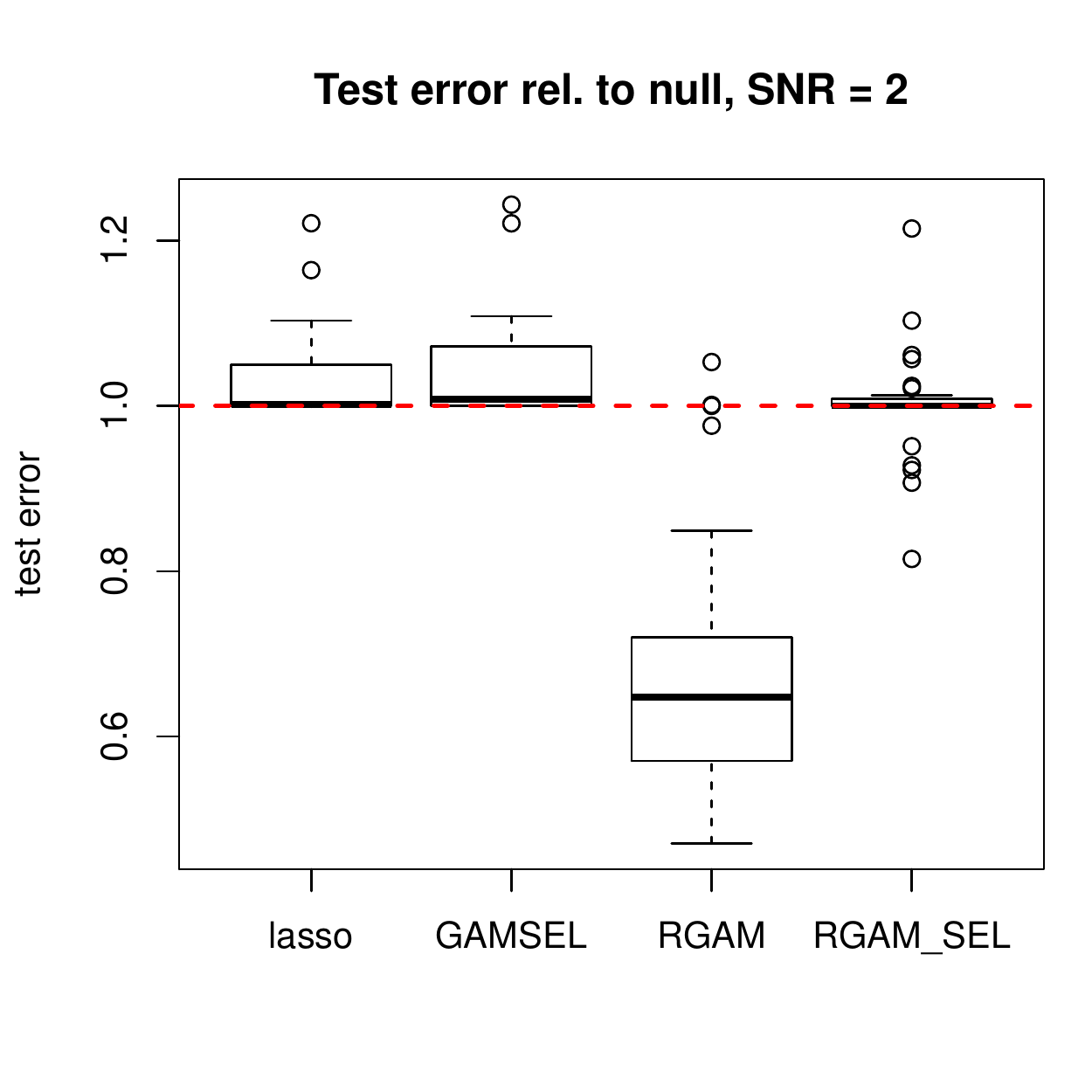}\includegraphics[width=2.0in]{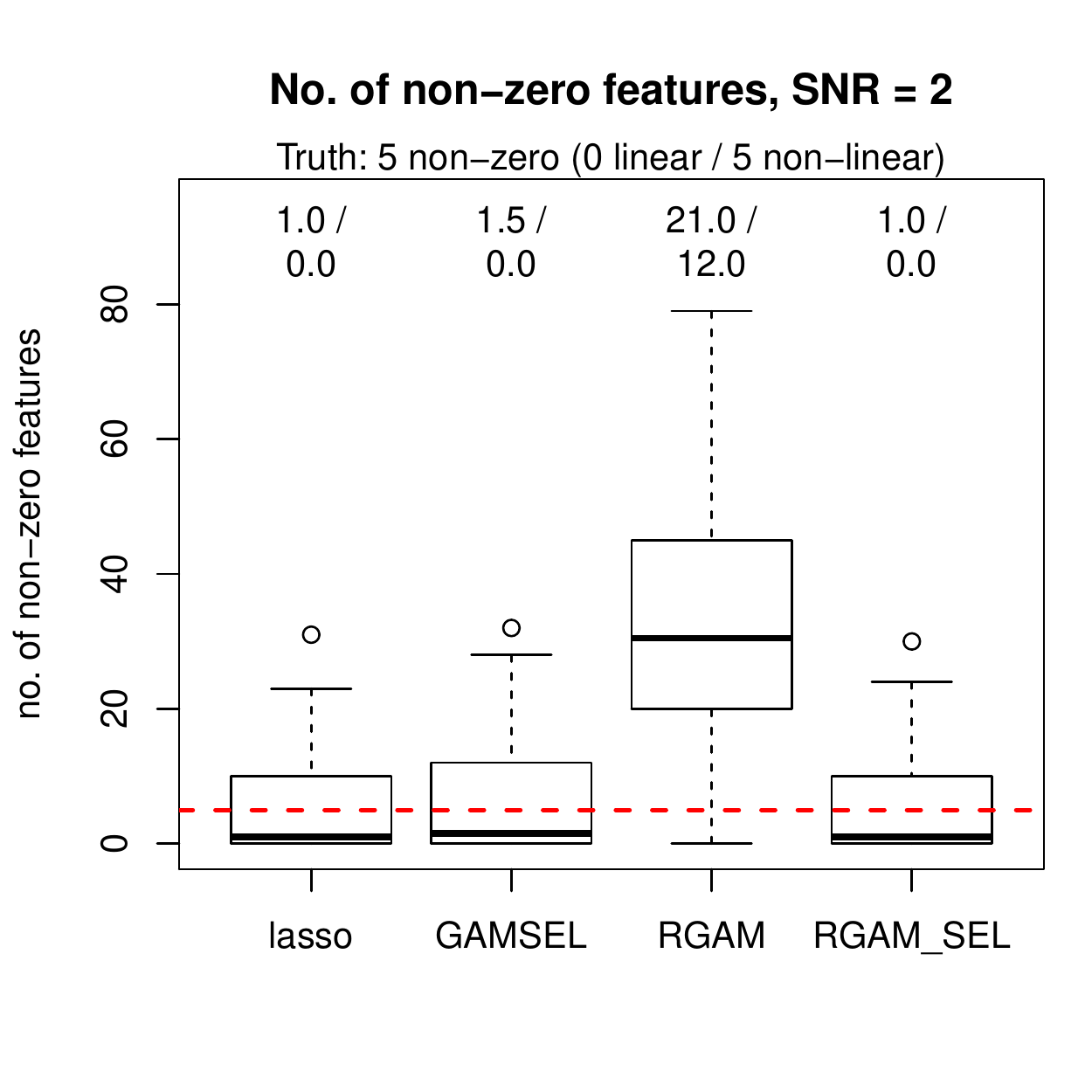}\includegraphics[width=2.0in]{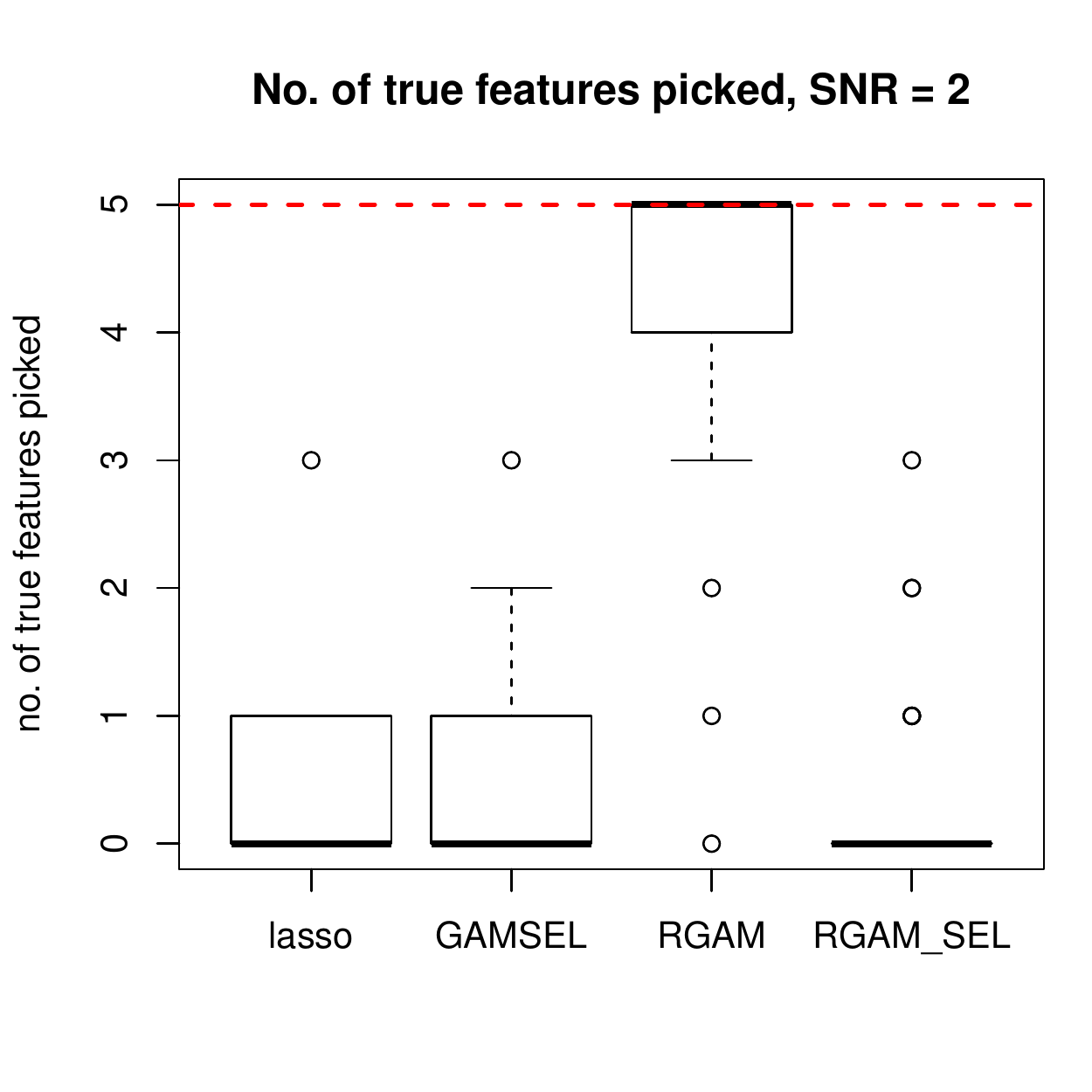}
\caption[fig:sim2]{\em Fully non-linear setting: $n = 100$, $p = 200$, $SNR = 2$. The left-most plot presents test error as a fraction of the null model's test error. The middle plot row presents the number of features each model selected, with the two numbers on top of the boxplots being the median number of linear components and non-linear components selected. The right-most plot presents the number of true features each method selected, with the total number of true features indicated by the dotted red line. Only RGAM is able to outperform the null model as there are no linear effects in the true signal.}
\label{fig:sim2}
\end{figure}

The results are shown in Figure \ref{fig:sim2}. Only RGAM is able to outperform the null model, i.e. mean of the responses in the training dataset. This is expected for the lasso since it only captures linear effects. GAMSEL can include a non-linear effect in a particular variable only if its corresponding linear effect is included too, and thus does not perform well either. Without linear effects in the signal, Step 1 of the RGAM algorithm cannot pick out the true features reliably. RGAM\_SEL thus cannot reliably pick out the correct non-linear features for Step 3 of the algorithm. RGAM circumvents this problem by constructing non-linear features for all $p$ features.

\subsection{Simulation 3: Large setting, hierarchical and non-hierarchical non-linear signals present}

In this setting, we have 1,000 observations and 500 features. The signal is $f(X_1, \dots, X_p) = \sum_{j=1}^{20} X_j + \sum_{j=1}^{20} \frac{3}{4}(5X_j^3 - 3X_j) + \sum_{j=21}^{28} (3X_j^2 - 1)$. The first 20 features have both linear and non-linear components featuring in the signal, while the next 8 features only have non-linear components in the signal. The SNR of the overall response is set to 1, with each of the three sums in the expression above having roughly equal SNR.

\begin{figure}[!htbp]
\centering
\includegraphics[width=2.0in]{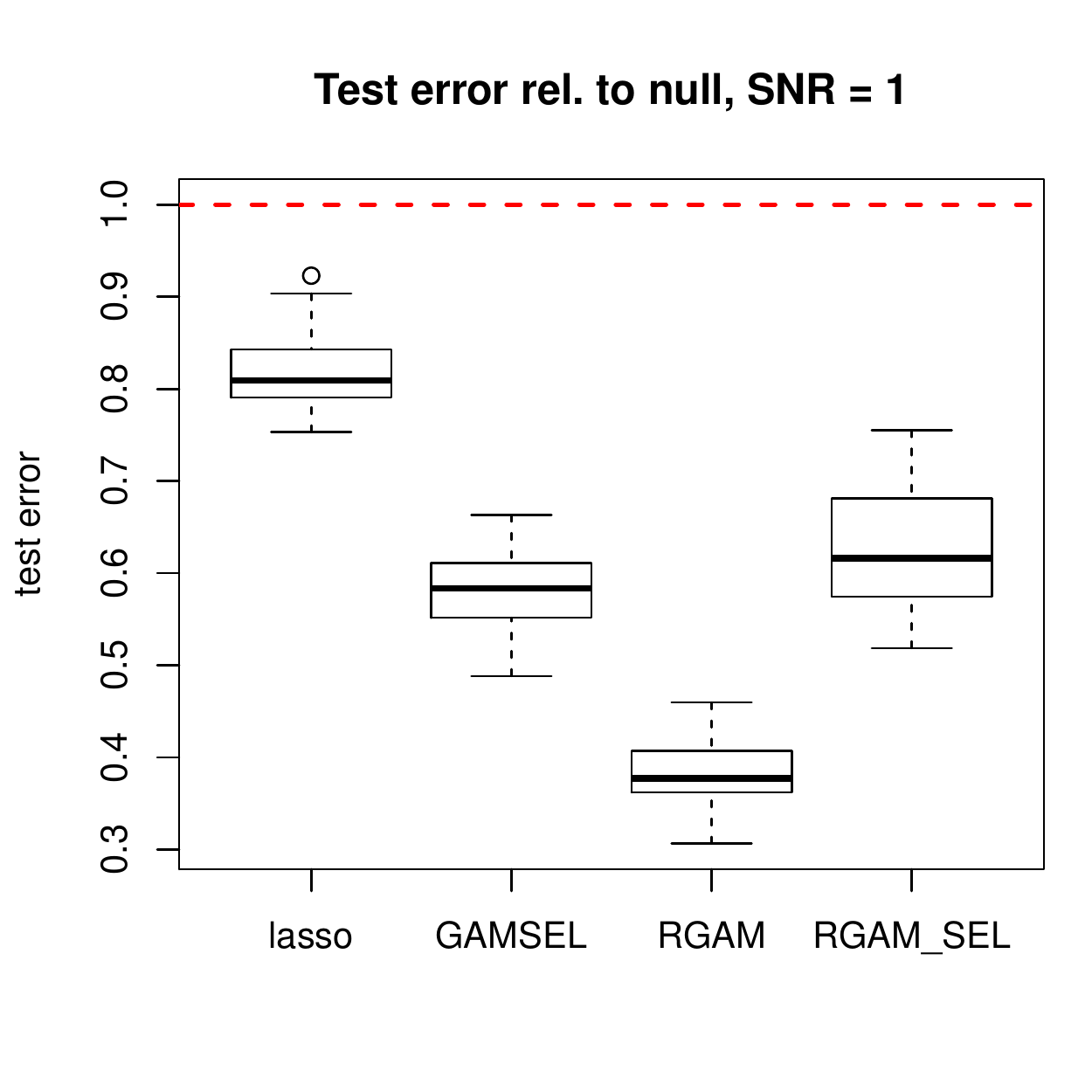}\includegraphics[width=2.0in]{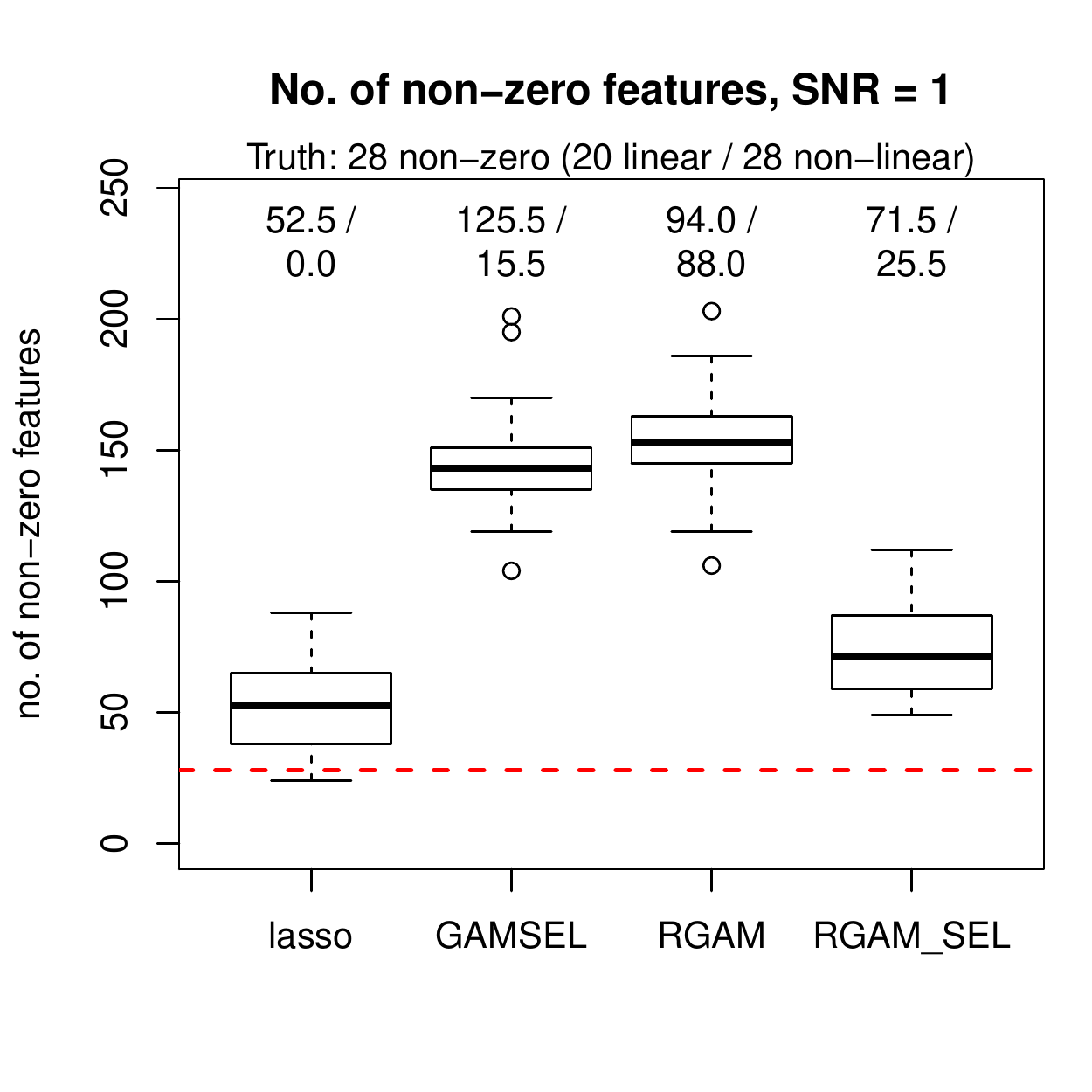}\includegraphics[width=2.0in]{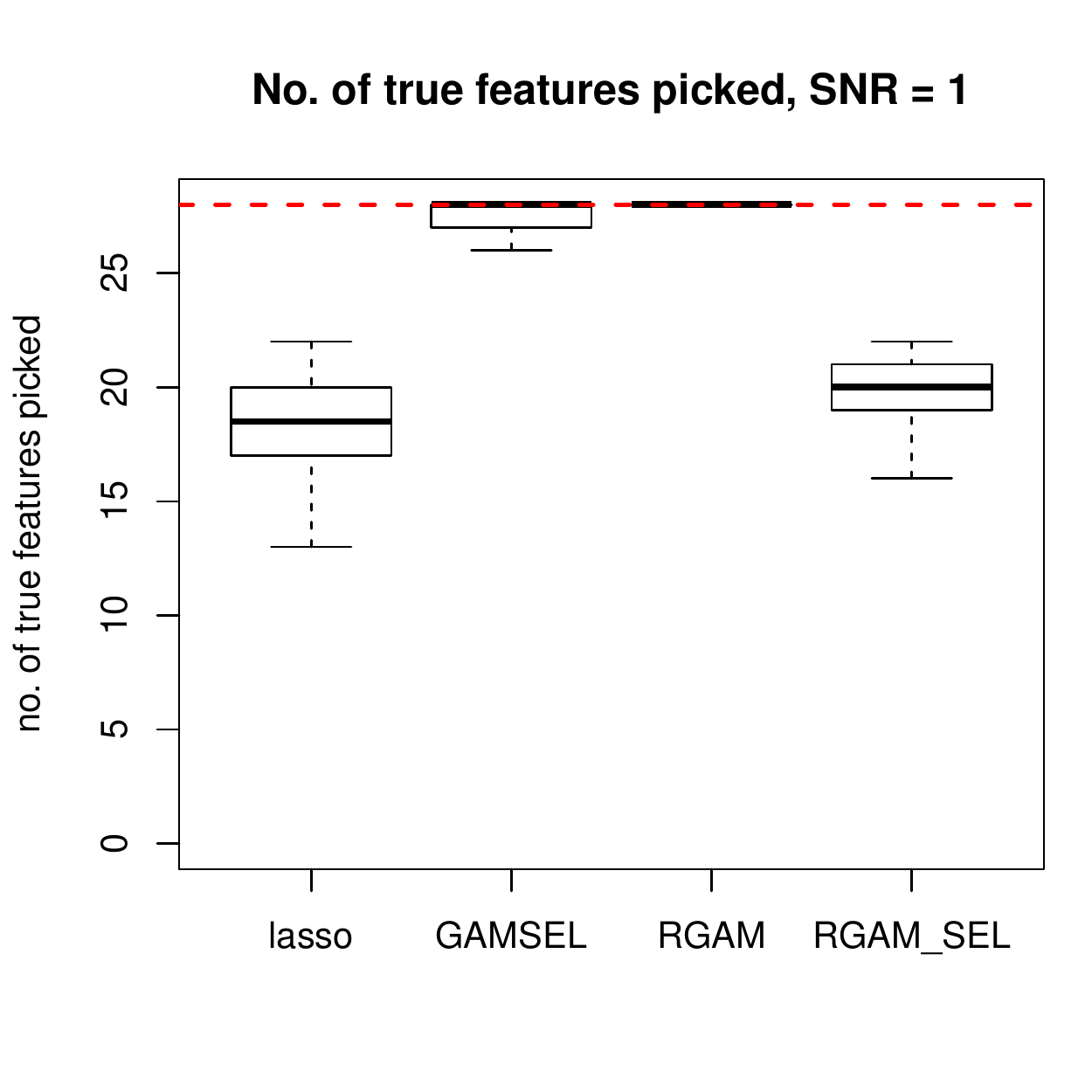}
\caption[fig:sim3]{\em Large setting, with hierarchical and non-hierarchical non-linear signals: $n = 1,000$, $p = 500$, $SNR = 1$. The left-most plot presents test error as a fraction of the null model's test error. The middle plot row presents the number of features each model selected, with the two numbers on top of the boxplots being the median number of linear components and non-linear components selected. The right-most plot presents the number of true features each method selected, with the total number of true features indicated by the dotted red line. RGAM performs best in terms of test error; RGAM\_SEL is comparable to GAMSEL but selects much fewer linear and non-linear components.}
\label{fig:sim3}
\end{figure}

The results are shown in Figure \ref{fig:sim3}. In this setting RGAM clearly outperforms all the other methods, and performs well despite low SNR. This is partially due to its ability to pick out the non-linear features that do not have a corresponding linear component in the signal. RGAM\_SEL exhibits roughly the same test error performance as GAMSEL, but selects much fewer linear and non-linear components in its predictive model.

\subsection{Prostate cancer dataset}

We apply RGAM to a microarray dataset from a prostate cancer study carried out by \cite{Singh2002}, and which was analyzed in \cite{Efron2012}. The data consists of expression levels for $6,033$ genes for $102$ men. $50$ men were normal control subjects while the remaining $52$ men were prostate cancer patients. The goal is to predict which subjects had prostate cancer based on the gene expression levels.

\begin{figure}[!htbp]
\centering
\includegraphics[width=3.0in]{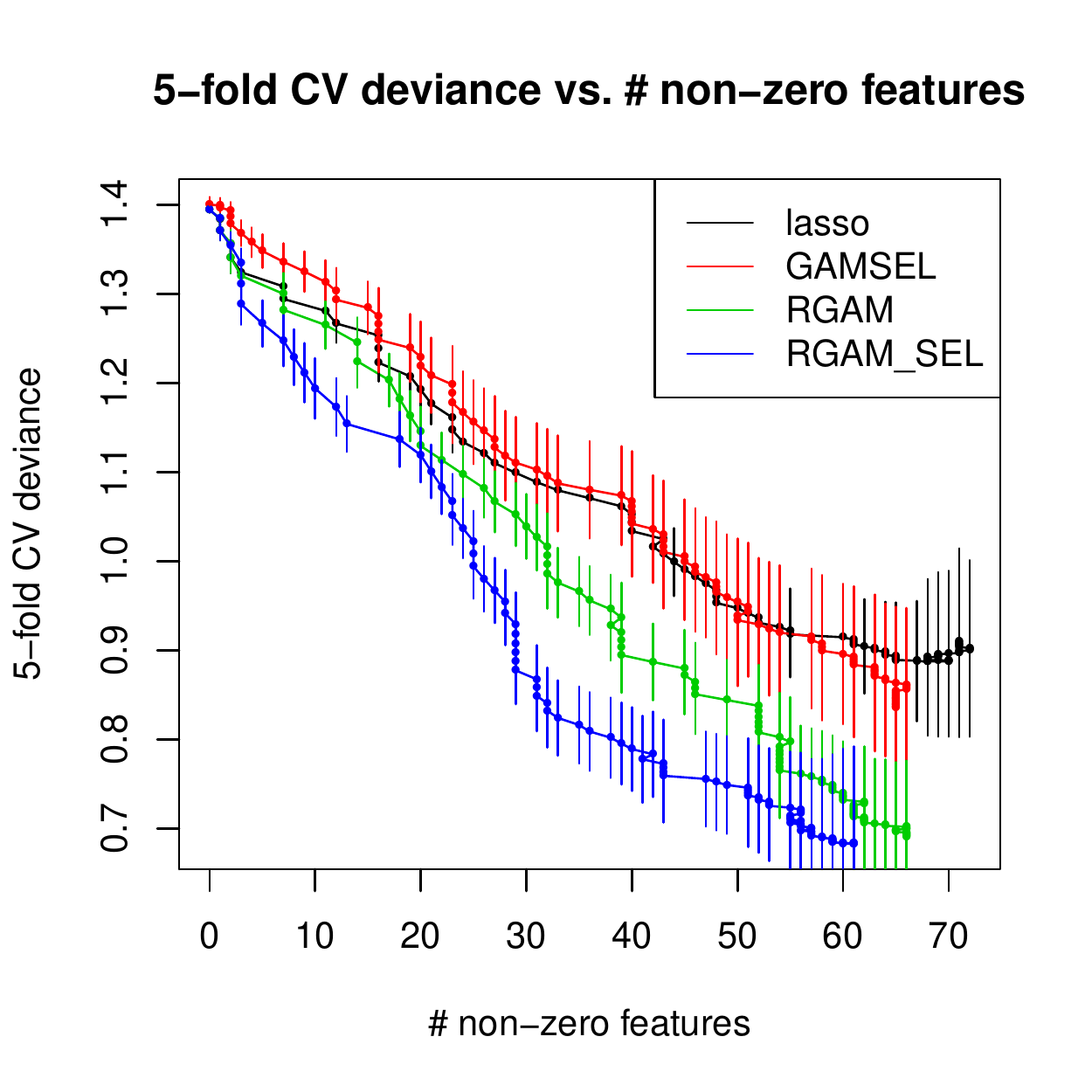}\includegraphics[width=3.0in]{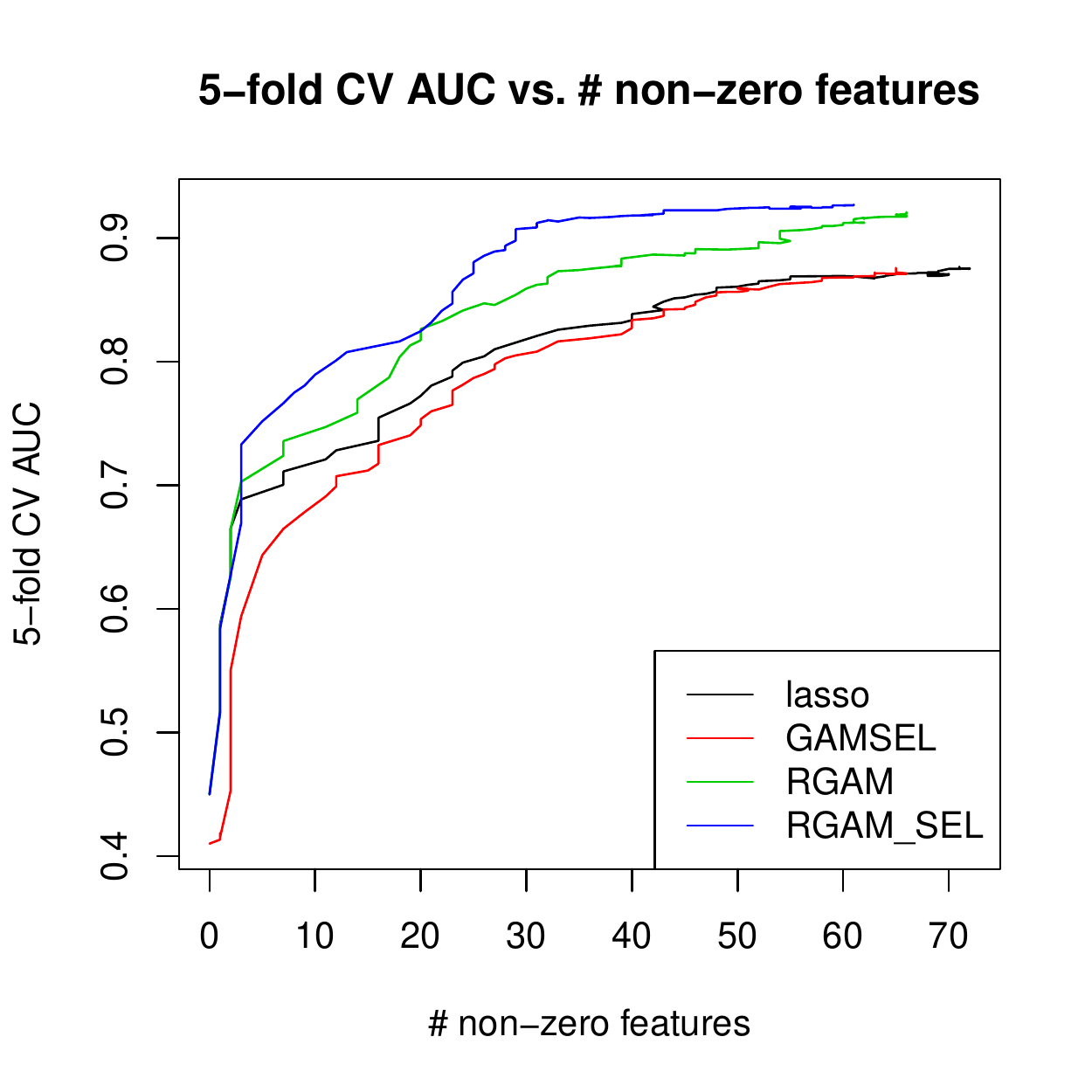}
\caption[fig:prostate]{\em 5-fold cross-validation results for the prostate cancer dataset. The left panel shows the cross-validated deviance (with error bars showing $\pm 1$ standard deviation) while the right panel shows the cross-validated area under the curve (AUC). The $x$-axis represents the number of features selected at each value in the lambda path.}
\label{fig:prostate}
\end{figure}

We compare RGAM's cross-validated performance with that of the lasso and GAMSEL. Each of these methods were run on a path of $\lambda$ values, with other hyperparameters set to their default values. The fitting times for GAMSEL, RGAM and RGAM\_SEL were $72$, $32$ and $3.5$ seconds respectively. The results are shown in Figure \ref{fig:prostate}. For the same model size, both versions of RGAM outperform the lasso and GAMSEL in terms of both cross-validated deviance and cross-validated area under the curve (AUC).

\section{Degrees of freedom}\label{sec:df}

When introducing RGAM, we claimed that, intuitively, the way in which the non-linear features are constructed in Step 2 gives the non-linear components less degrees of freedom than giving Step 3 $d$ spline basis functions for each $X_j$. The degrees of freedom measures the flexibility of the fit: the larger the degrees of freedom, the more closely the fit matches the response values. We explore this claim in more detail here.

Given a vector of response values $y$ with corresponding fits $\hat{y}$, \cite{Efron1986} defines the degrees of freedom as
\begin{equation}
\text{df}(\hat{y}) = \frac{\sum_i \text{Cov}(y_i, \hat{y}_i)}{\sigma^2}.
\end{equation}

We can estimate this quantity via Monte Carlo simulation. Consider the model
\begin{equation}\label{eqn:sim-model}
y^* = \mu + \sigma z,
\end{equation}

where $z \sim \mathcal{N}(0,1)$ and $\mu$ is considered fixed. For $b = 1, \dots, B$, we generate a new response vector $y^{*b}$ according to \eqref{eqn:sim-model}. We fit a predictive model to this data, generating predictions $\hat{y}^{*b}$. This gives us the Monte Carlo estimate
\begin{align*}
\text{df} &\approx \sum_{i=1}^n \widehat{\text{Cov}}(\hat{y}_i^*, y_i^*) / \sigma^2, \\
\widehat{\text{Cov}}(\hat{y}_i^*, y_i^*) &= \frac{1}{B} \sum_{b=1}^B [\hat{y}_i^{*b} - a_i] [y_i^{*b} - \mu_i],
\end{align*}
where the $a_i$'s can be any fixed known constants (usually taken to be 0).

\begin{figure}[!htbp]
\centering
\includegraphics[width=5.5in]{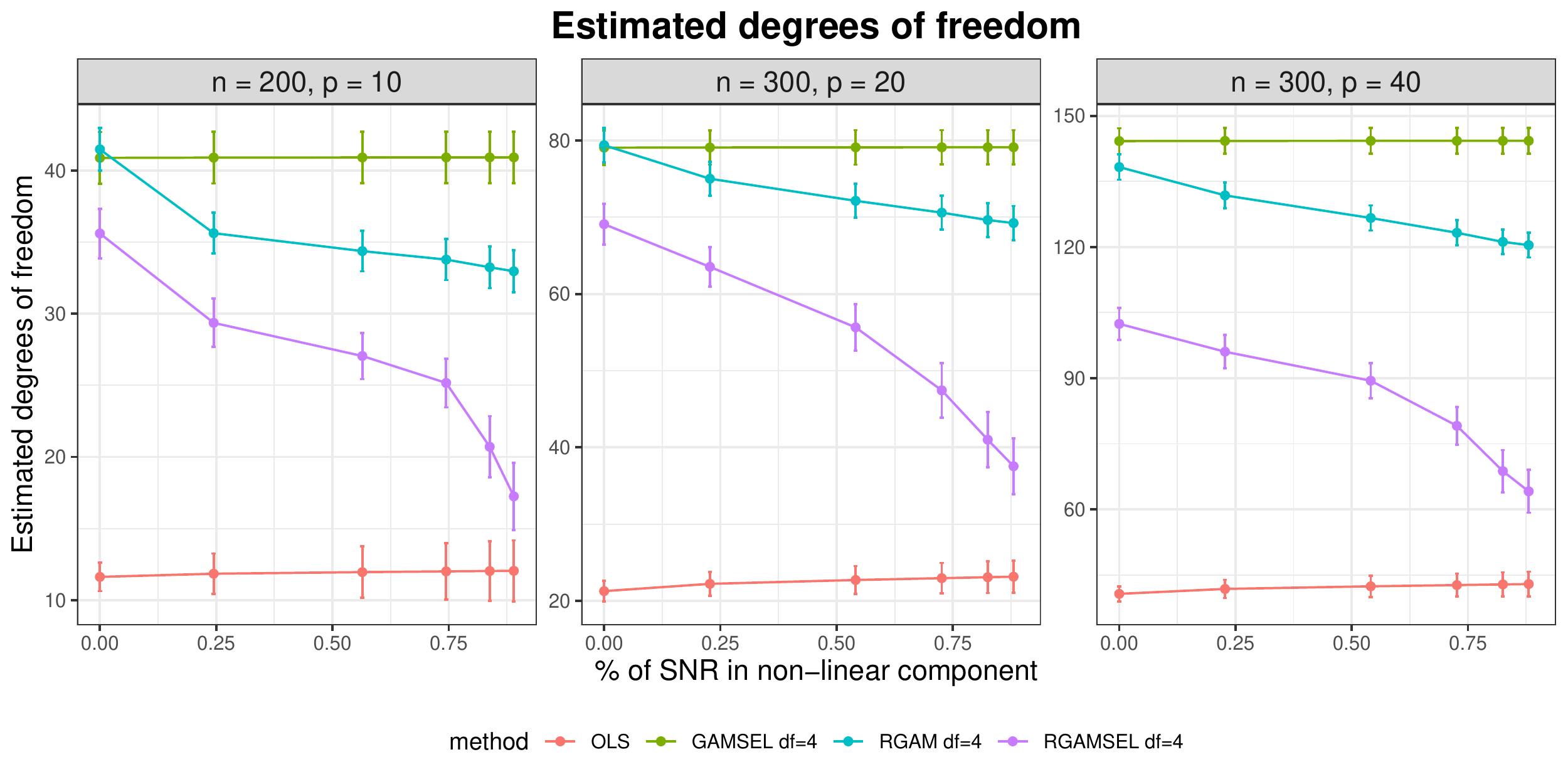}
\caption[fig:df]{\em Estimated degrees of freedom across three different $(n, p)$ settings. The points are the Monte Carlo estimates from $B = 100$ simulation runs, and the error bars are $\pm 1$ standard deviation for the sample mean. For unpenalized GAMSEL, each feature was given 6 basis functions. Unpenalized RGAM has smaller degrees of freedom than unpenalized GAMSEL. As the non-linear component of the true signal increases (in terms of SNR), unpenalized RGAM's degrees of freedom appears to decrease.}
\label{fig:df}
\end{figure}

We compare the unpenalized versions of RGAM and GAMSEL, i.e. setting the hyperparameter $\lambda = 0$. Figure \ref{fig:df} shows the estimated degrees of freedom for the unpenalized procedures (with degrees of freedom $d = 4$) and OLS of $y$ on the $X_j$'s for three different settings. (For GAMSEL, each feature was given 6 basis functions; the default value for \texttt{gamsel()} is 10.) As predicted in theory, OLS on the $X_j$'s (with intercept) has $p + 1$ degrees of freedom. The degrees of freedom for unpenalized GAMSEL seems to be relatively constant at roughly $p$ times the value of the degrees of freedom hyperparameter, even as the non-linear component's contribution to the SNR of the signal changes. Unpenalized RGAM has roughly the same degrees of freedom when the true underlying signal is completely linear. As the proportion of SNR in the true underlying signal coming from the non-linear component increases, RGAM's degrees of freedom decreases. We currently do not have a good explanation for this phenomenon. The degrees of freedom for unpenalized RGAM\_SEL is substantially lower than both that of unpenalized GAMSEL and unpenalized RGAM.

\section{Discussion}\label{sec:discussion}

In this paper we introduced reluctant generalized additive modeling (RGAM), a three-step algorithm for fitting sparse GAMs. The model's prediction is allowed to vary linearly or non-linearly with each input variable. RGAM is guided by the reluctant non-linear selection principle, preferring linear effects over non-linear effects, only including the latter if they add to predictive performance. Unlike existing methods for sparse GAMs, RGAM can be extended easily to binary, count and survival data.

The three-step framework of Algorithm \ref{alg:rgam} is extremely flexible. As previously noted, one may replace the lasso method in Steps 1 and 3 with a regression method of one's choosing. We note that Step 2 is highly customizable as well: we seek to model the residual in this step, and we can use any method to do so. In our software implementation we model the residual with a cubic smoothing spline for each $X_j$; other spline methods could be used. If we believe that there are discontinuities in the relationship between the response $y$ and $X_j$, we could model the residual as a piece-wise constant function of $X_j$. There are even more possibilities if we are willing to allow interactions between different input variables. For example, we could combine RGAM with reluctant interaction modeling by adding the interaction terms chosen by reluctant interaction modeling in Step 3 of Algorithm \ref{alg:rgam}. Another possibility is to fit the residual to random trees, much like a random forest; Step 3 then selects the most appropriate linear effects and trees for the final model. We leave the implementation and exploration of these more complex methods for future work.

An R language package {\tt relgam} which implements RGAM is available on the CRAN repository.

\medskip

{\bf Acknowledgements:} We would like to thank Jacob Bien and Hugo Yu for helpful comments. Robert Tibshirani was supported by NIH grant 5R01 EB001988-16 and NSF grant 19 DMS1208164.

\appendix

\section{Full details of simulation study}\label{app:simstudy}

In all the simulations that follow, the feature values $X_{ij}$ are independent draws from the $\text{Unif}[-1,1]$ distribution. The response is $y_i = \mu_i + \eps_i = f(X_{i1}, \dots, X_{ip}) + \eps_i$, where $f$ is a function that depends on the simulation and the $\eps_i$'s are independent $\mathcal{N}(0, \sigma^2)$ draws. The data generating process for the signal is such that the linear and non-linear components are orthogonal. $\sigma^2$ is set so that the data has the desired signal-to-noise ratio (SNR).

We compare the following methods across a range of settings:

\begin{enumerate}
\item The null model, i.e. mean of the training responses,
\item The lasso \citep{Tibshirani1996},
\item Generalized additive model selection (GAMSEL) \citep{Chouldechova2015},
\item Reluctant generalized additive modeling (RGAM), where non-linear features are constructed for all $p$ main effects, and
\item RGAM where non-linear features are only constructed for the main effects which are in the active set after Step 1 of Algorithm \ref{alg:rgam} (denoted RGAM\_SEL).
\end{enumerate}

For all methods, 5-fold cross-validation was performed to select the hyperparameter $\lambda$ only: default values were used for all other hyperparameters. Each boxplot is the result of 30 simulation runs. The test error metric is mean-squared error where the target is the true signal value, i.e. $\mathbb{E}[(\hat{y}_{test} - \mu_{test})^2]$ instead of $\mathbb{E}[(\hat{y}_{test} - y_{test})^2]$. (With this test error metric, the oracle which knows the data generating model would have a test error of $0$.) Test error is estimated using 5,000 test points.

For each simulation setting, the results are presented in a $3 \times 3$ panel. Each row of the panel corresponds to the one of three SNR values: 1, 2 or 5. In each row, the left-most plot presents test error as a fraction of the test error achieved by the null model (denoted by the dotted red line). The middle plot in each row presents the number of features each model selected, with the two numbers on top of the boxplots being the median number of linear components and non-linear components selected. The number of true features is indicated by the dotted red line, and the number of true linear and non-linear components is in the plot's subtitle. (Note that the number of non-zero features is not necessarily the sum of the number of non-zero linear and non-linear components: this is because a feature can have both a linear and non-linear component.) The right-most plot in each row presents the number of true features each method selected, with the total number of true features indicated by the dotted red line.

\textbf{Note:} The simulations that follow are not presented in the same order as in the main text.

\pagebreak
\subsection{Fully linear signal}

$n = 100$, $p = 200$, $\mu = \sum_{j=1}^{10} X_j$.

\begin{figure}[!htbp]
\centering
\includegraphics[width=2.0in]{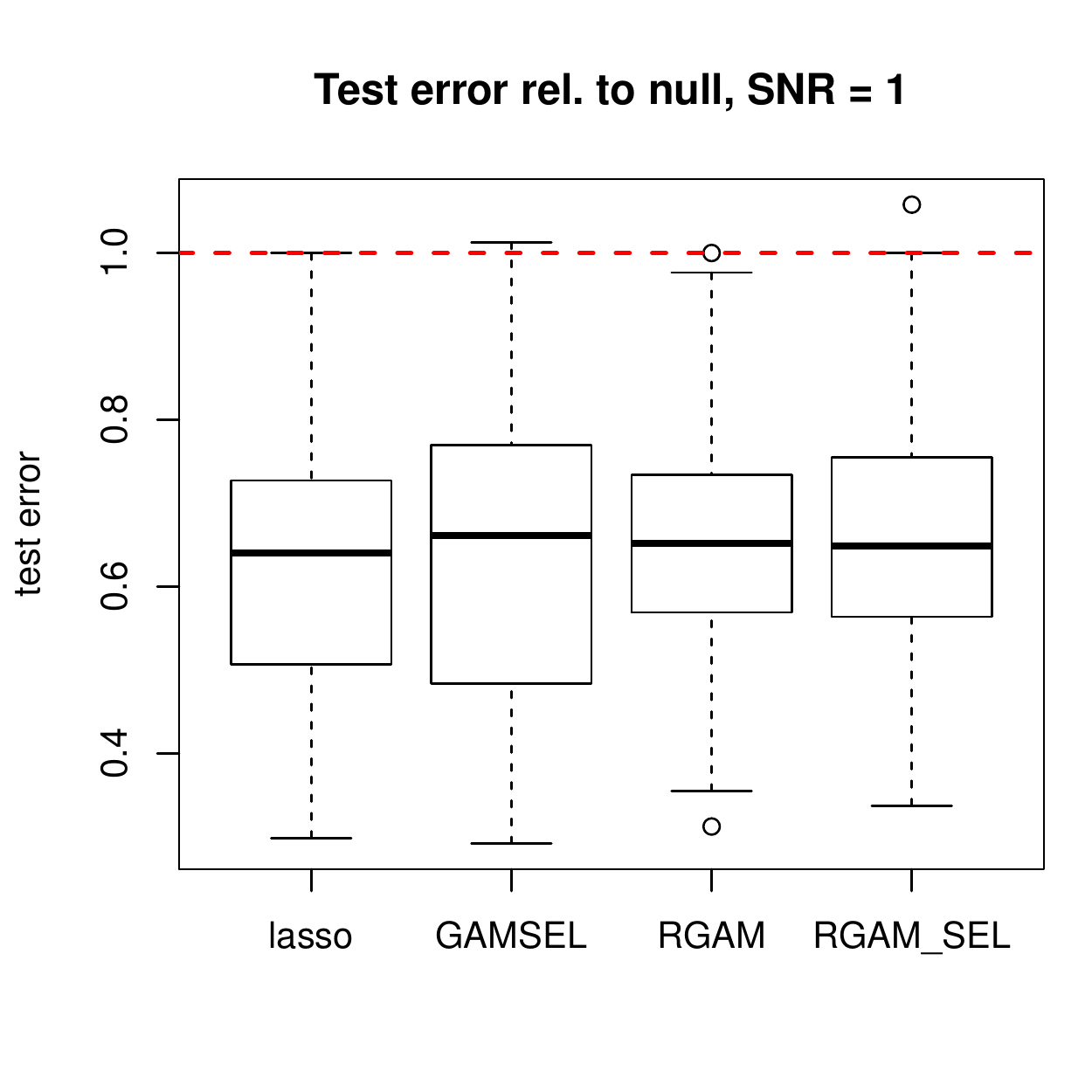}\includegraphics[width=2.0in]{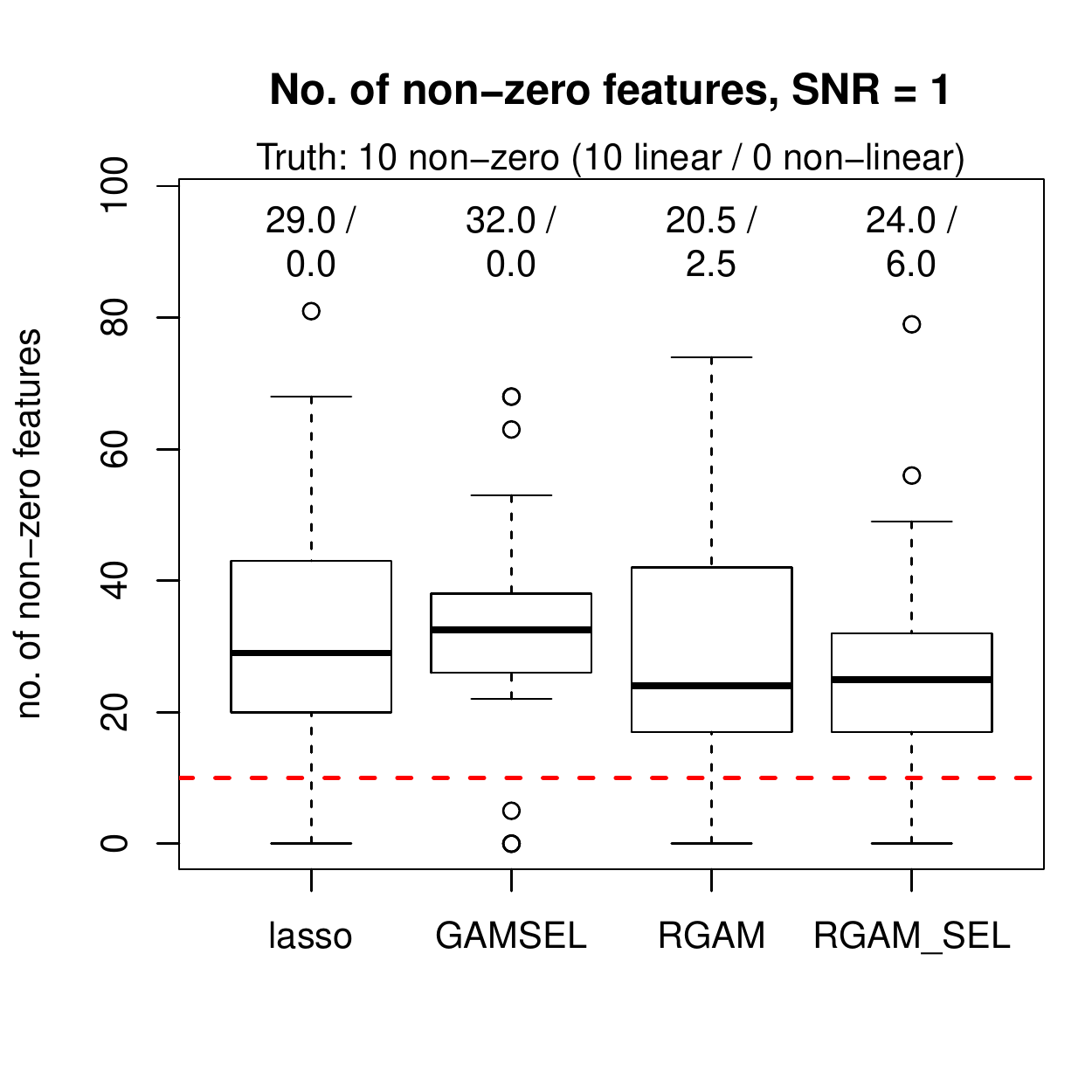}\includegraphics[width=2.0in]{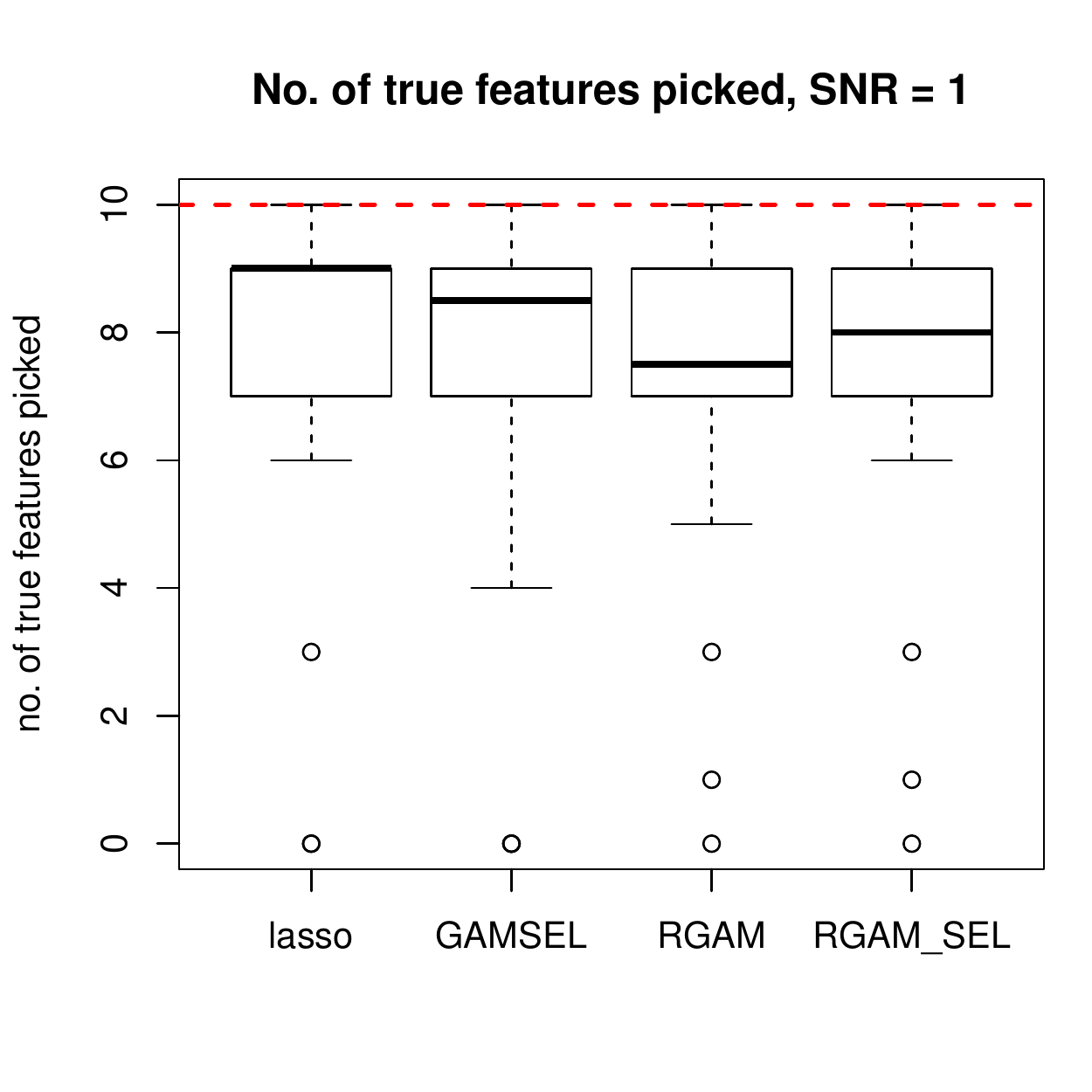}

\includegraphics[width=2.0in]{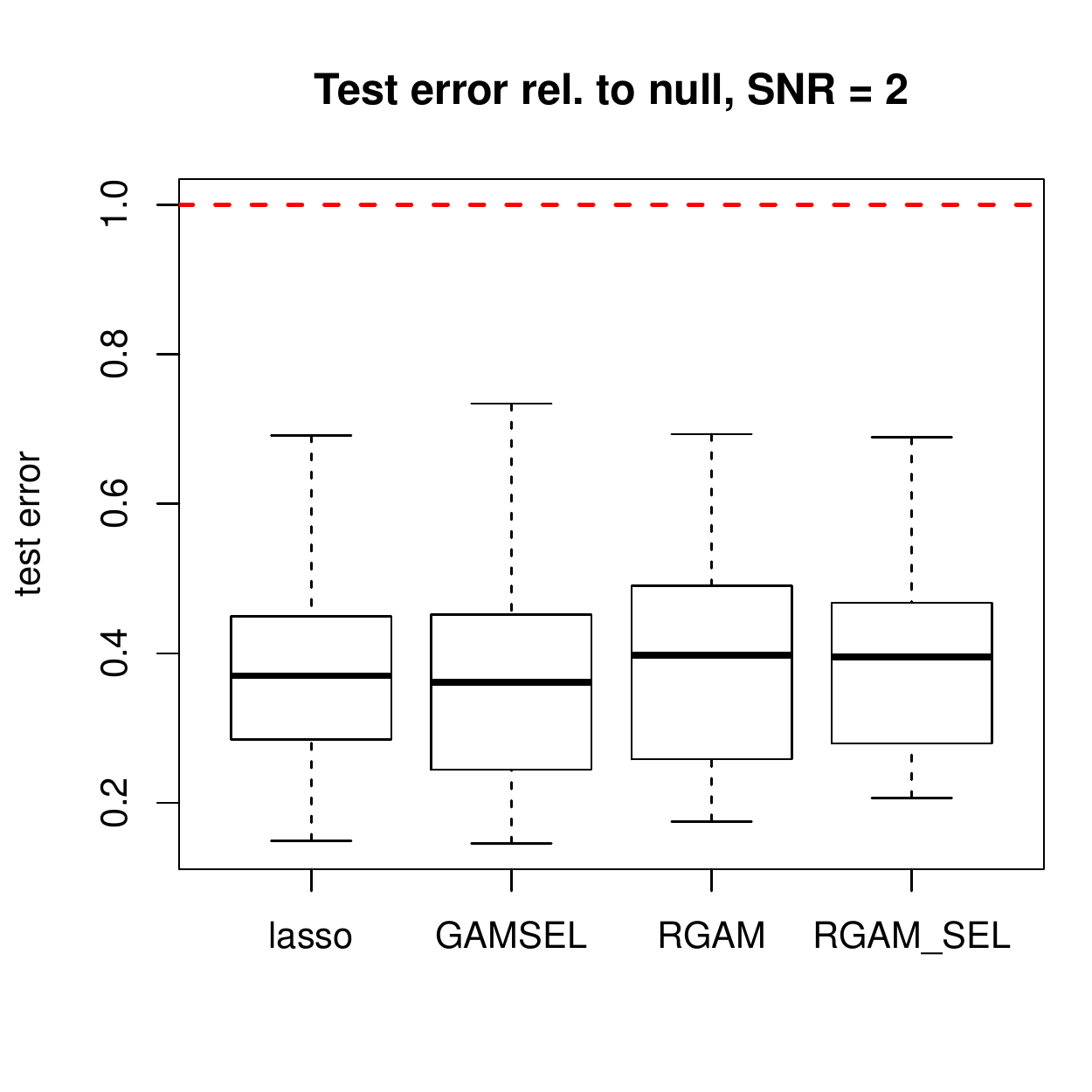}\includegraphics[width=2.0in]{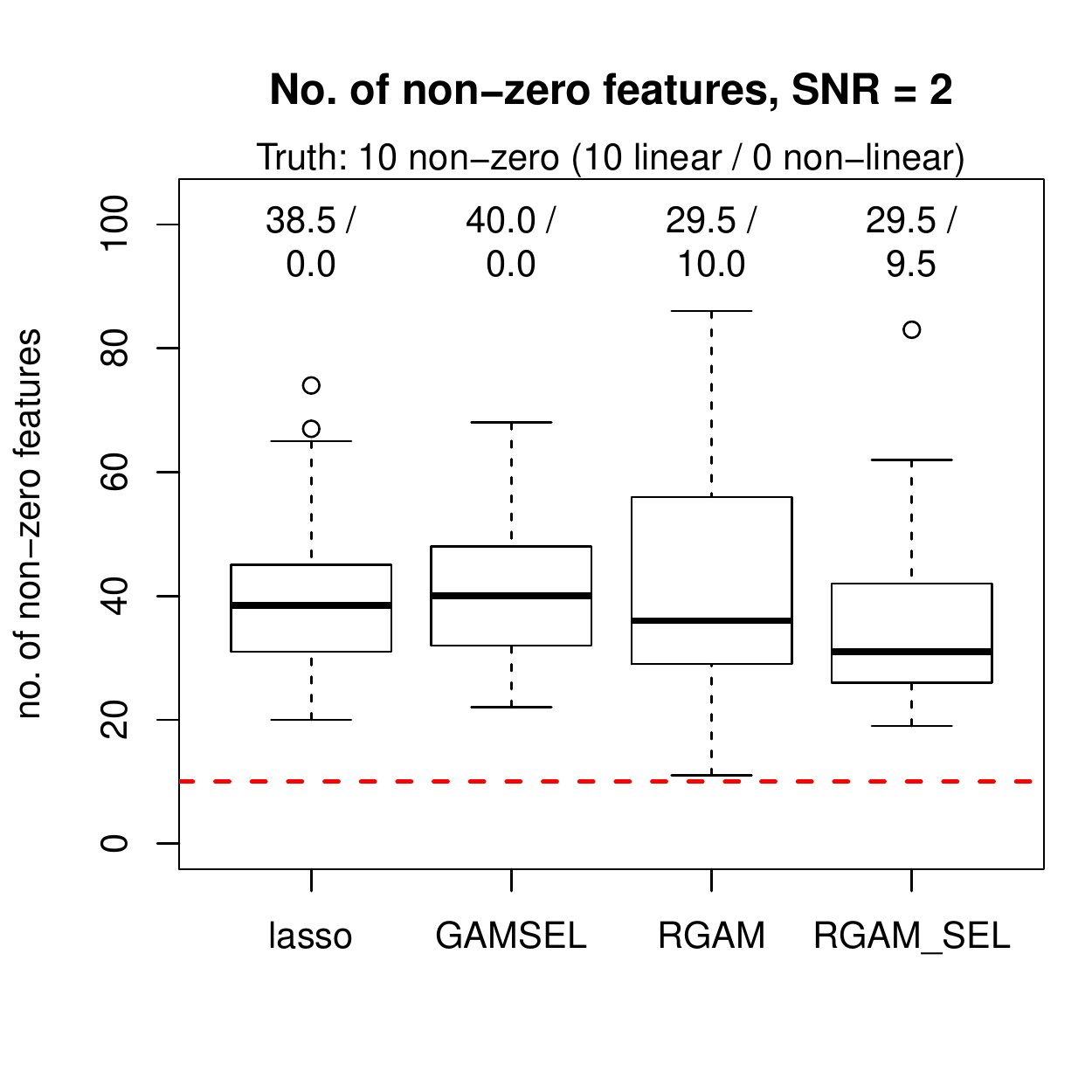}\includegraphics[width=2.0in]{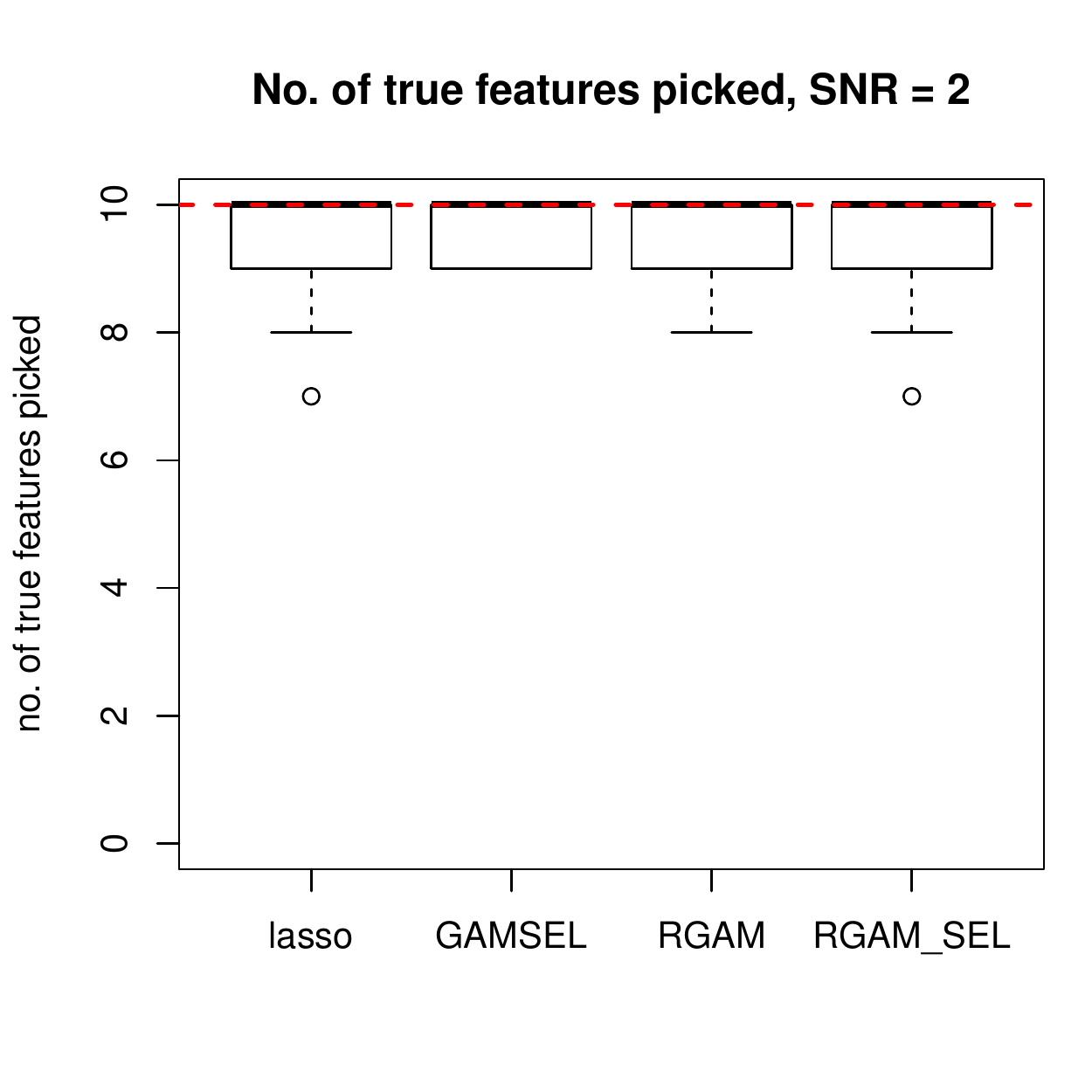}

\includegraphics[width=2.0in]{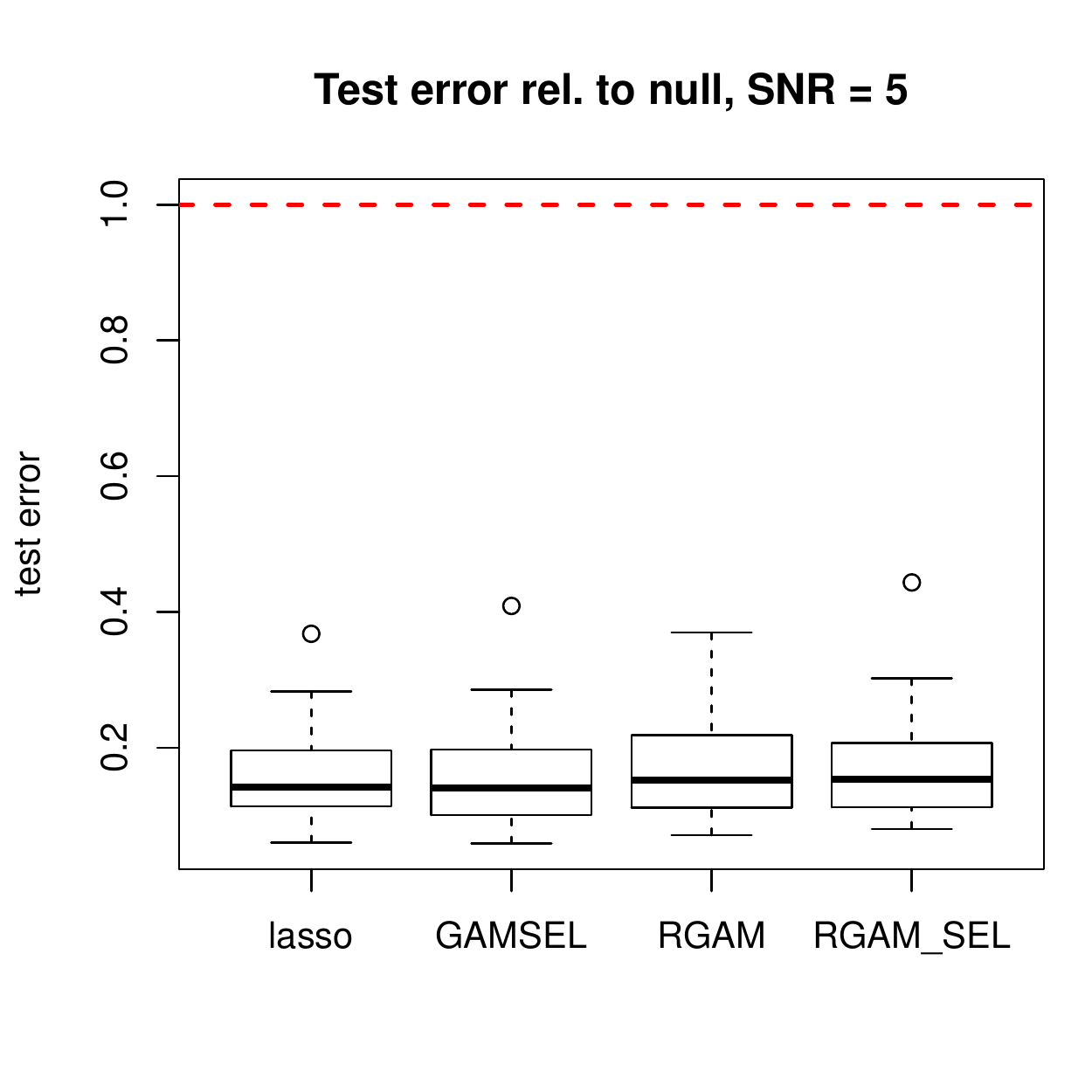}\includegraphics[width=2.0in]{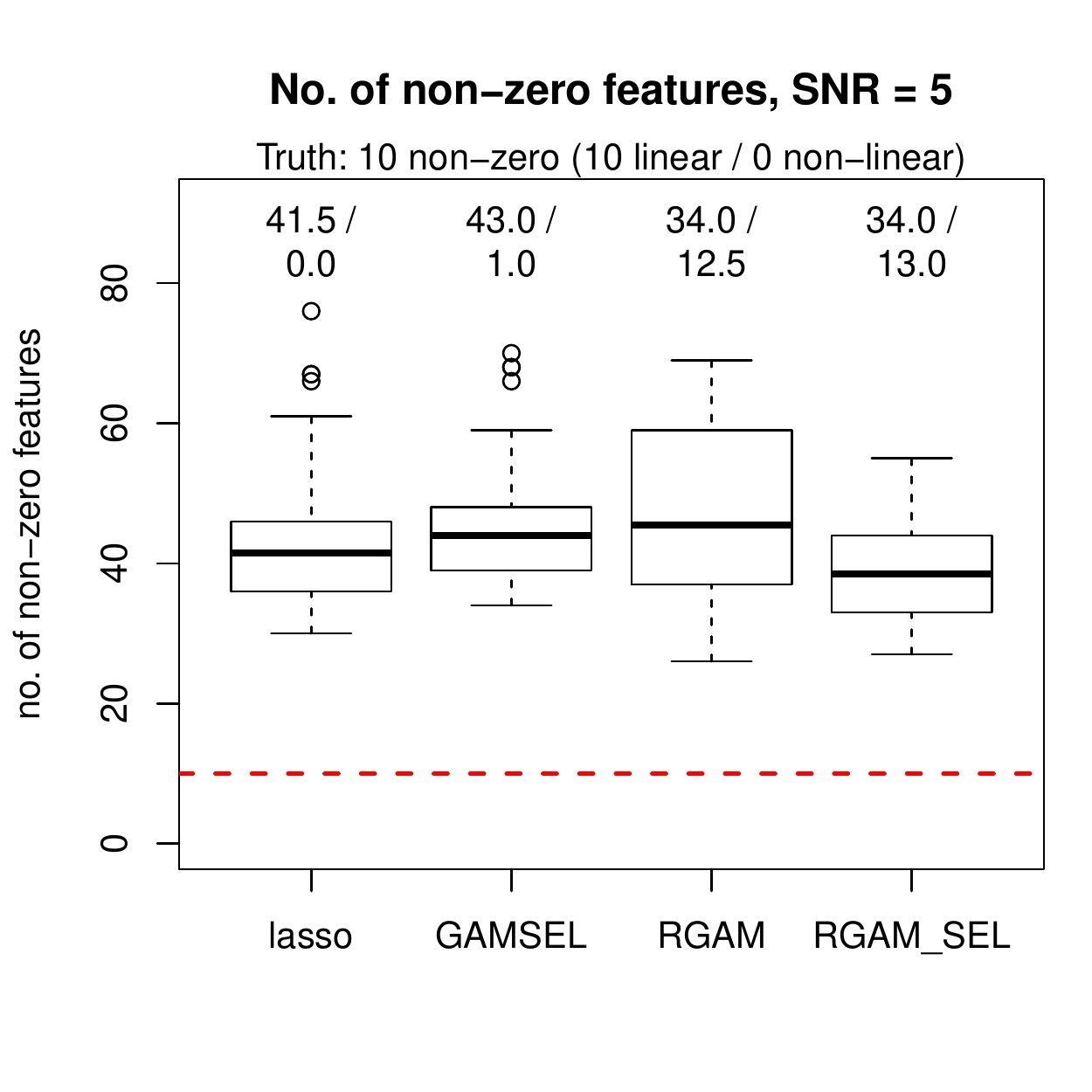}\includegraphics[width=2.0in]{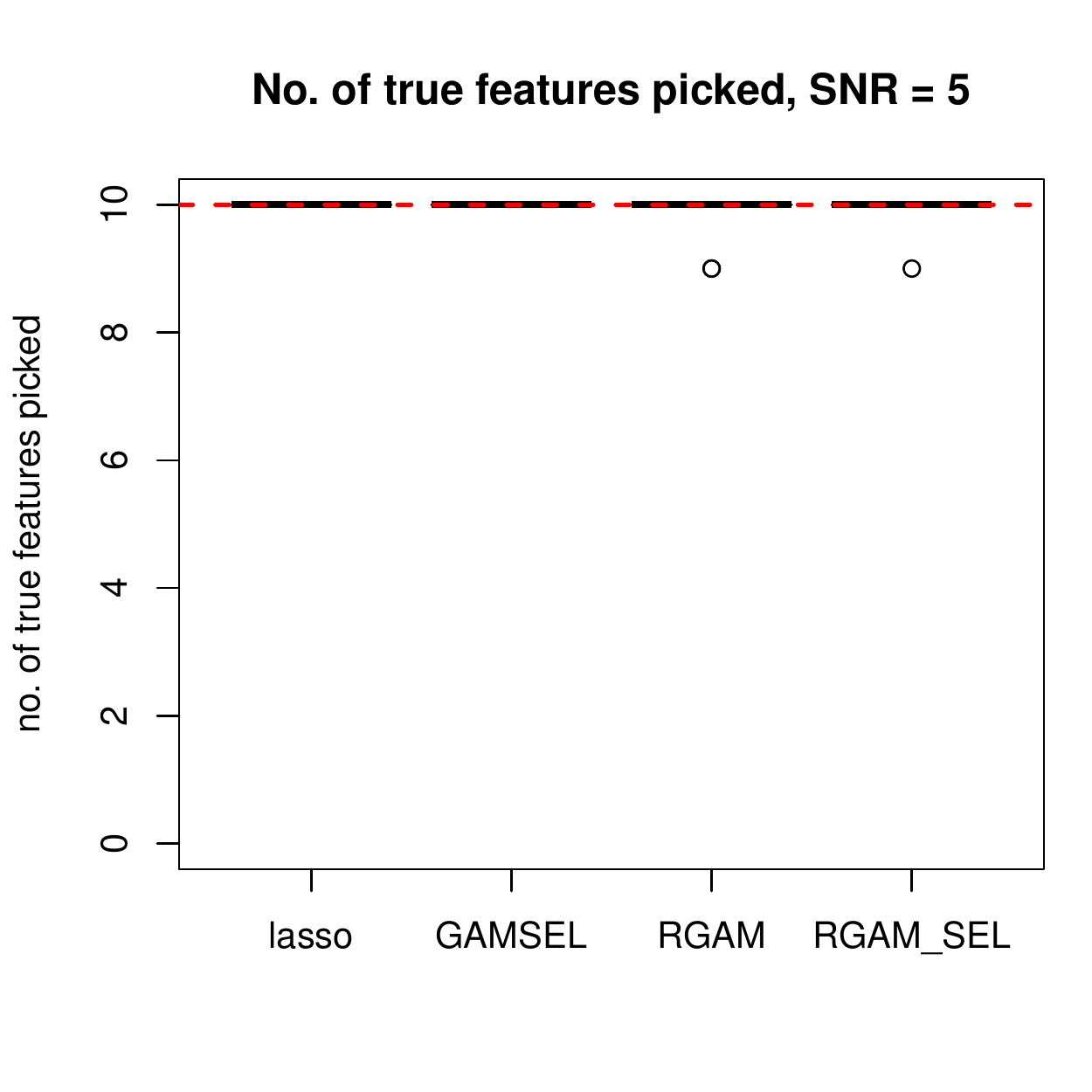}
\end{figure}

\pagebreak
\subsection{Hierarchical setting, linear and non-linear signals}

$n = 100$, $p = 200$, $\mu = \sum_{j=1}^5 [X_j + \frac{2}{3} (3X_j^2 - 1)]$.

\begin{figure}[!htbp]
\centering
\includegraphics[width=2.0in]{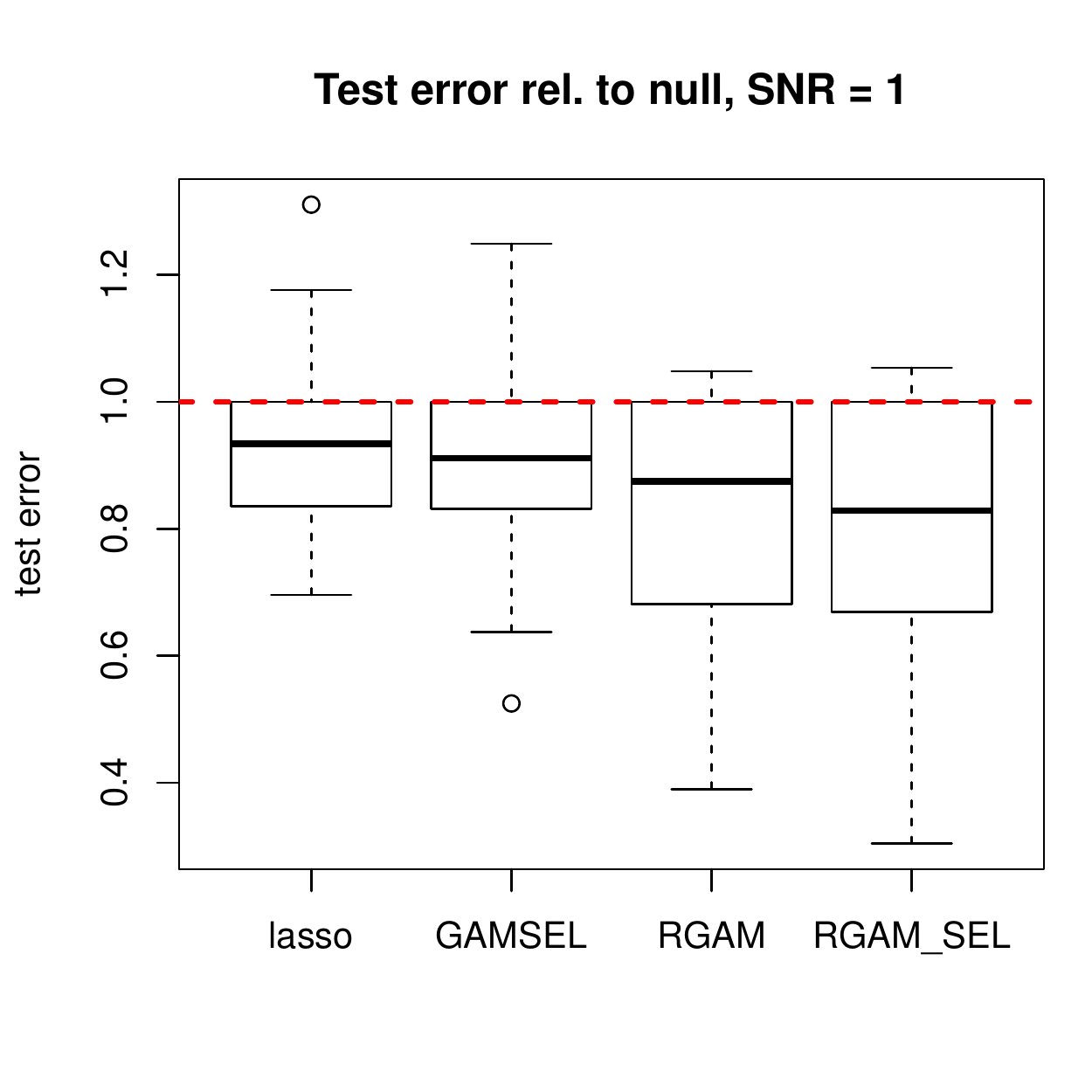}\includegraphics[width=2.0in]{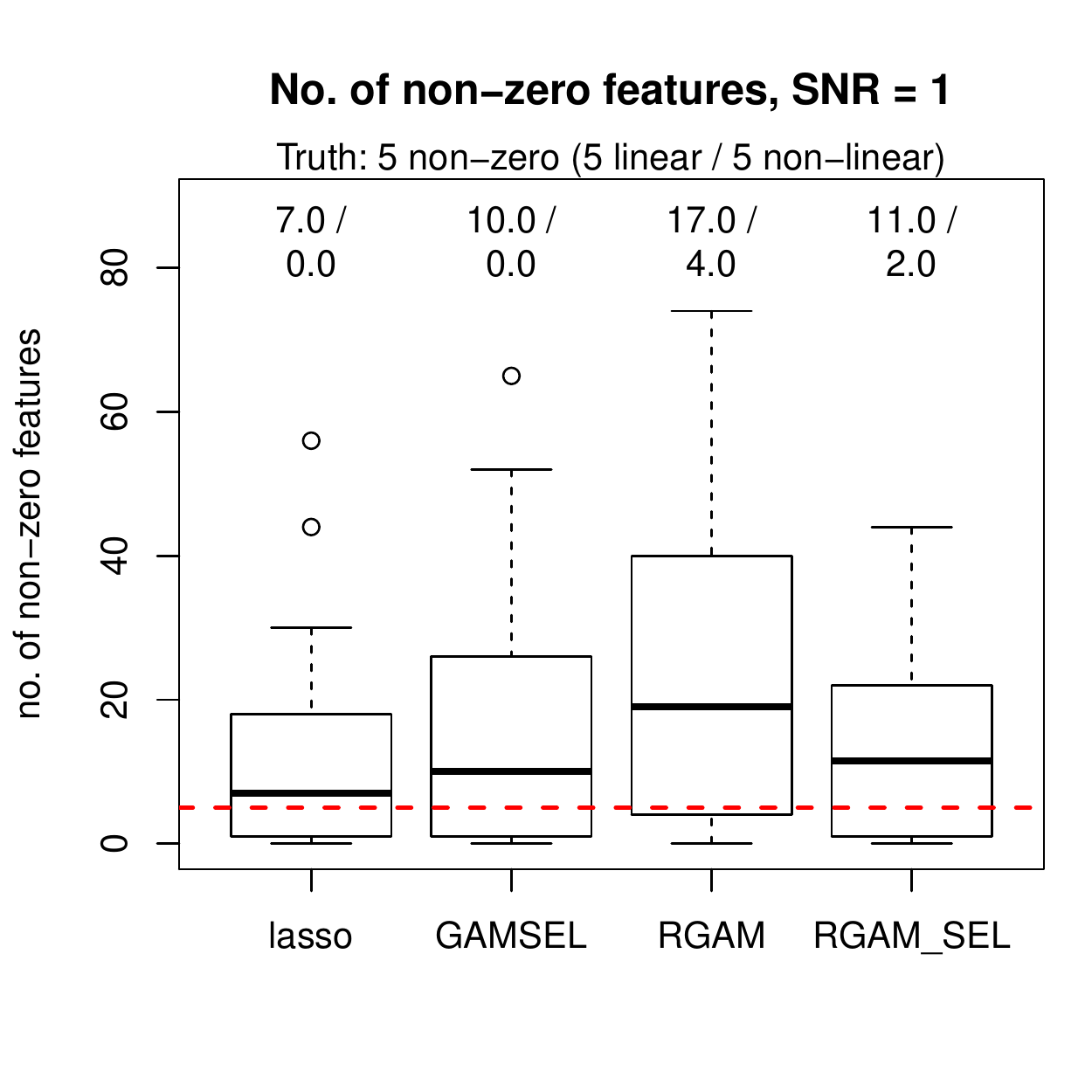}\includegraphics[width=2.0in]{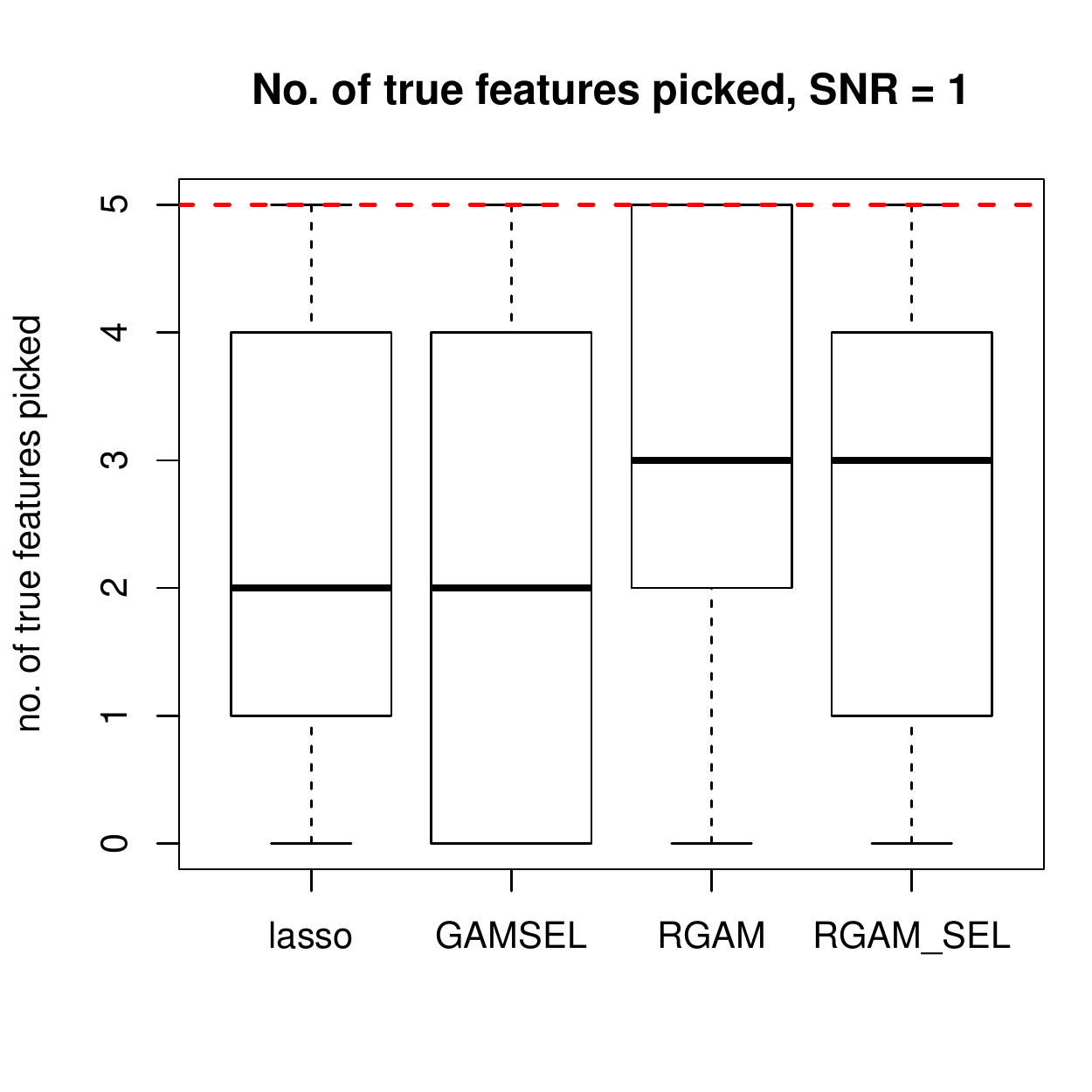}

\includegraphics[width=2.0in]{setup2_snr2_errrel}\includegraphics[width=2.0in]{setup2_snr2_nzfeat}\includegraphics[width=2.0in]{setup2_snr2_nztruefeat}

\includegraphics[width=2.0in]{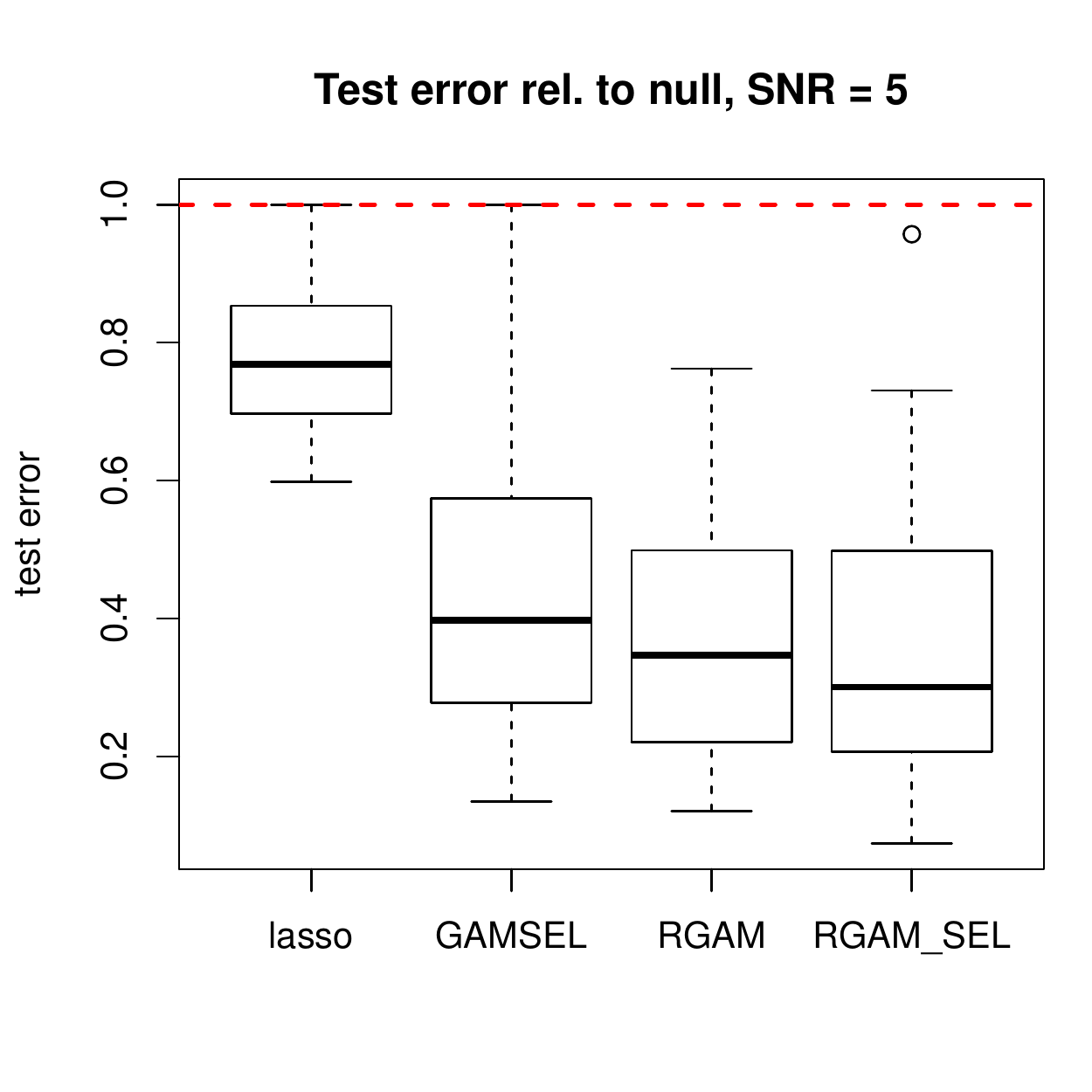}\includegraphics[width=2.0in]{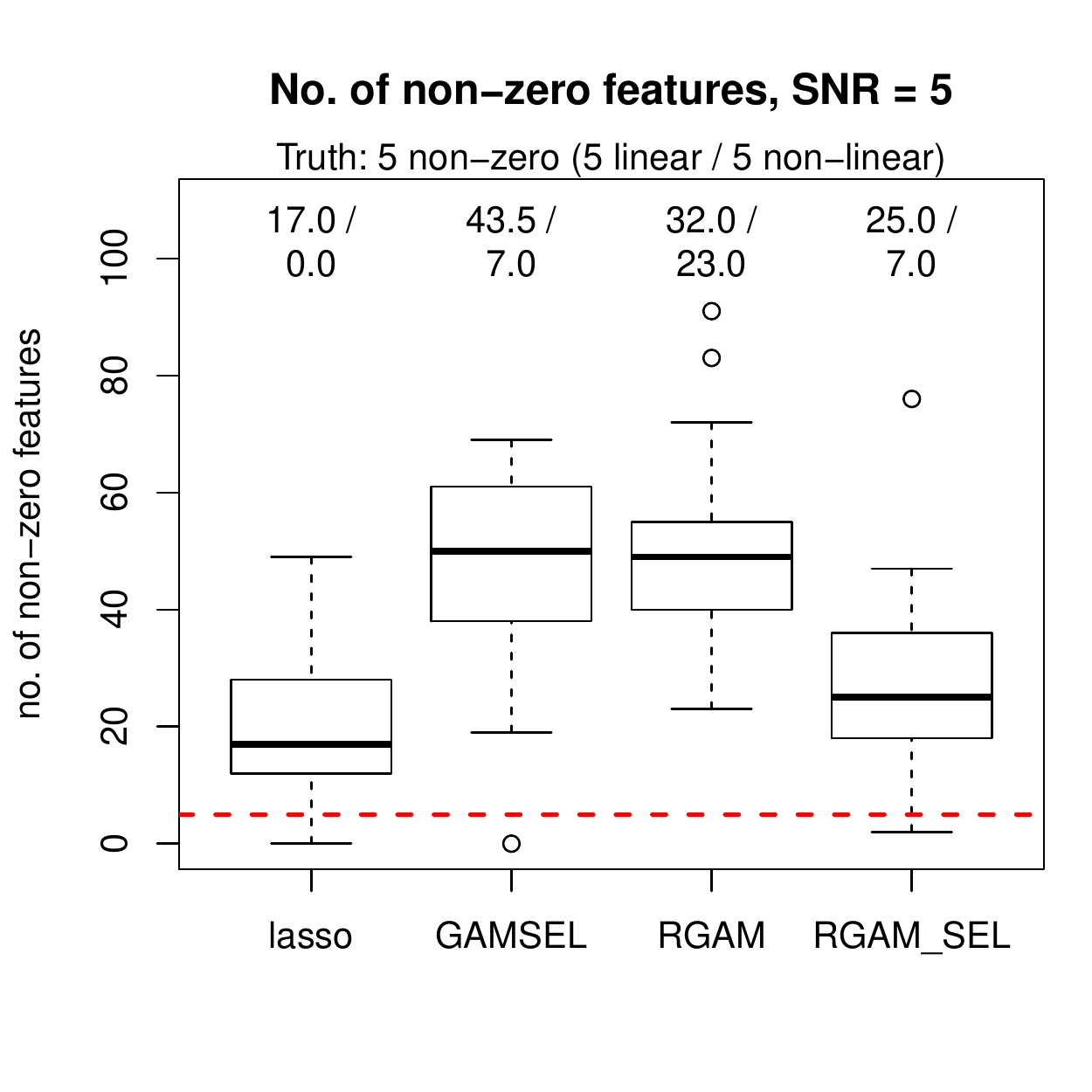}\includegraphics[width=2.0in]{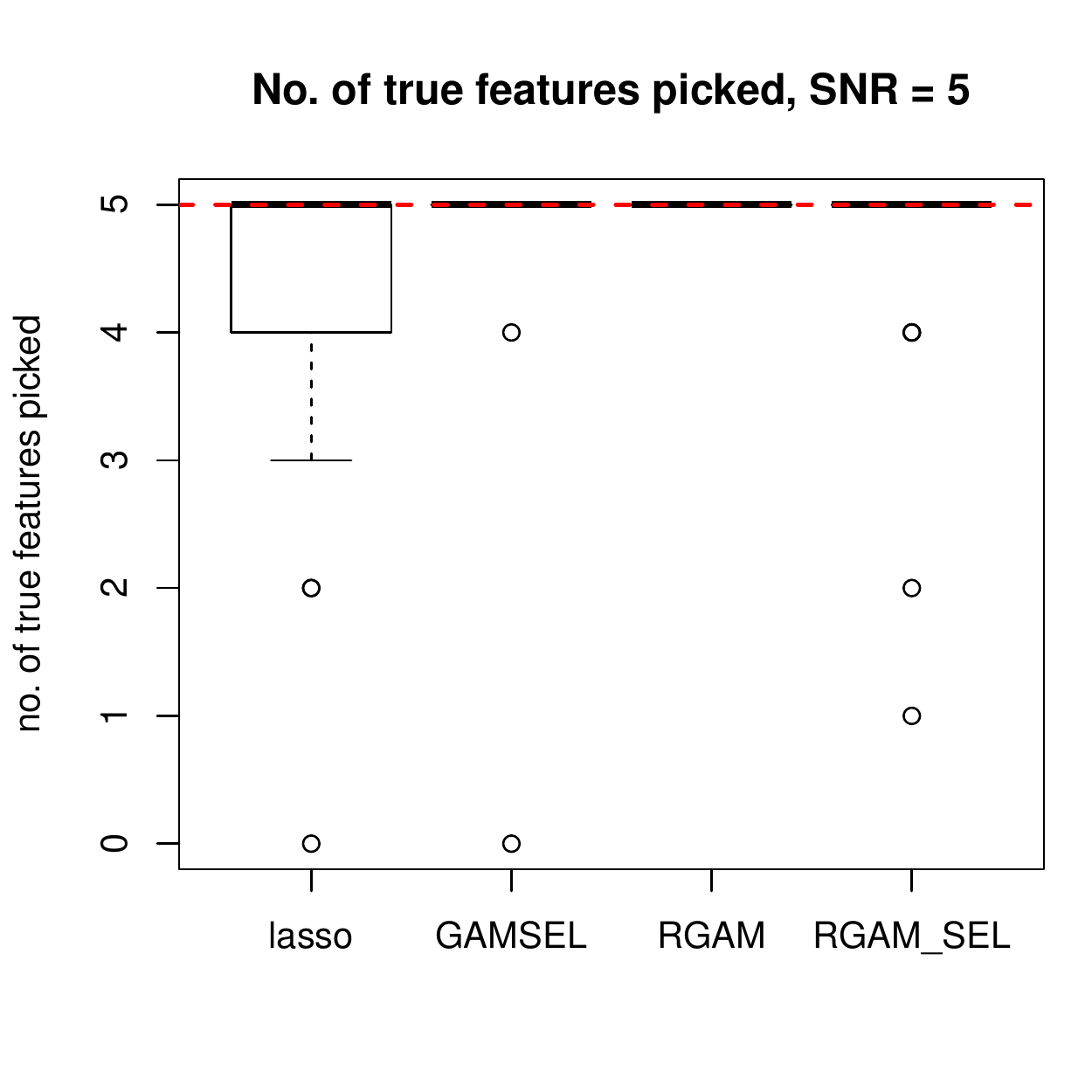}
\end{figure}

\pagebreak
\subsection{Fully non-linear signal}

$n = 100$, $p = 200$, $\mu = \sum_{j=1}^5 2(5X_j^3 - 3X_j)$.

\begin{figure}[!htbp]
\centering
\includegraphics[width=2.0in]{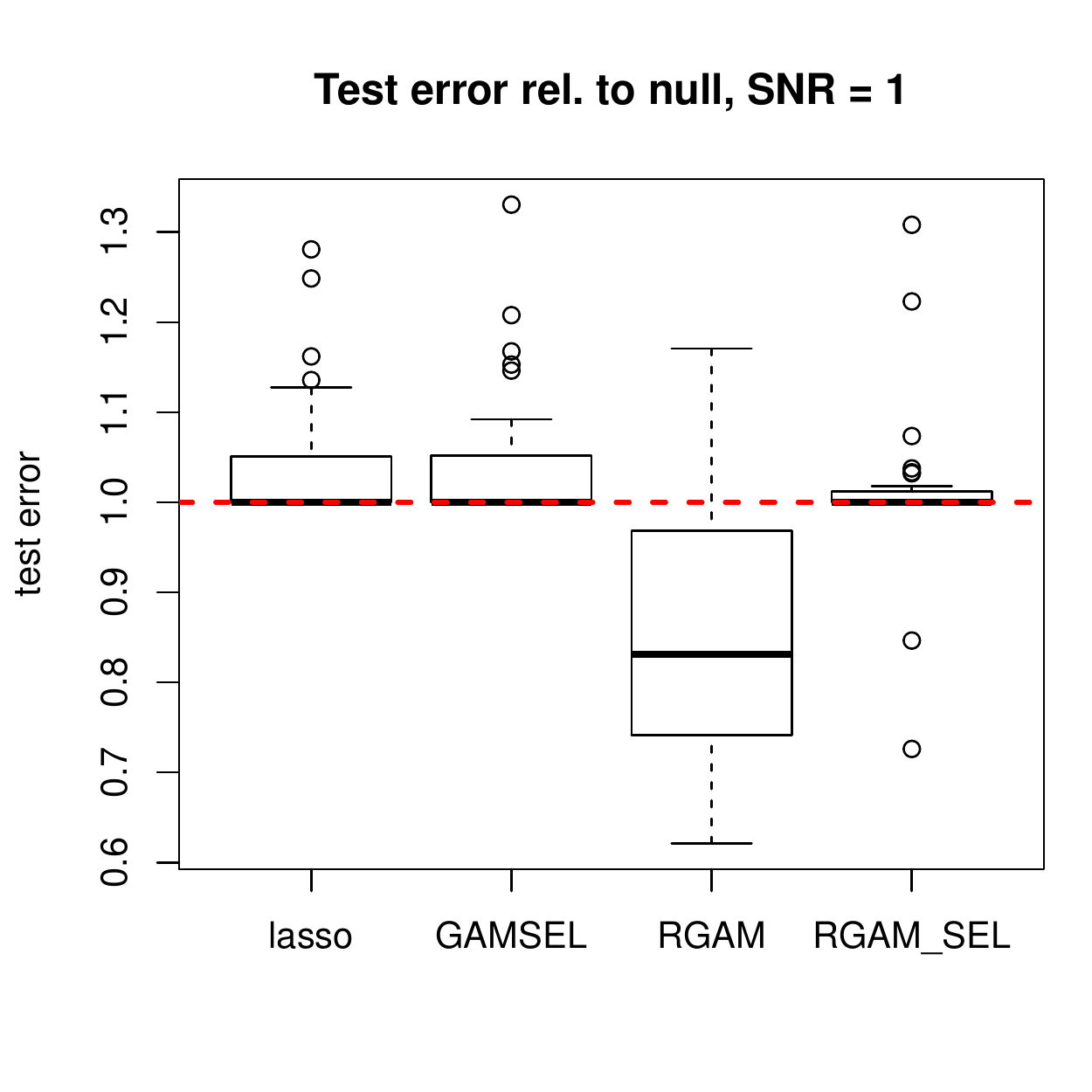}\includegraphics[width=2.0in]{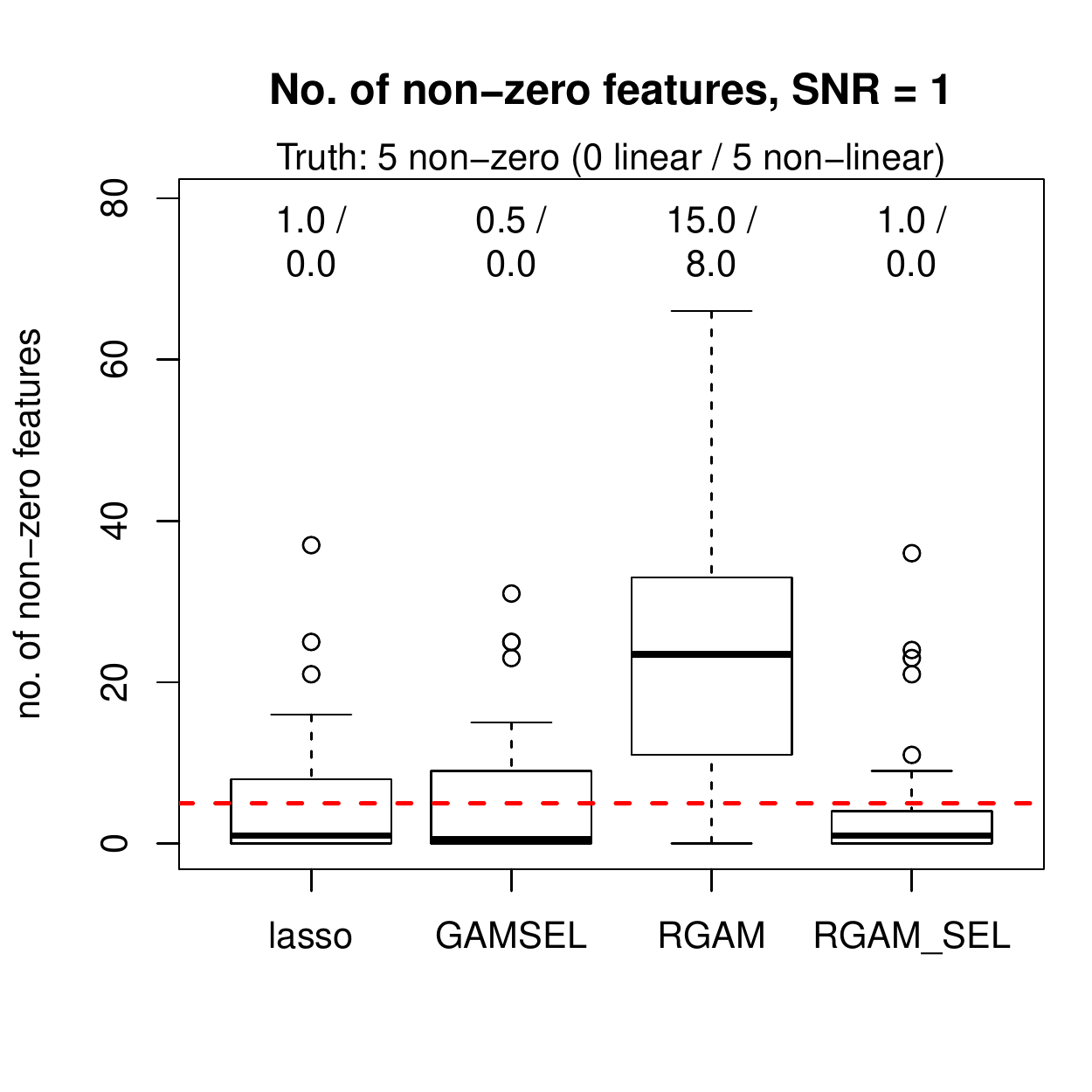}\includegraphics[width=2.0in]{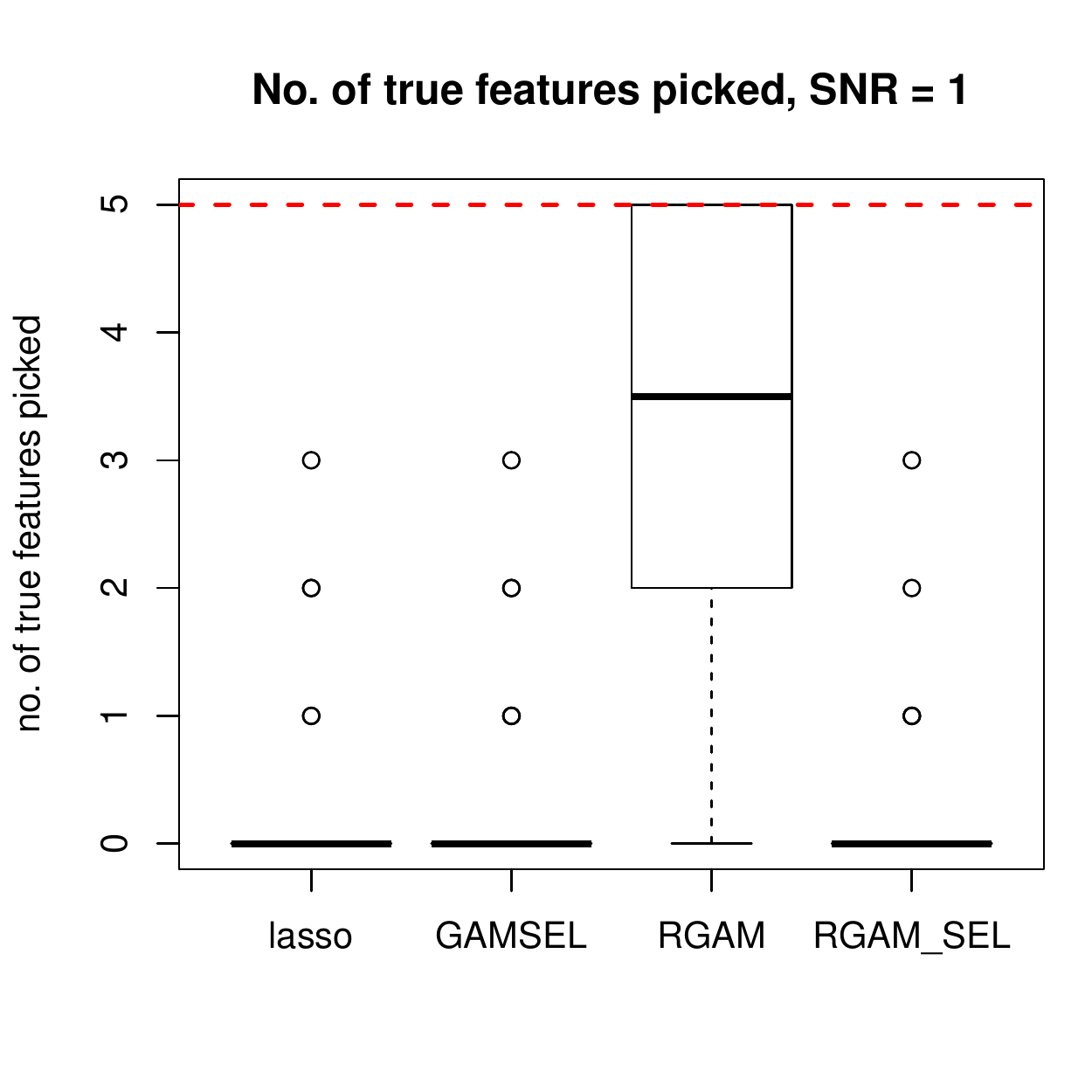}

\includegraphics[width=2.0in]{setup3_snr2_errrel}\includegraphics[width=2.0in]{setup3_snr2_nzfeat}\includegraphics[width=2.0in]{setup3_snr2_nztruefeat}

\includegraphics[width=2.0in]{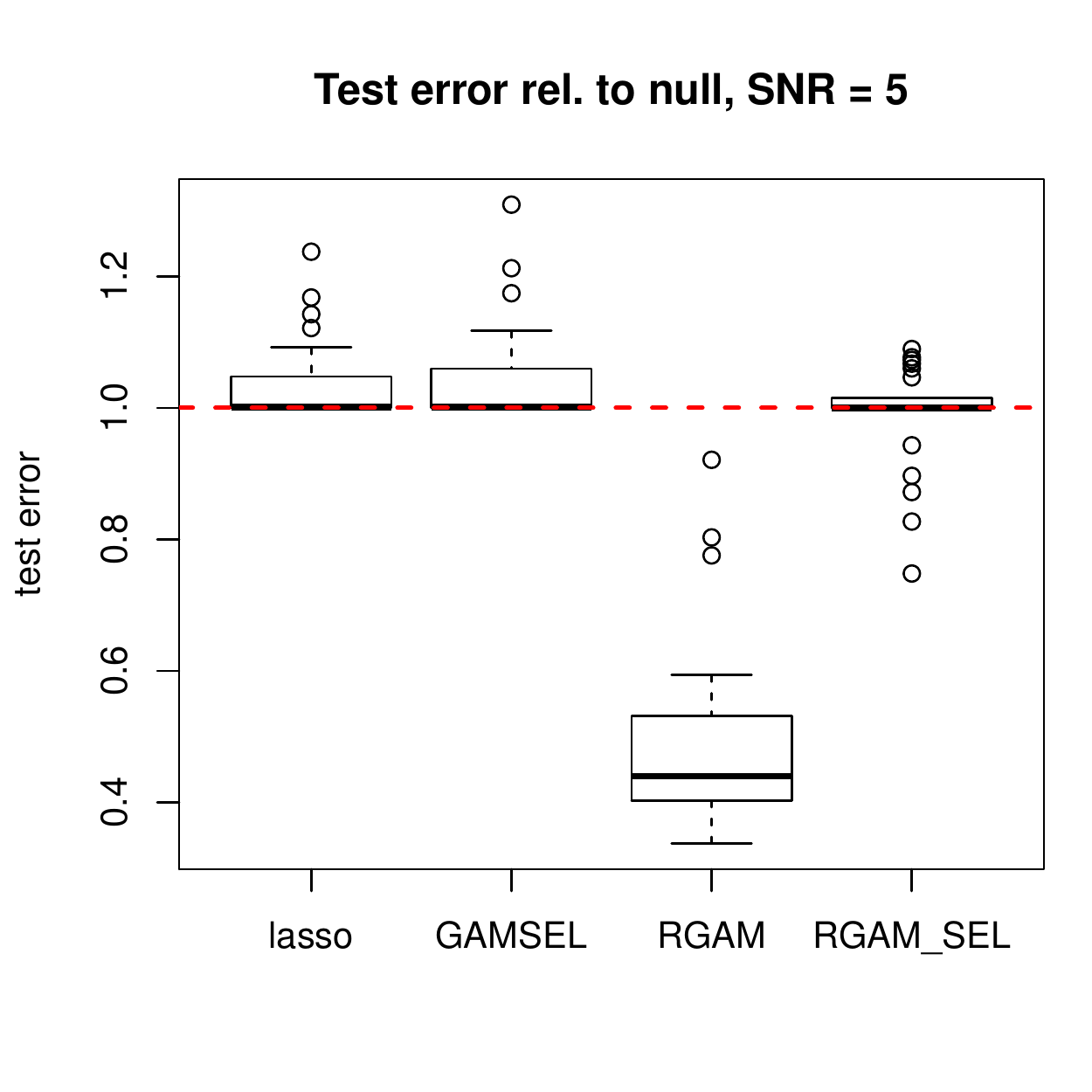}\includegraphics[width=2.0in]{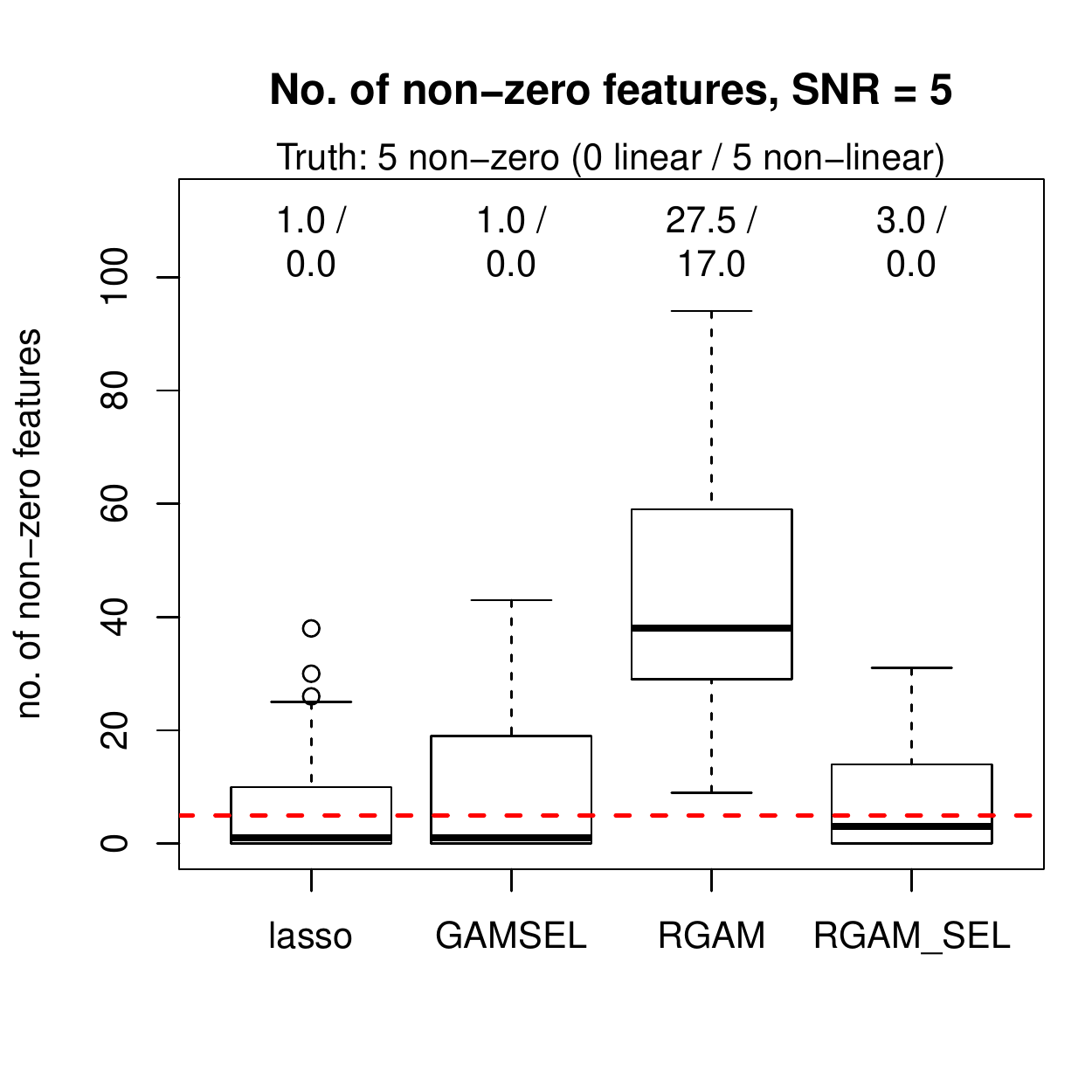}\includegraphics[width=2.0in]{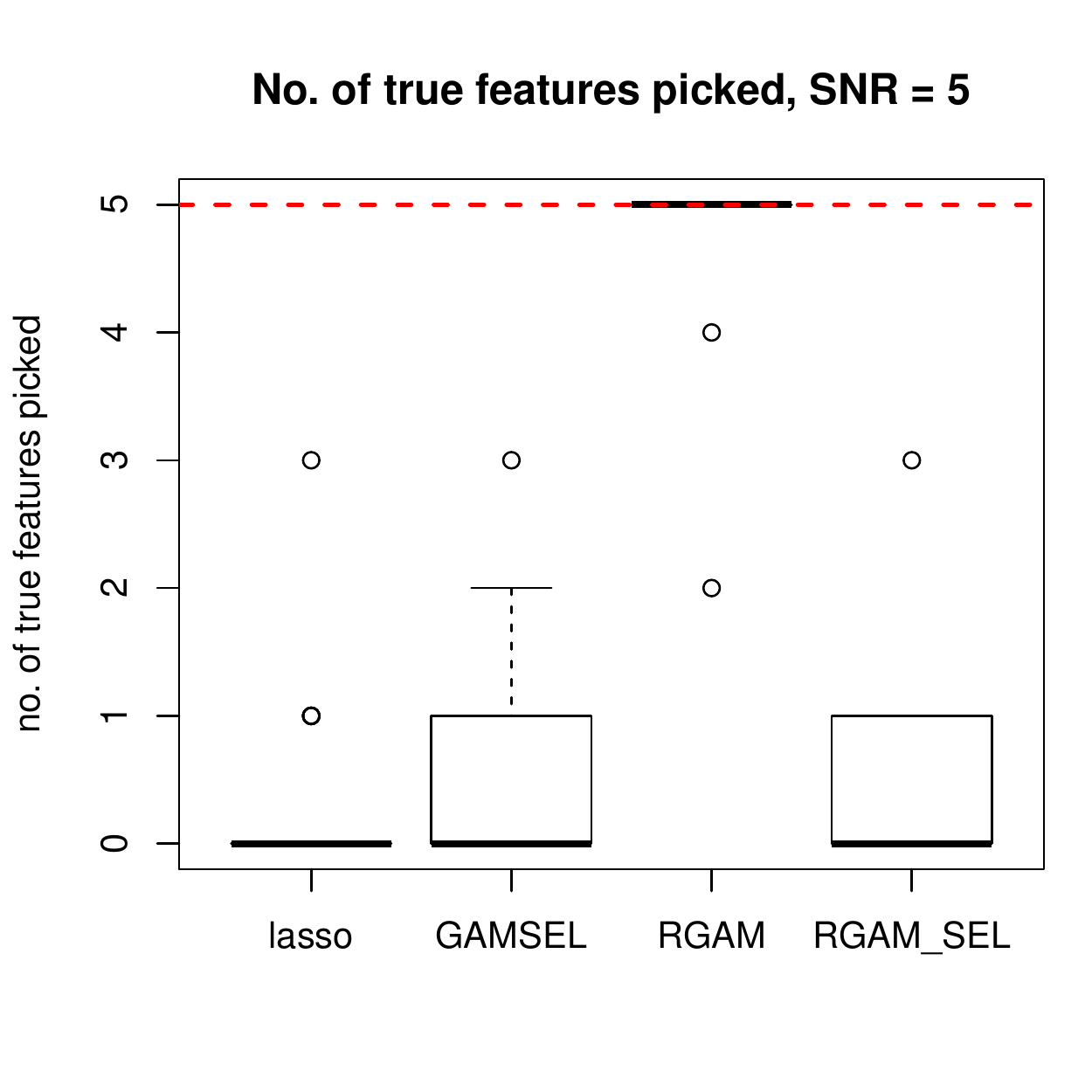}
\end{figure}

\pagebreak
\subsection{Non-hierarchical setting, linear and non-linear signals}

$n = 100$, $p = 200$, $\mu = \sum_{j=1}^5 X_j + \sum_{j=6}^{10} \frac{2}{3} (3 X_j^2 - 1)$.

\begin{figure}[!htbp]
\centering
\includegraphics[width=2.0in]{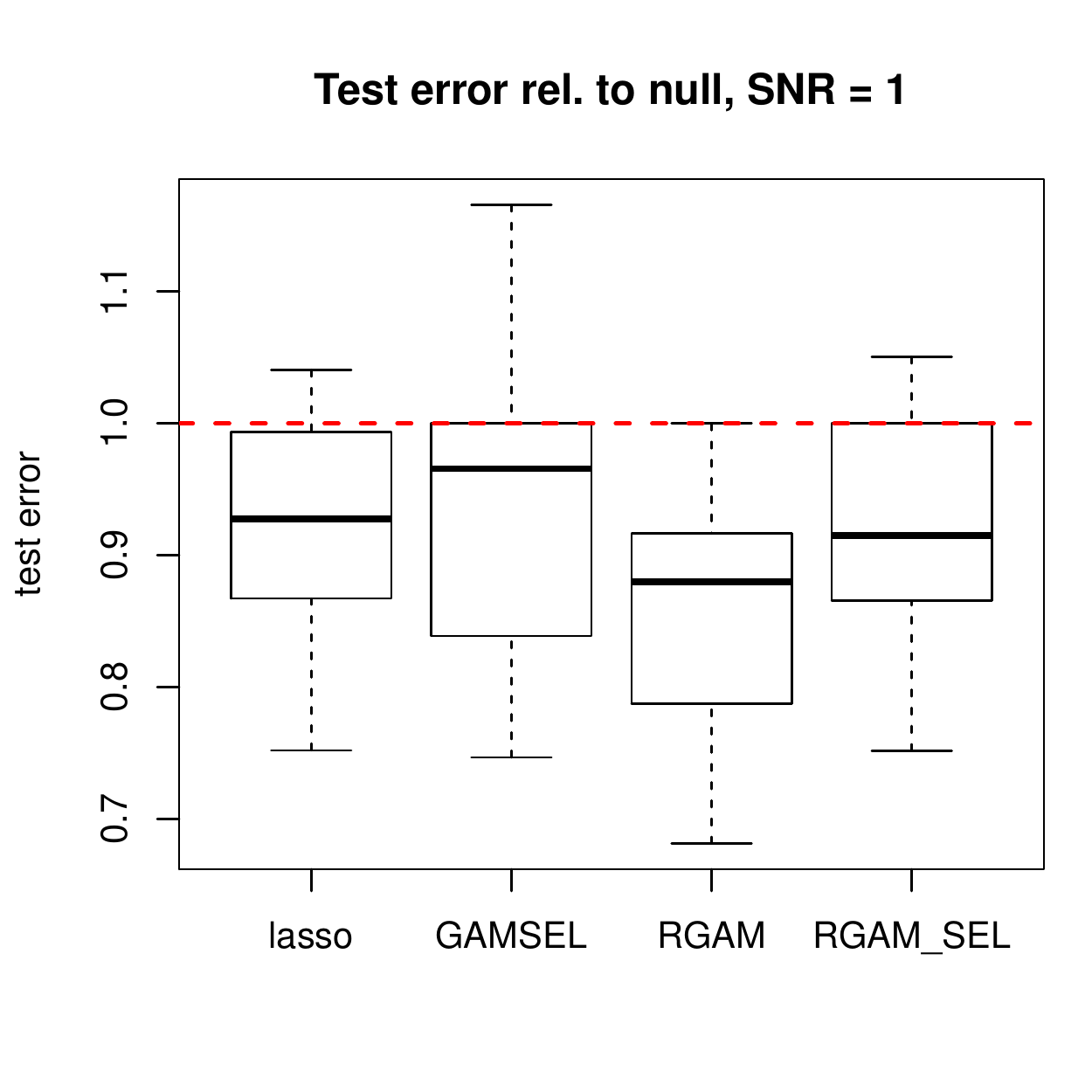}\includegraphics[width=2.0in]{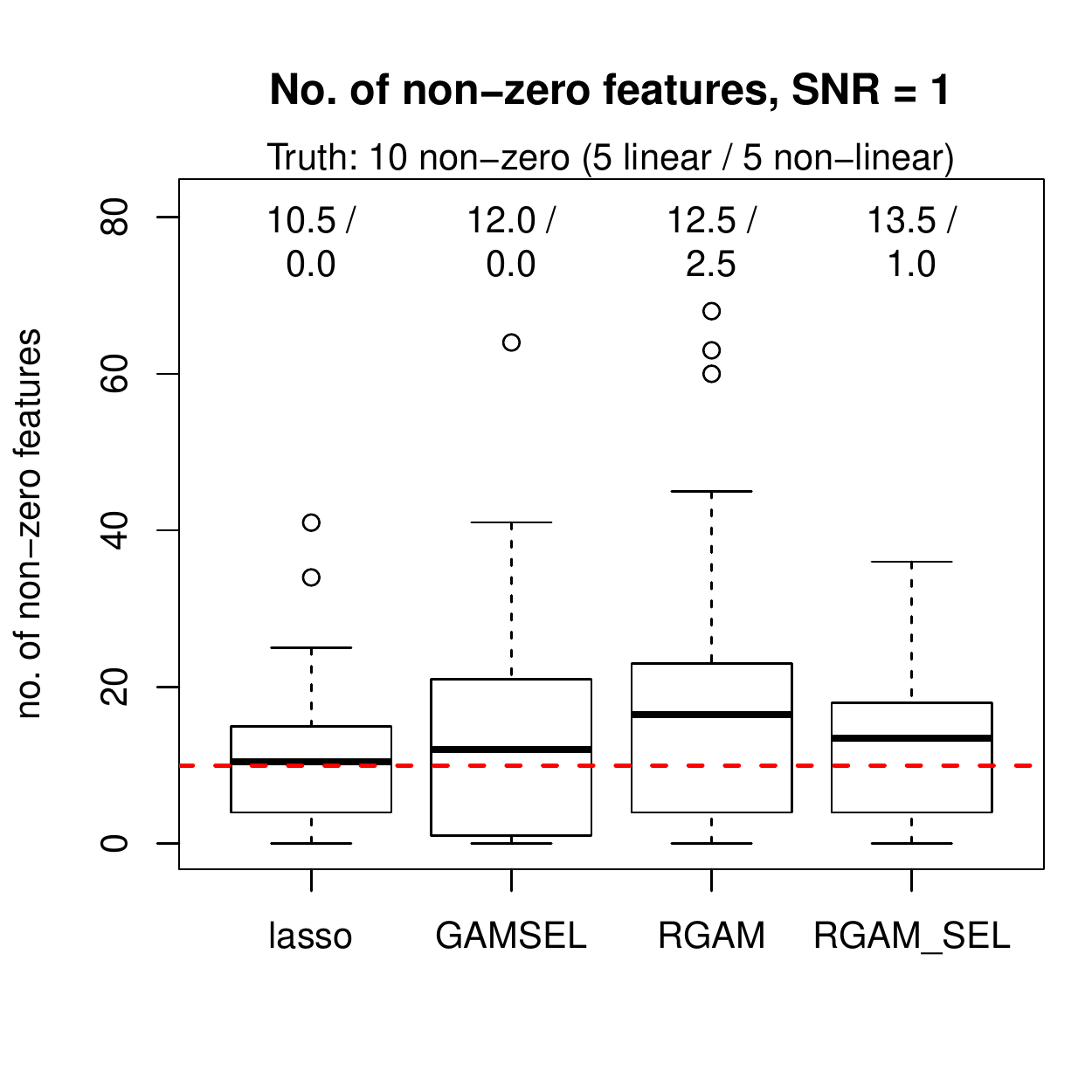}\includegraphics[width=2.0in]{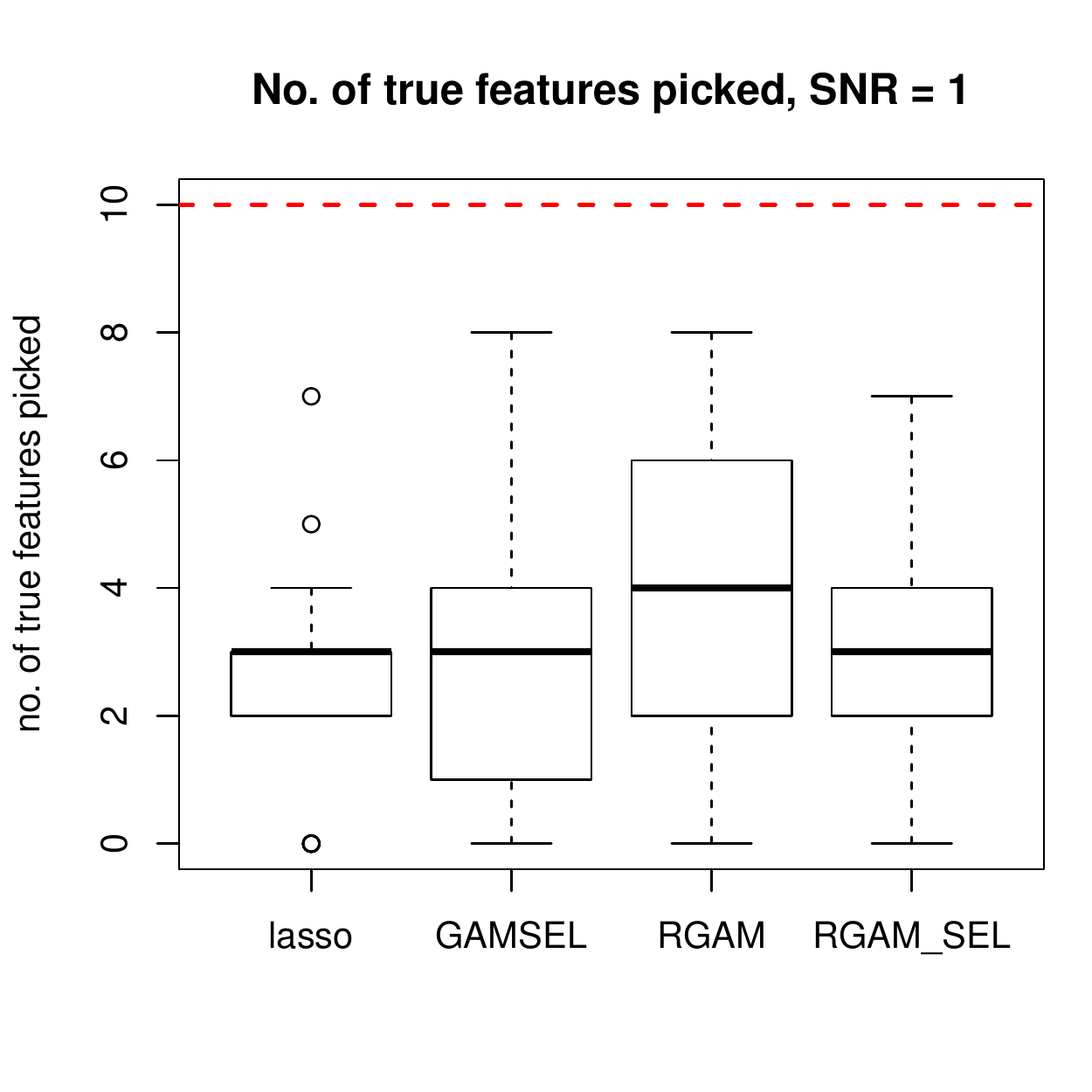}

\includegraphics[width=2.0in]{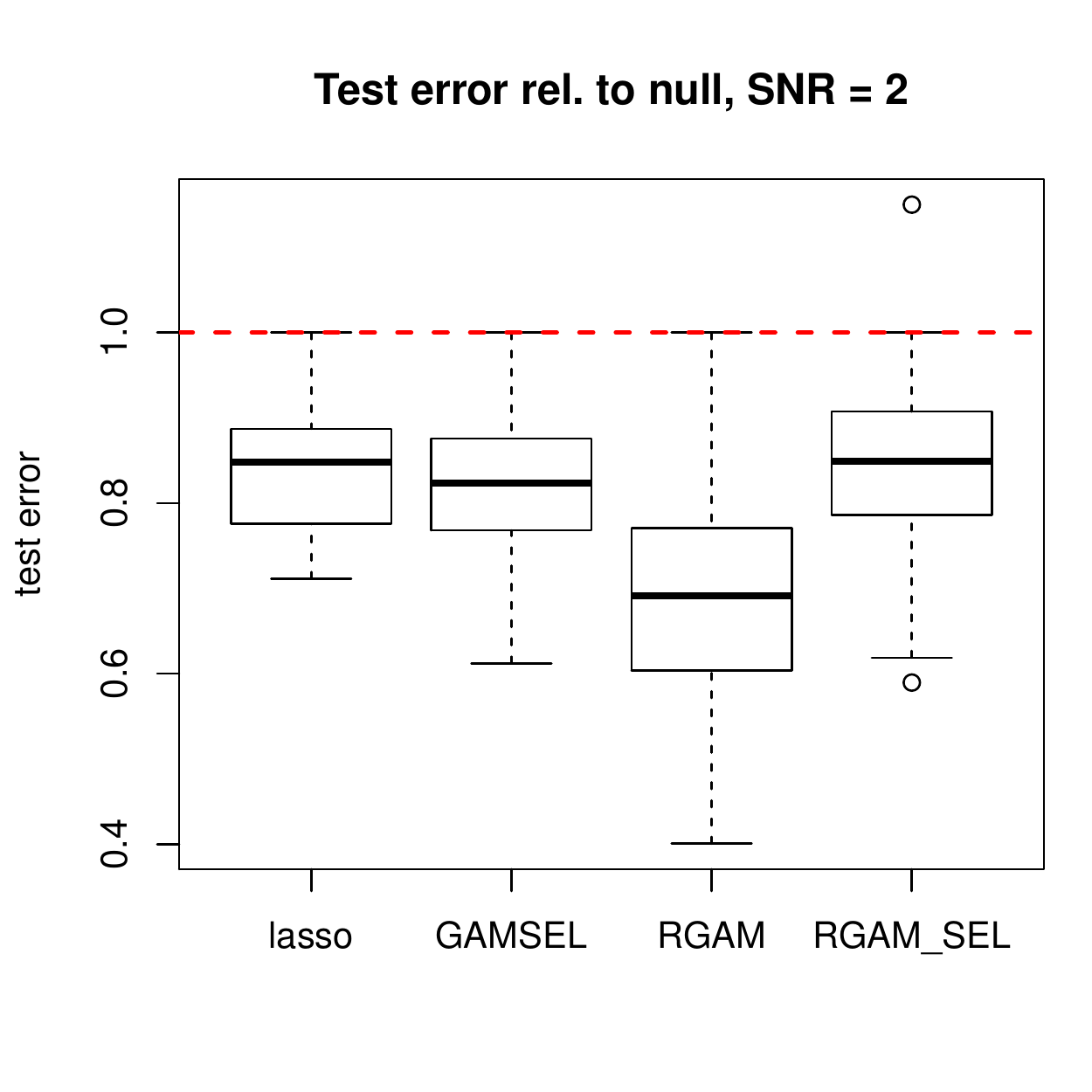}\includegraphics[width=2.0in]{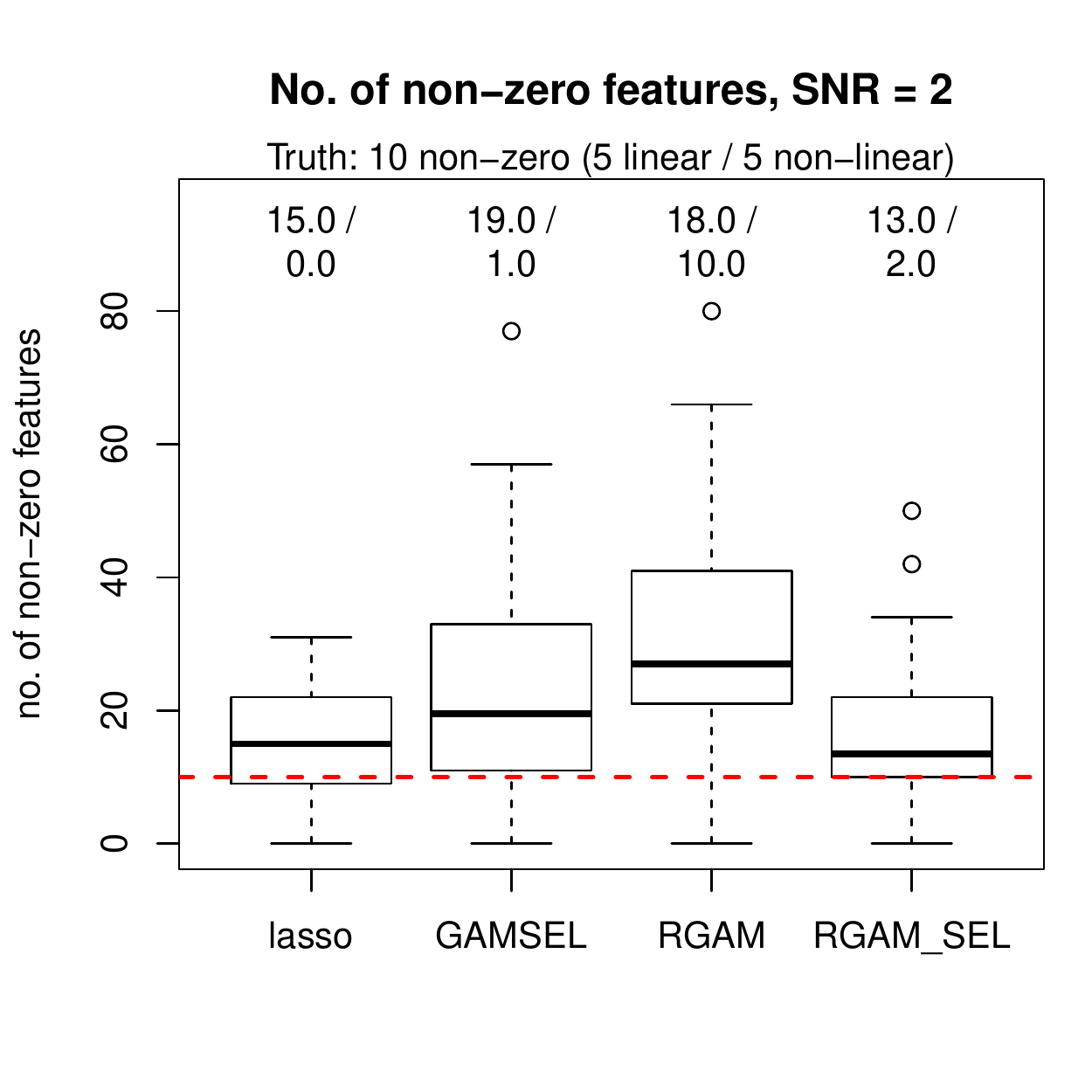}\includegraphics[width=2.0in]{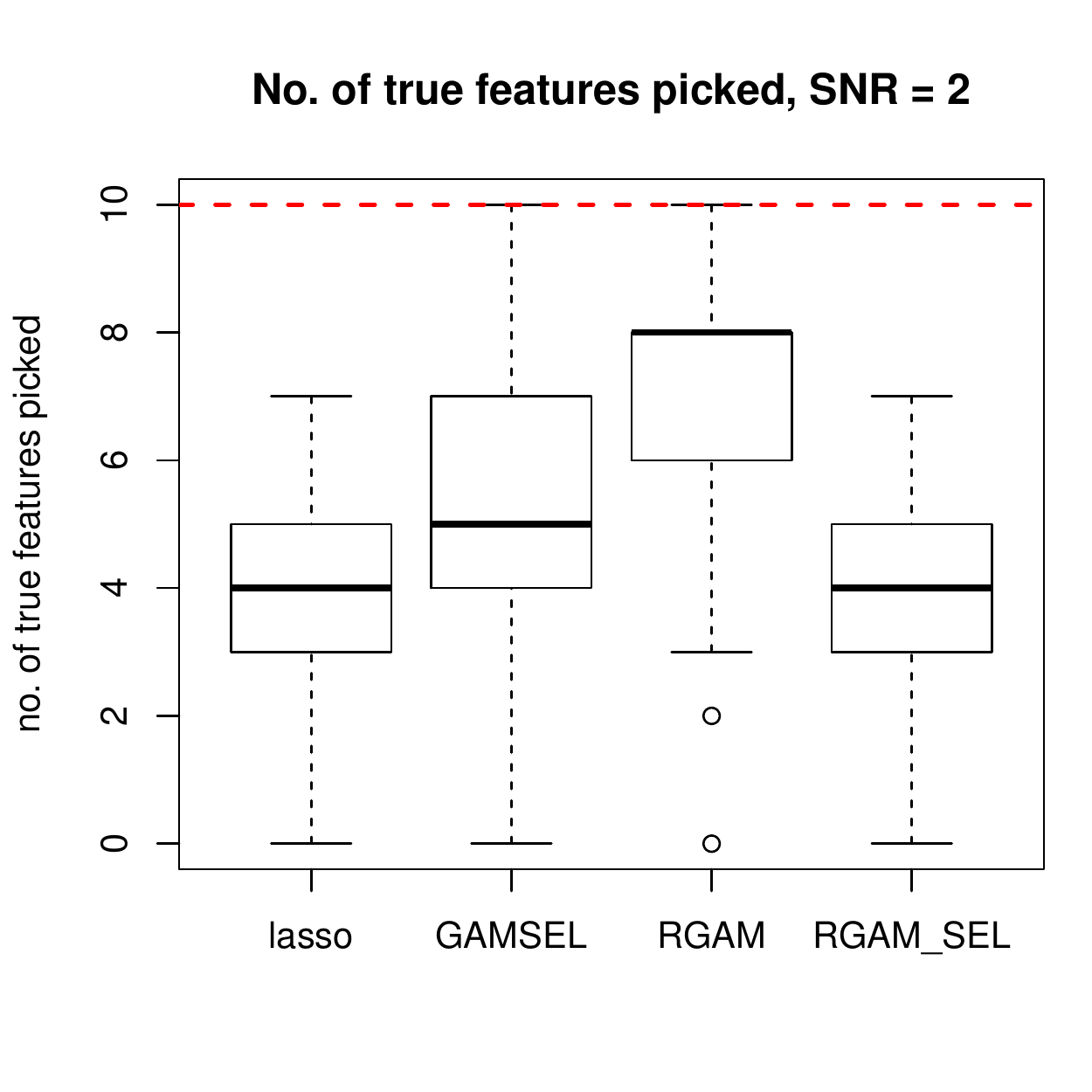}

\includegraphics[width=2.0in]{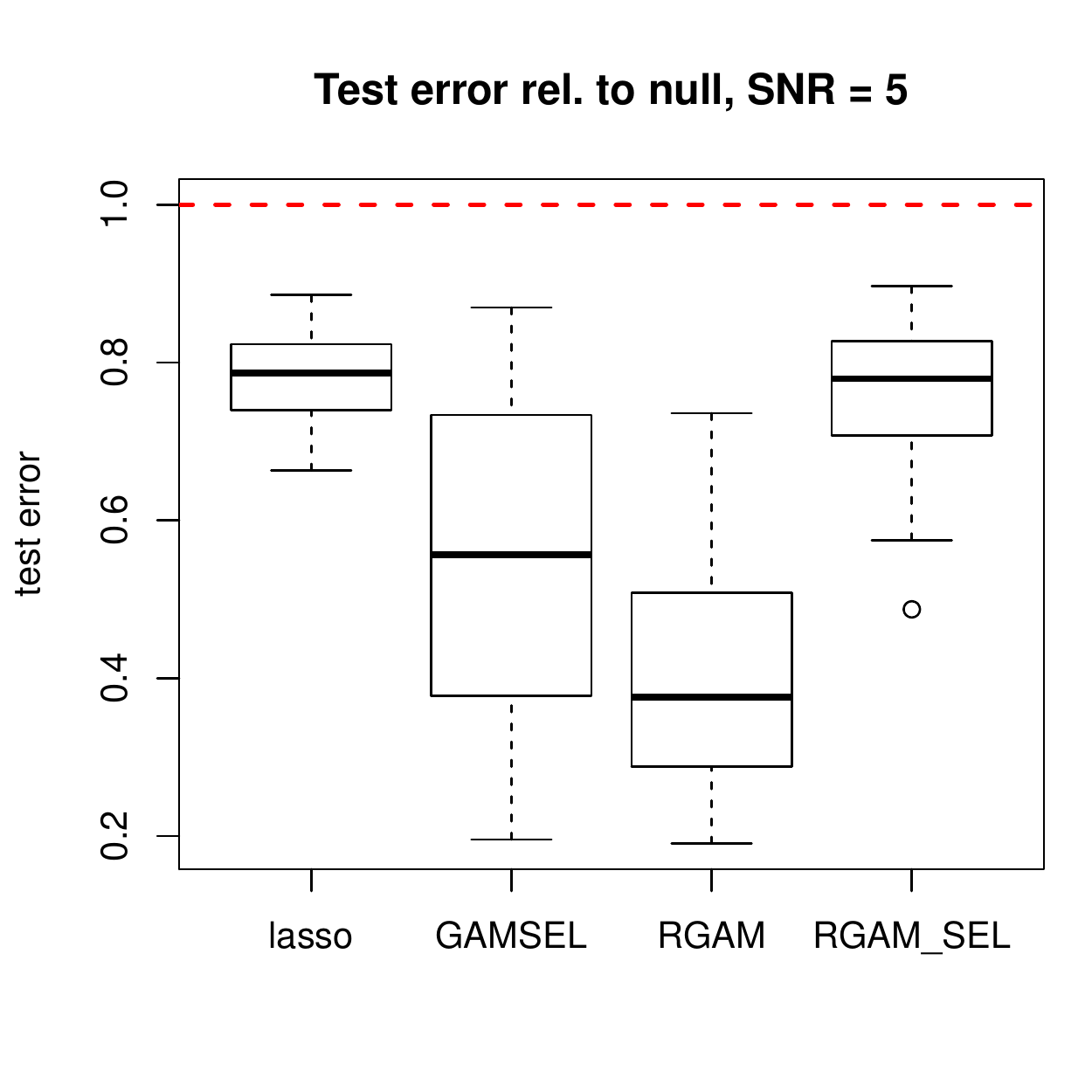}\includegraphics[width=2.0in]{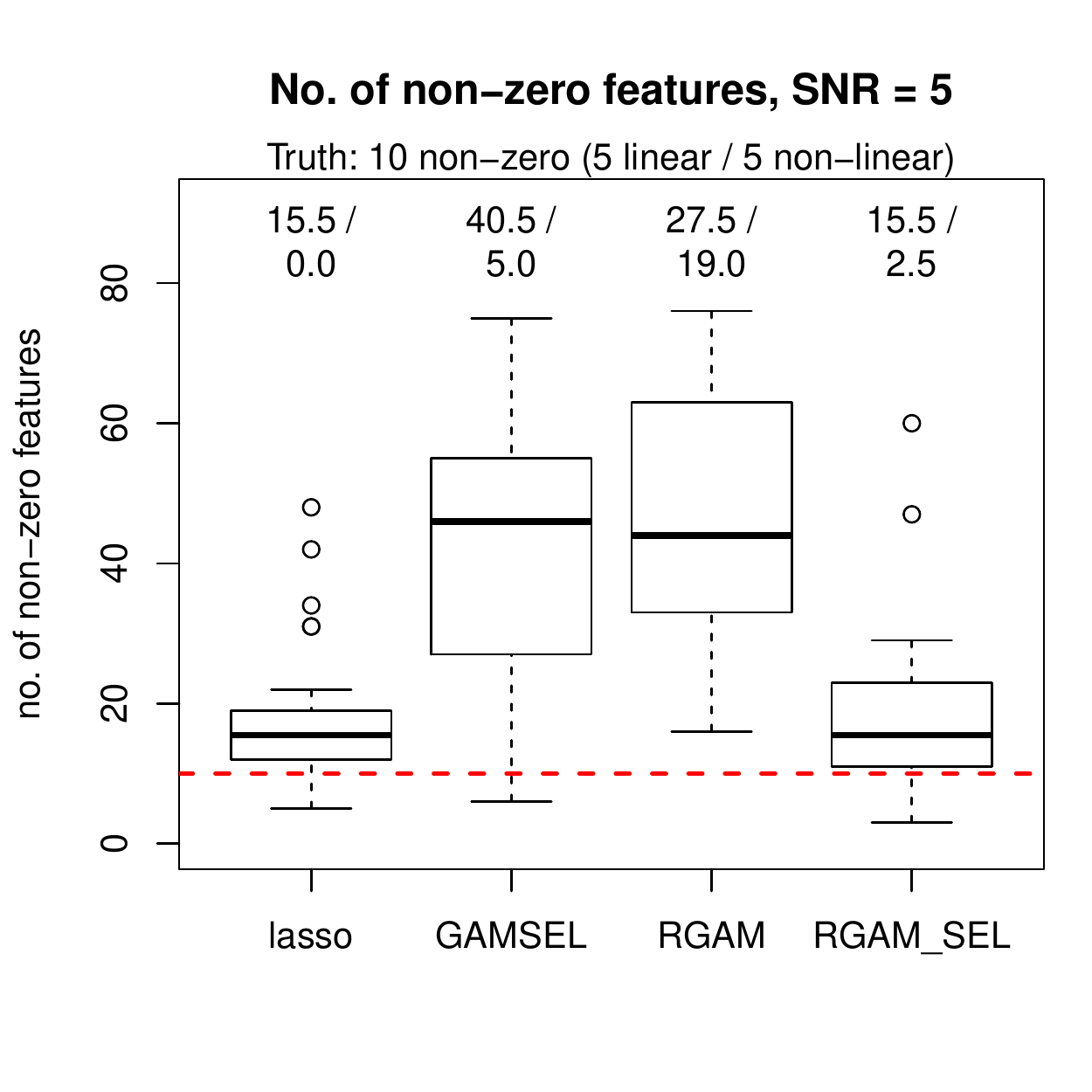}\includegraphics[width=2.0in]{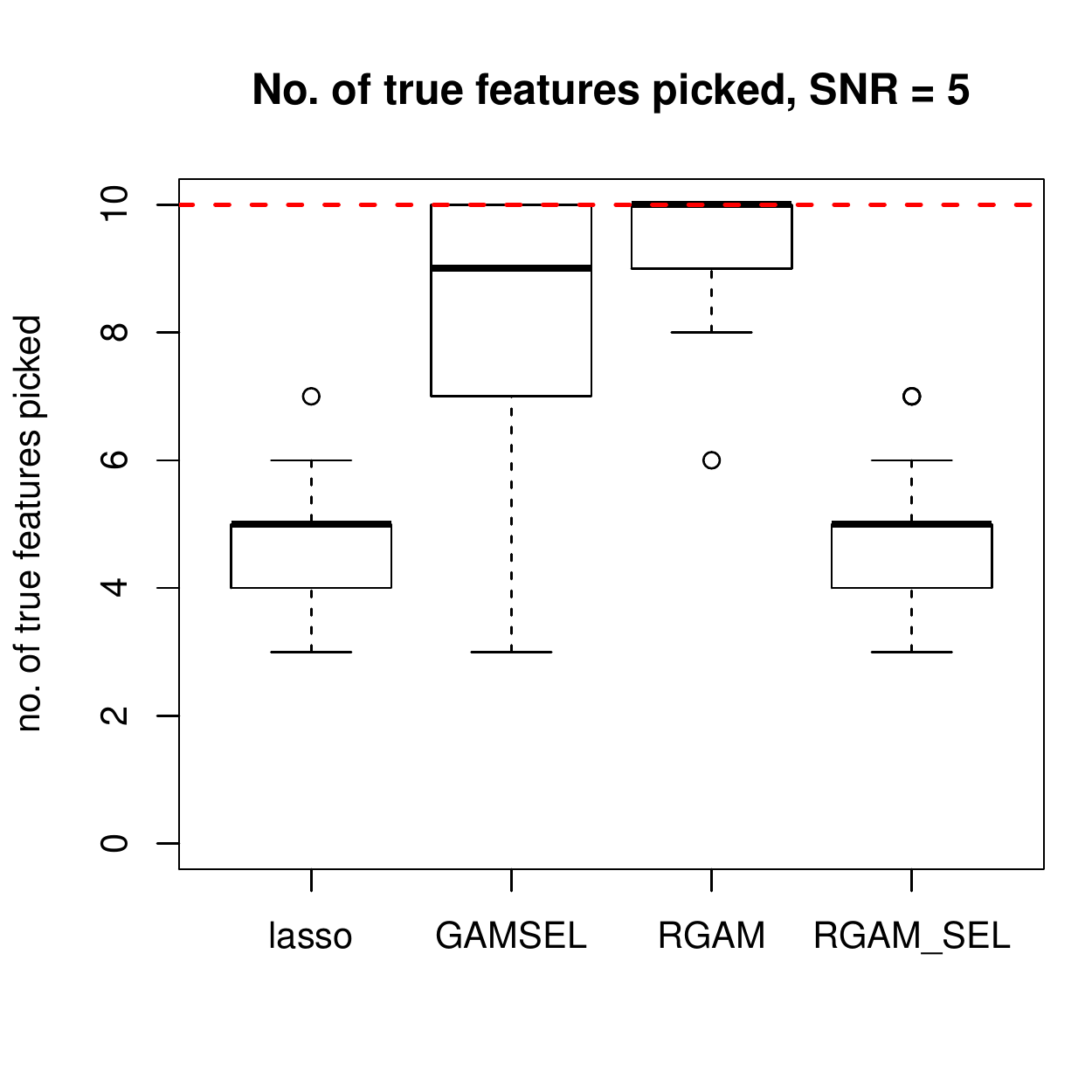}
\end{figure}

\pagebreak
\subsection{Mixed setting}

$n = 100$, $p = 200$, $\mu = \sum_{j=1}^5 [X_j + \frac{3}{4}(5X_j^3 - 3X_j)] + \sum_{j=6}^8 0.85 (3 X_j^2 - 1)$.

\begin{figure}[!htbp]
\centering
\includegraphics[width=2.0in]{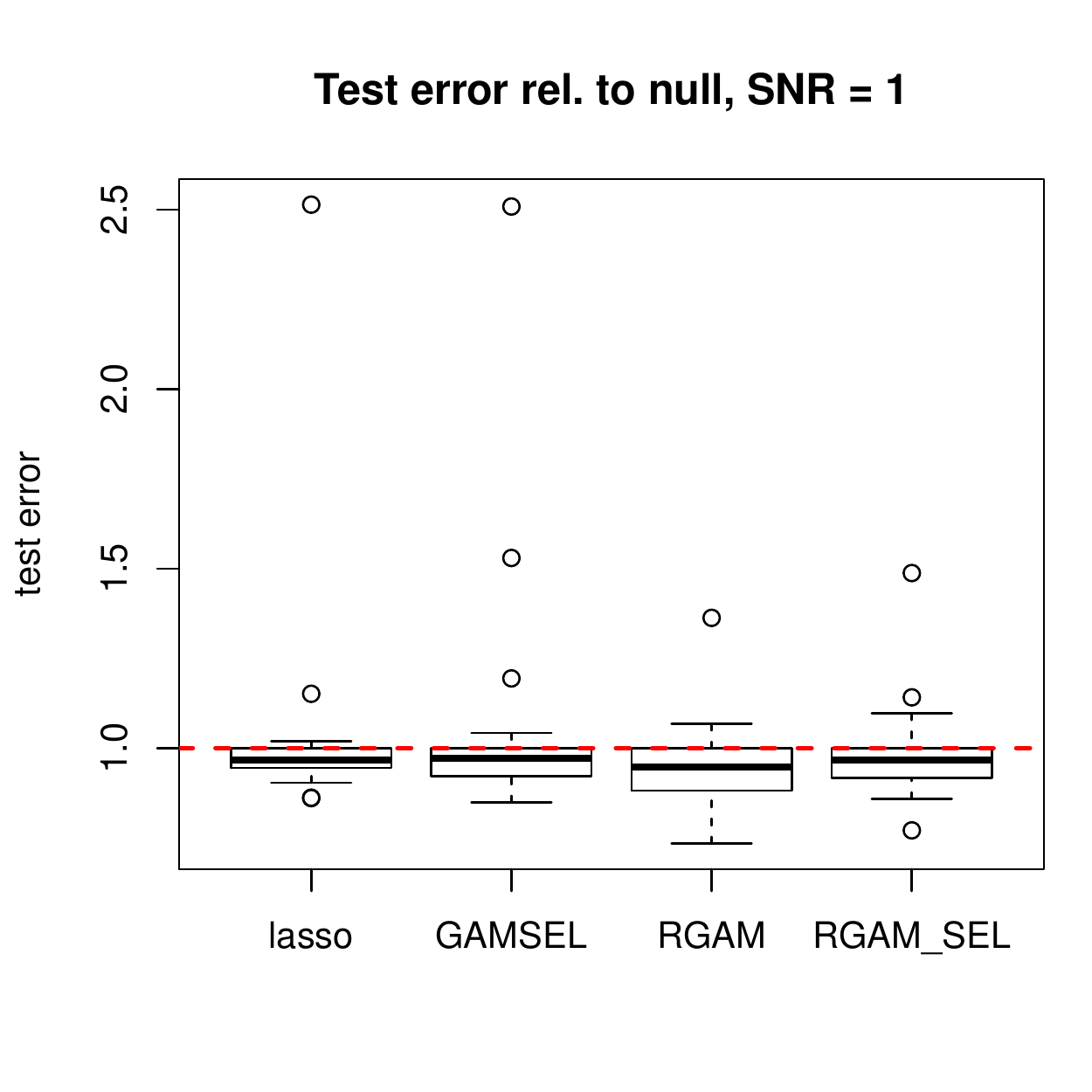}\includegraphics[width=2.0in]{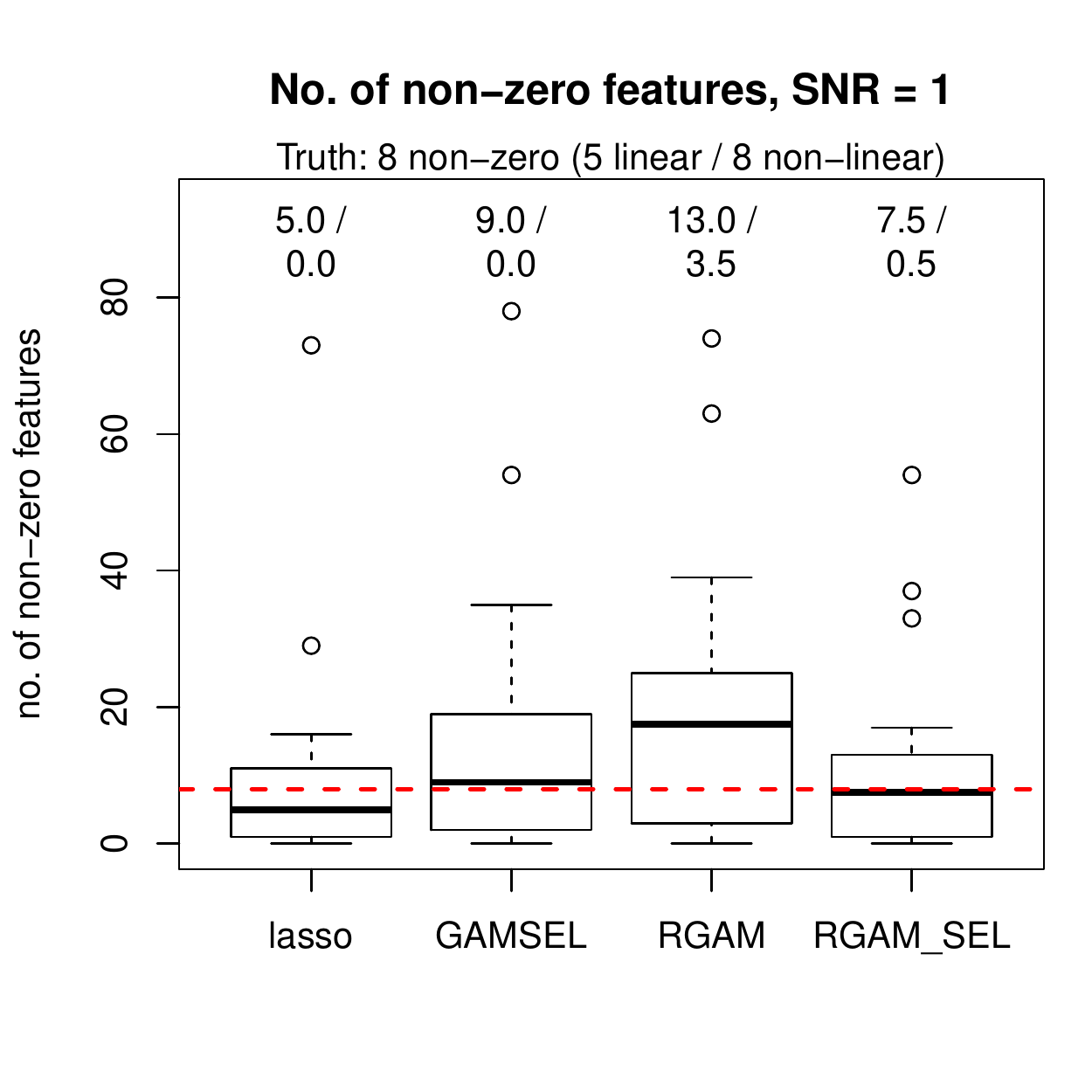}\includegraphics[width=2.0in]{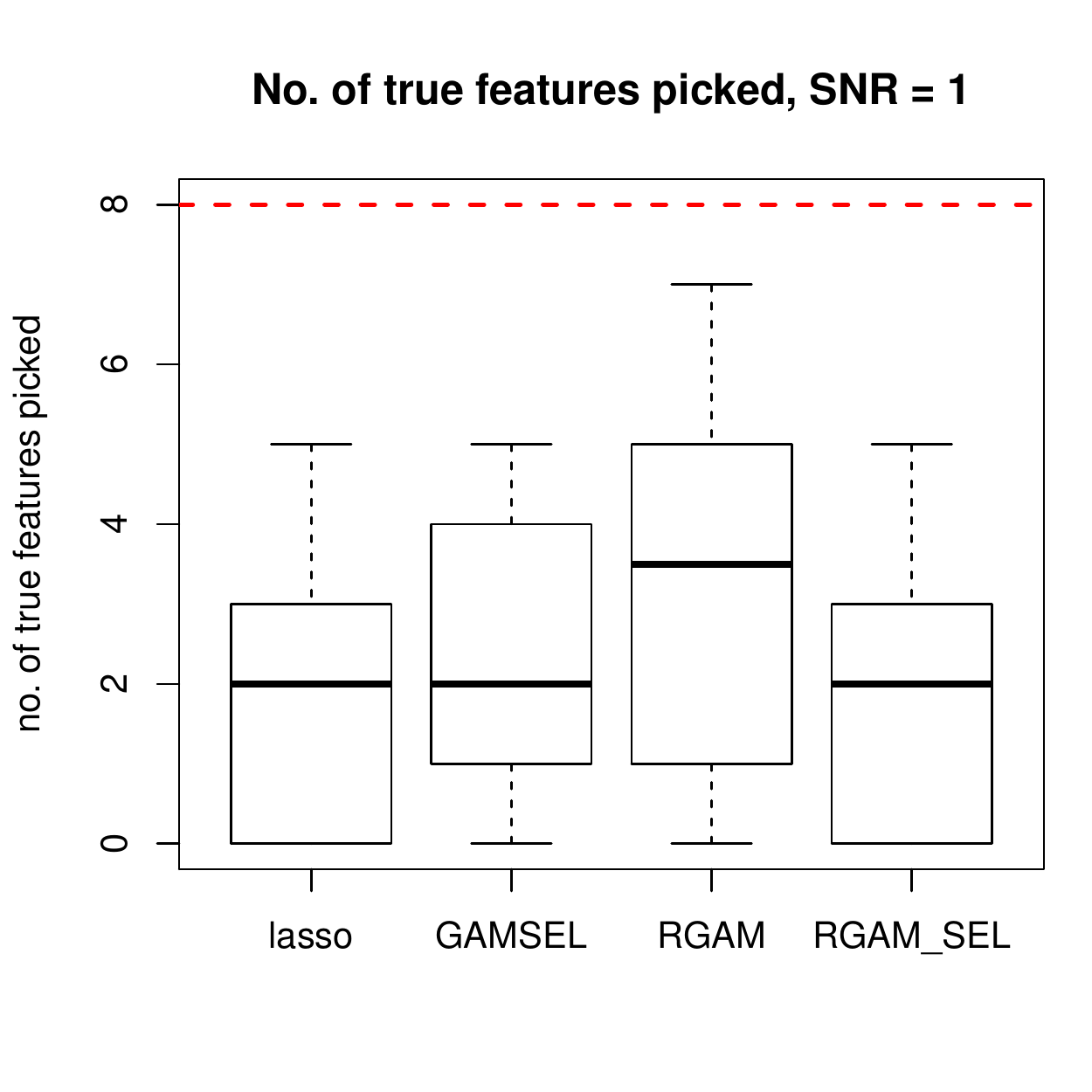}

\includegraphics[width=2.0in]{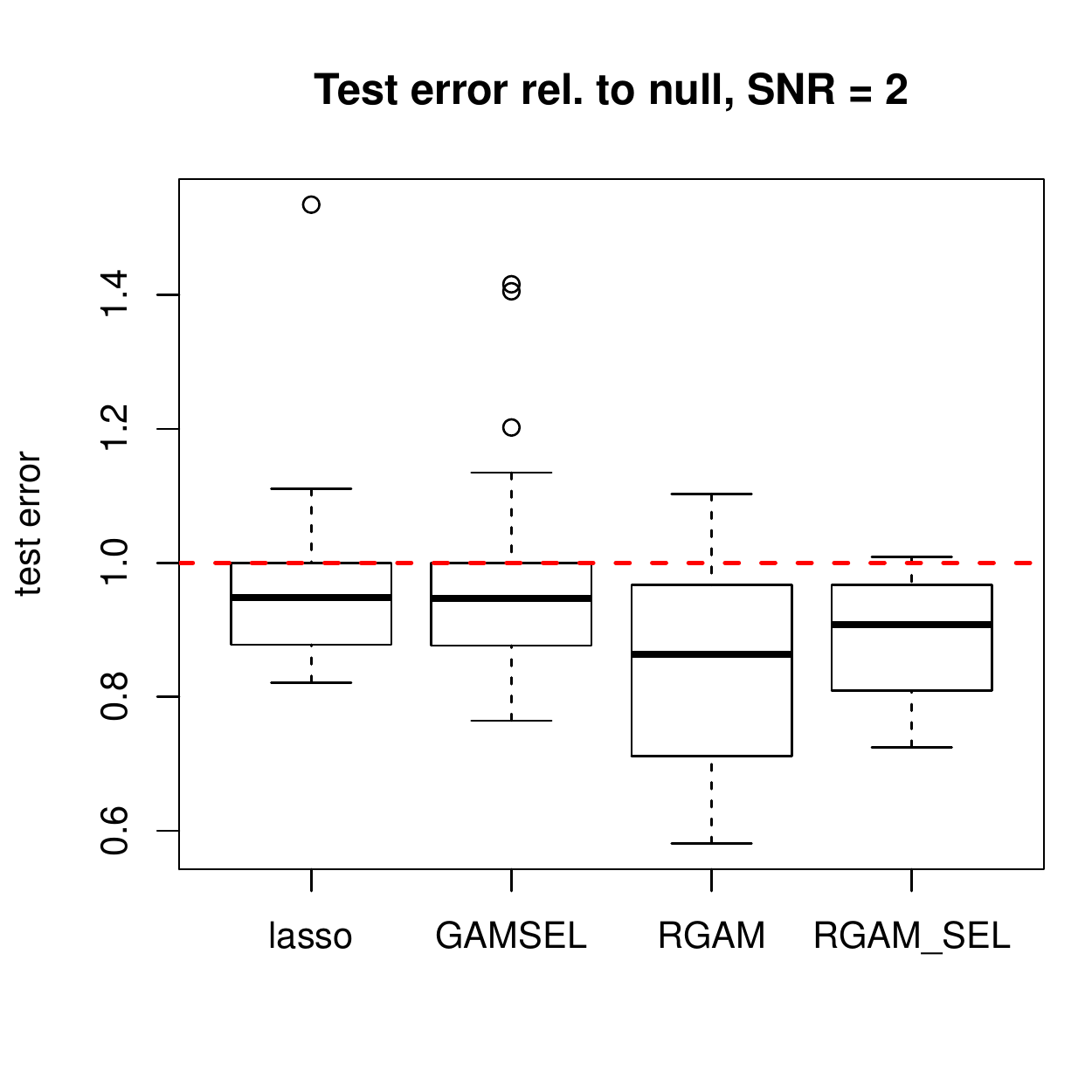}\includegraphics[width=2.0in]{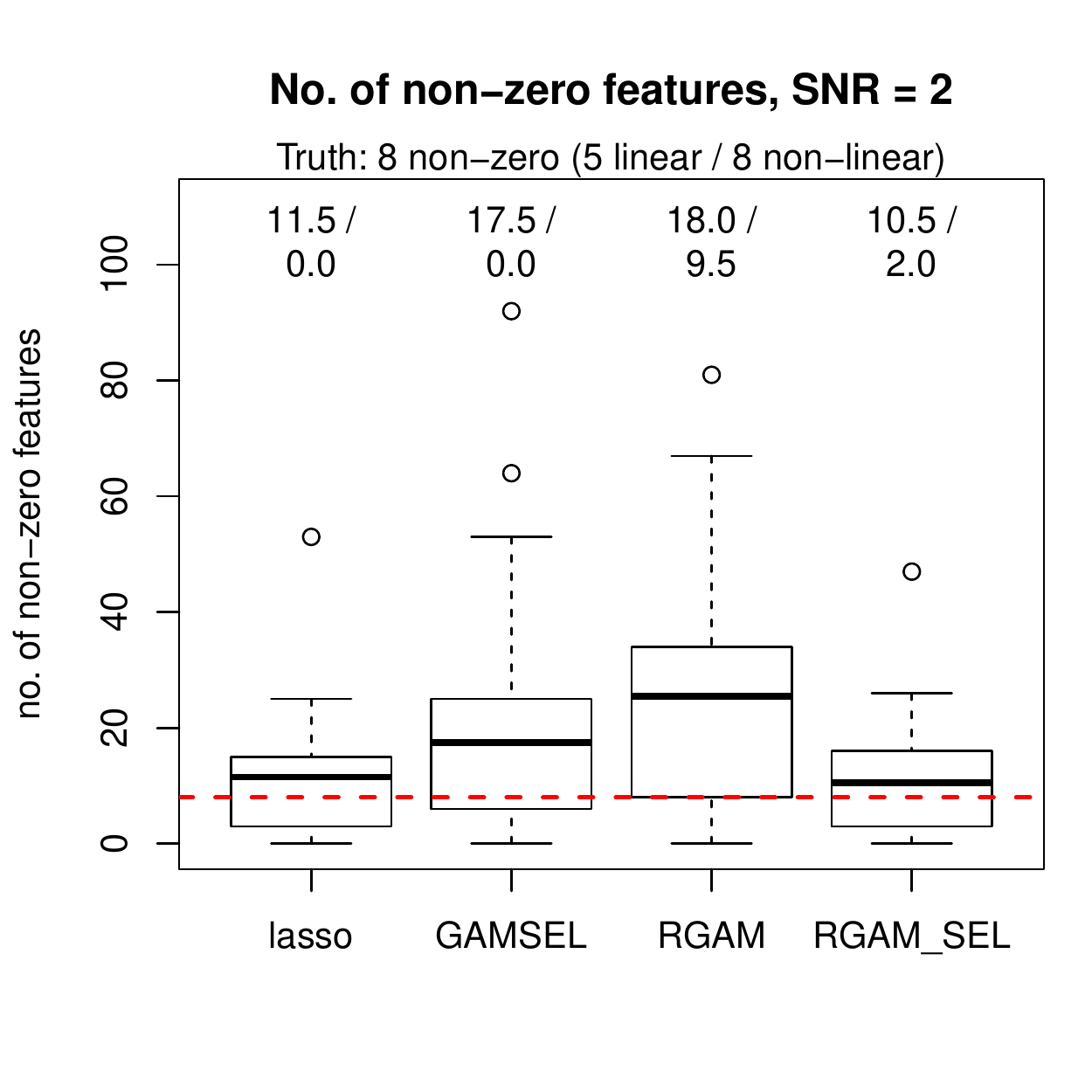}\includegraphics[width=2.0in]{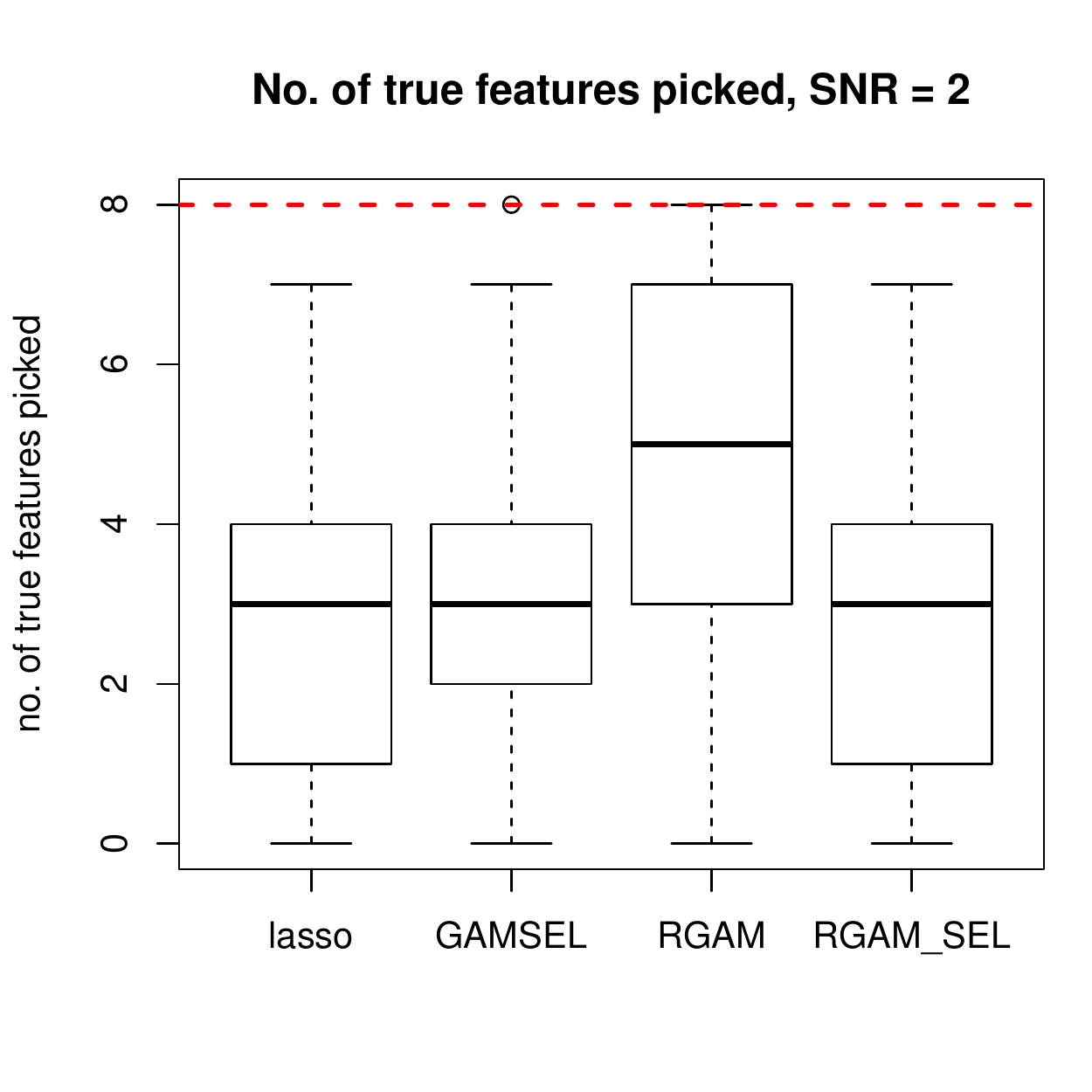}

\includegraphics[width=2.0in]{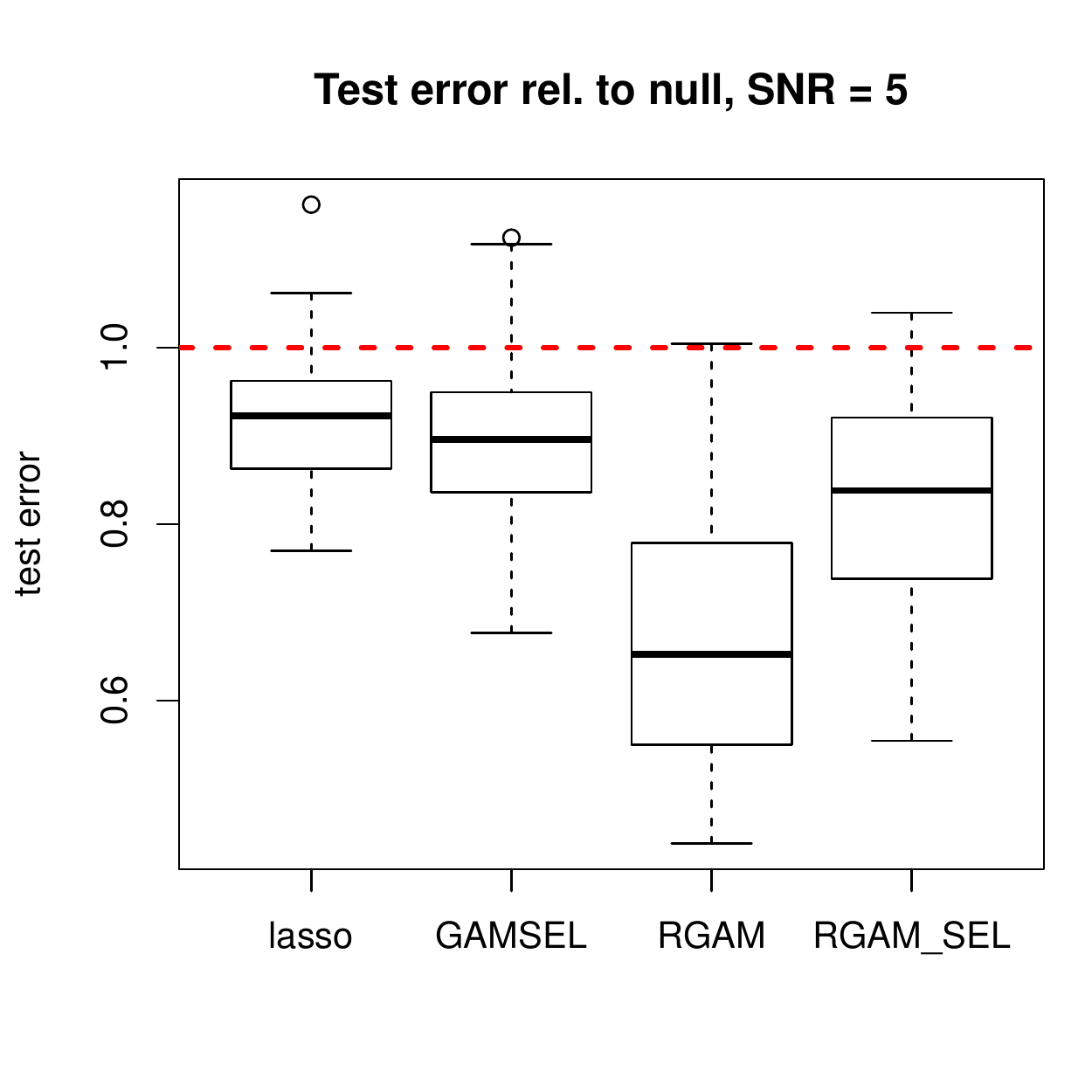}\includegraphics[width=2.0in]{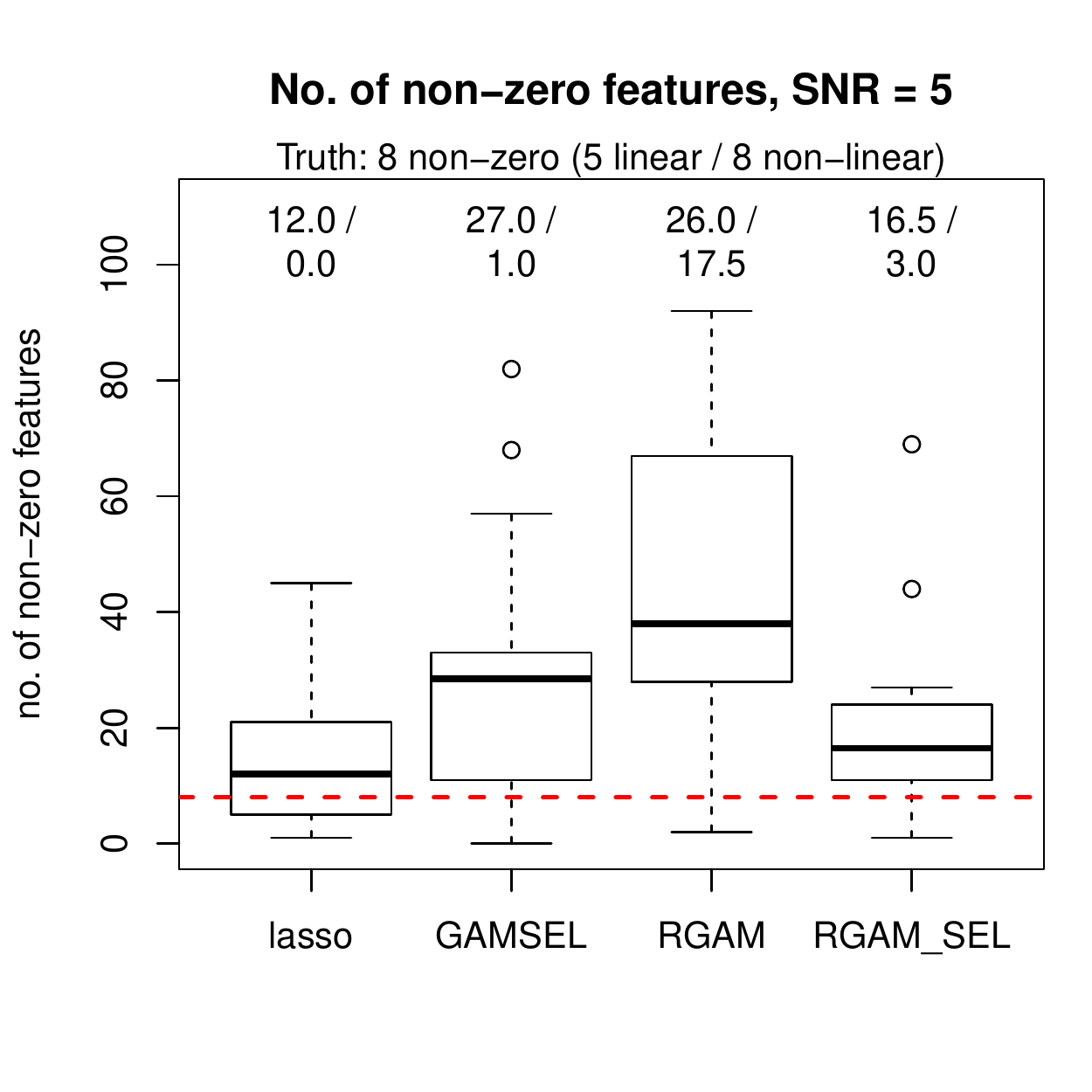}\includegraphics[width=2.0in]{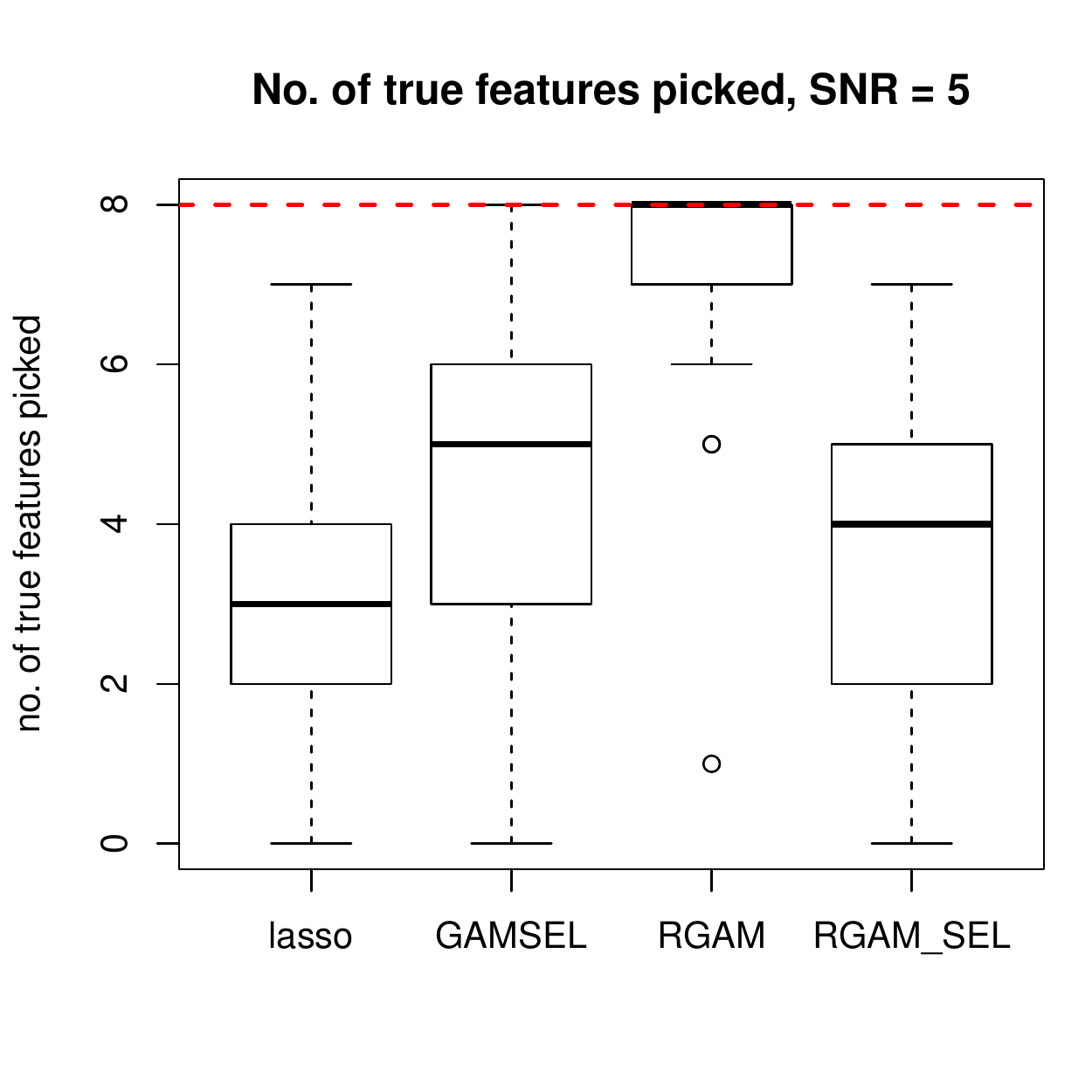}
\end{figure}

\pagebreak
\subsection{Mixed setting, large $n$ and $p$}

$n = 1,000$, $p = 500$, $\mu = \sum_{j=1}^{20} X_j + \sum_{j=1}^{20} \frac{3}{4}(5X_j^3 - 3X_j) + \sum_{j=21}^{28} (3X_j^2 - 1)$.

\begin{figure}[!htbp]
\centering
\includegraphics[width=2.0in]{setup6_snr1_errrel}\includegraphics[width=2.0in]{setup6_snr1_nzfeat}\includegraphics[width=2.0in]{setup6_snr1_nztruefeat}

\includegraphics[width=2.0in]{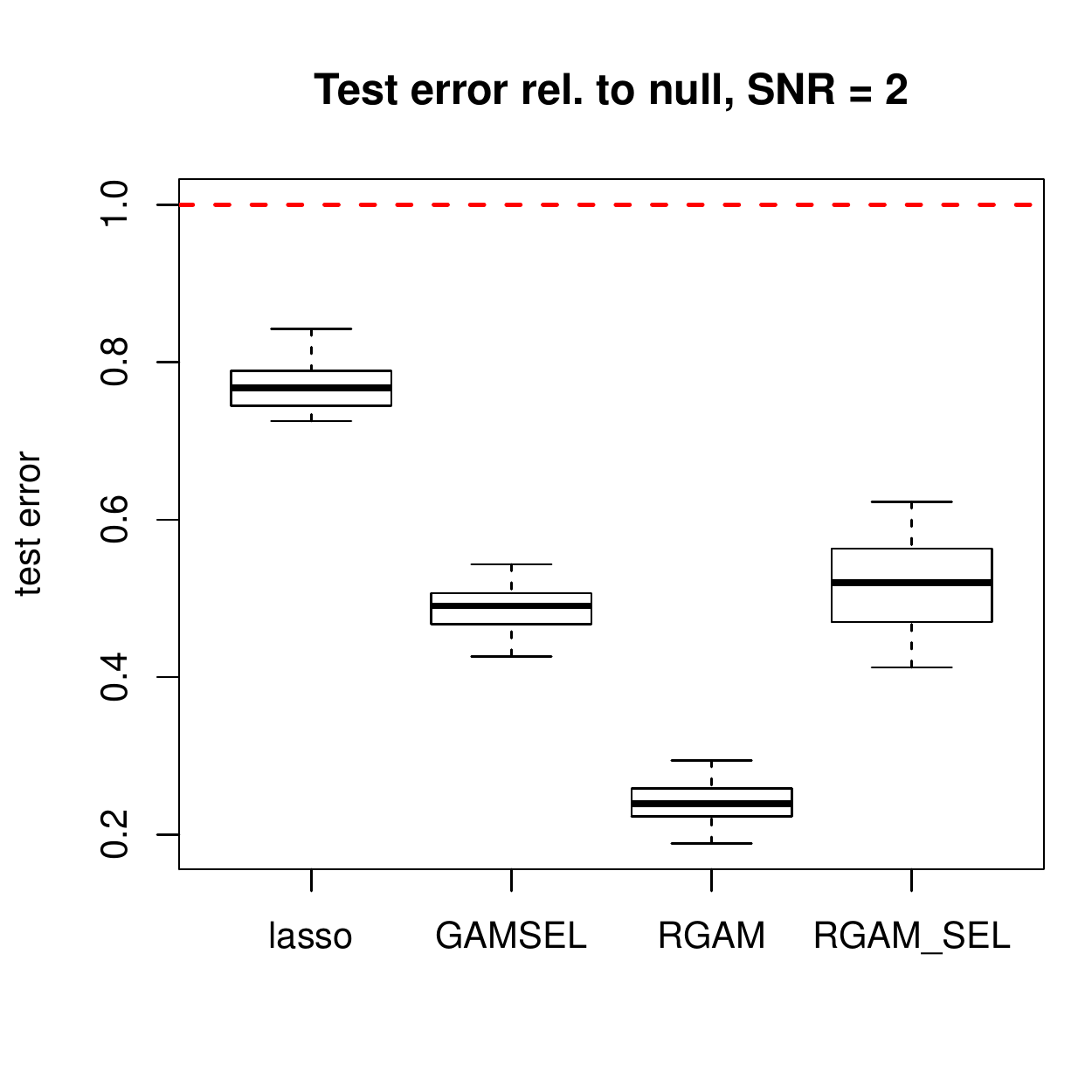}\includegraphics[width=2.0in]{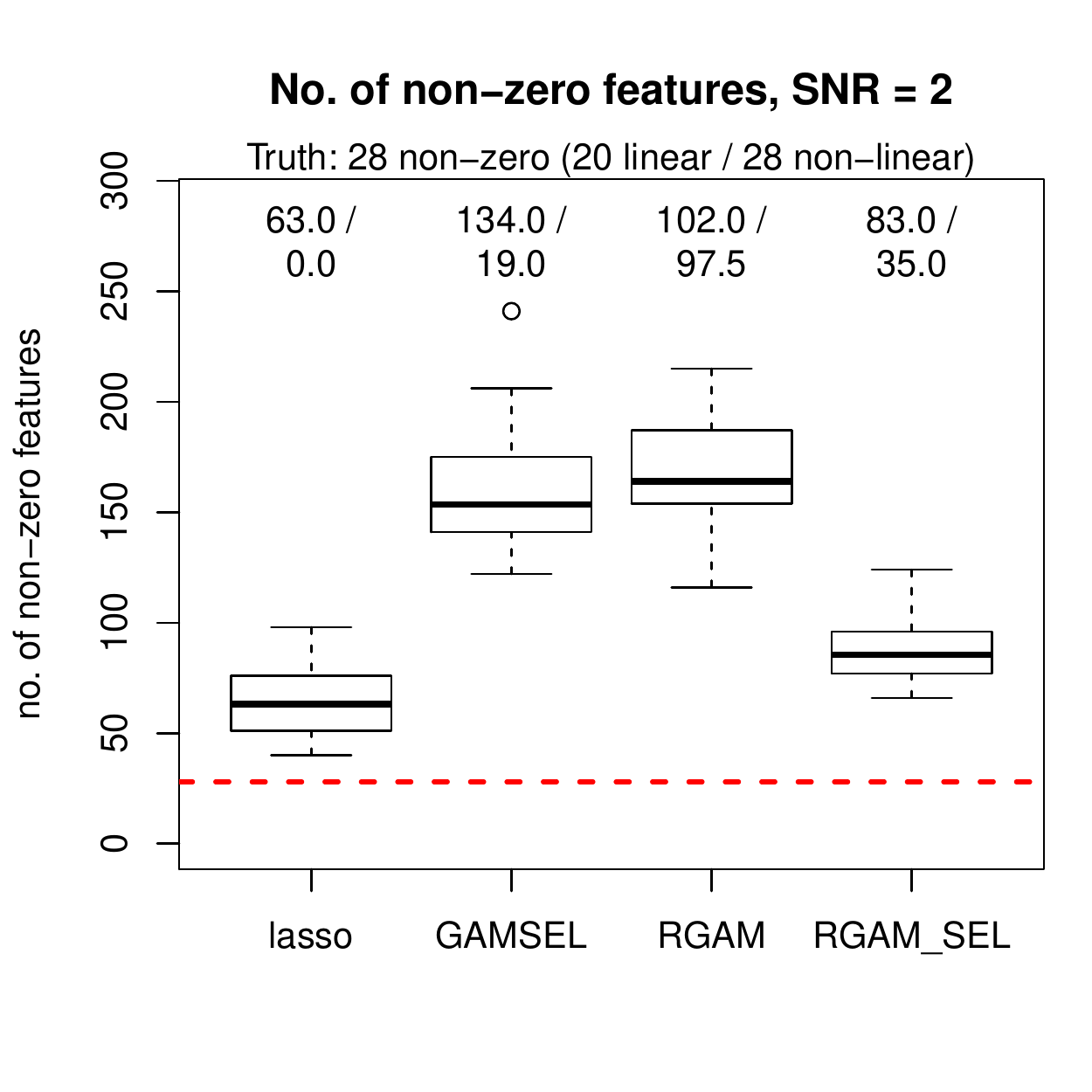}\includegraphics[width=2.0in]{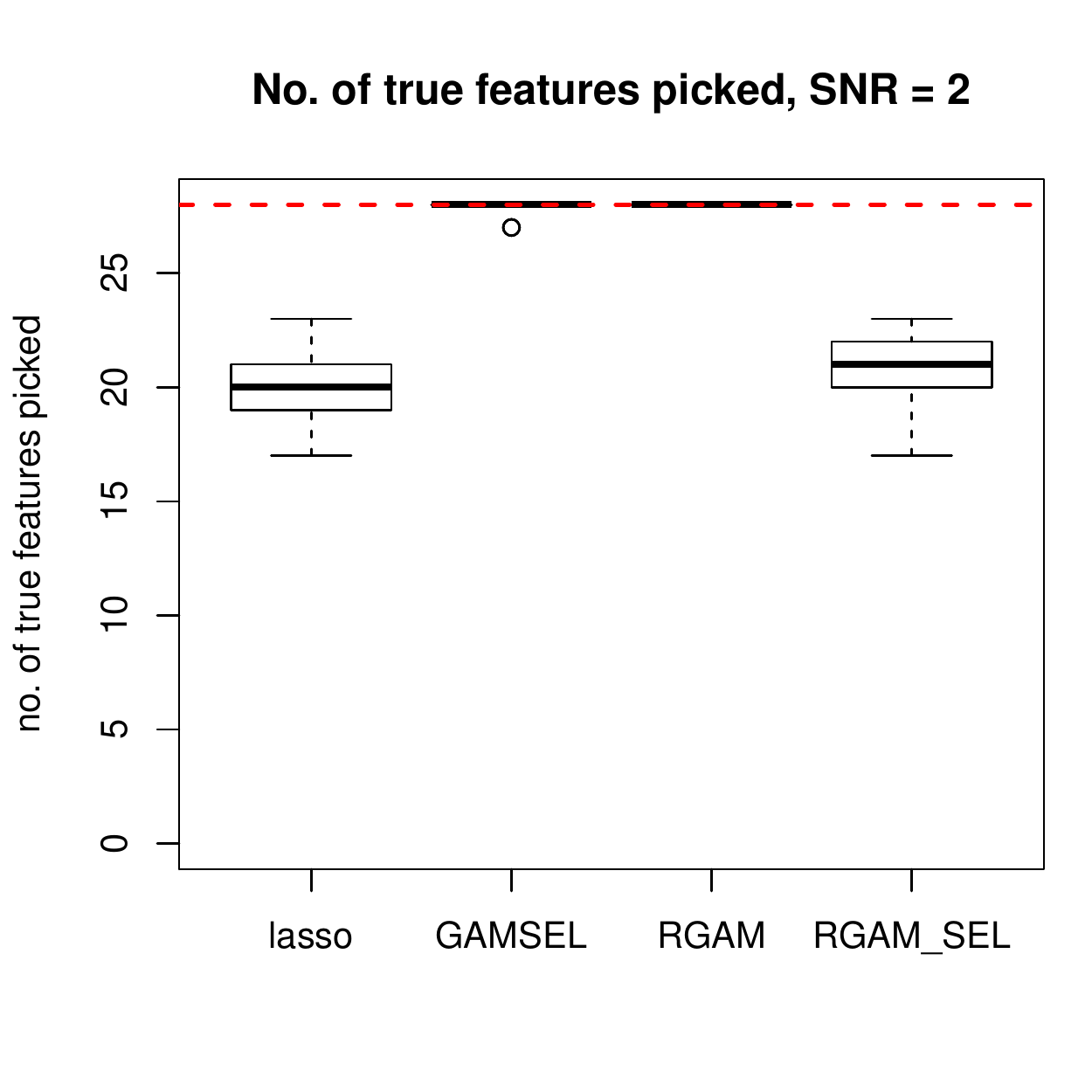}

\includegraphics[width=2.0in]{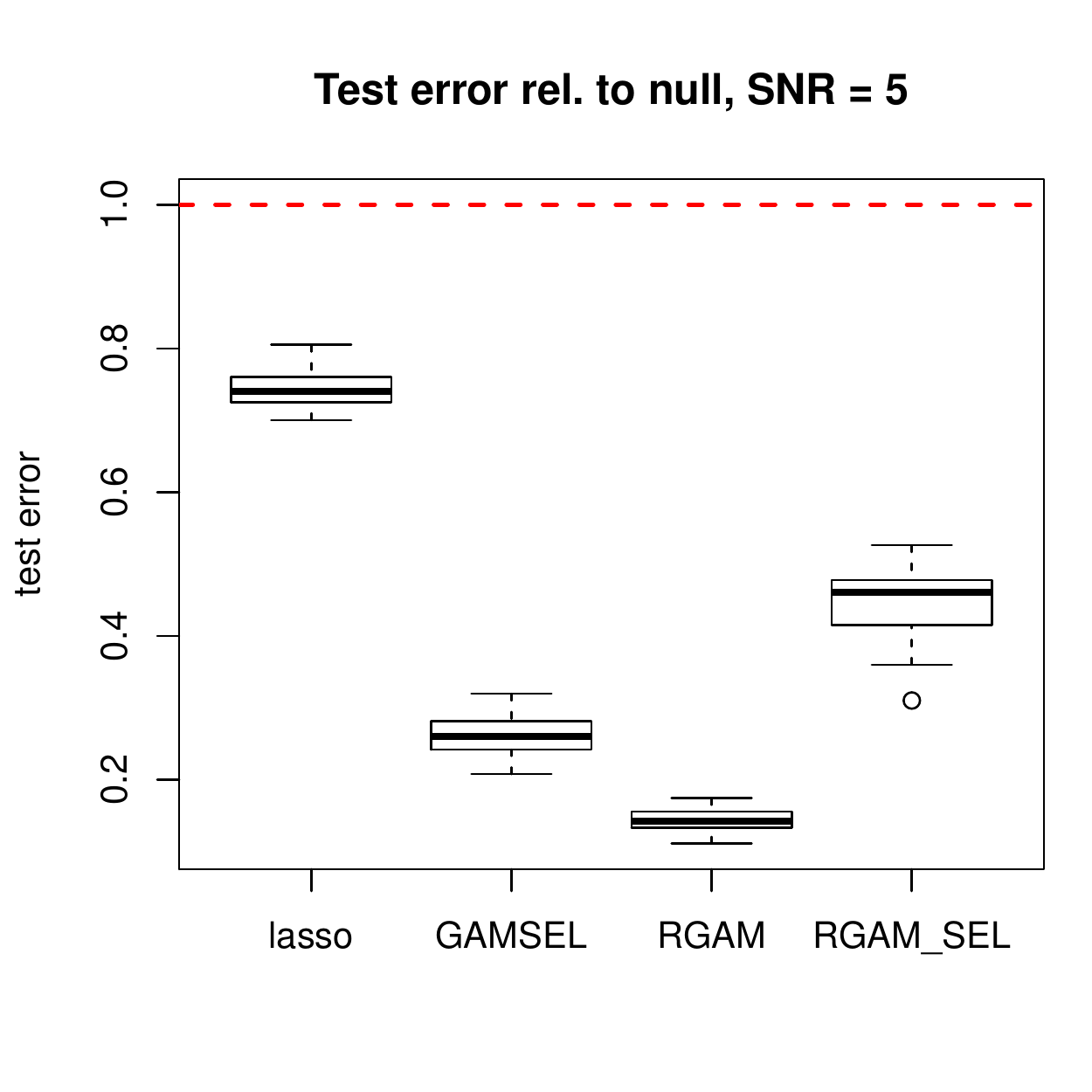}\includegraphics[width=2.0in]{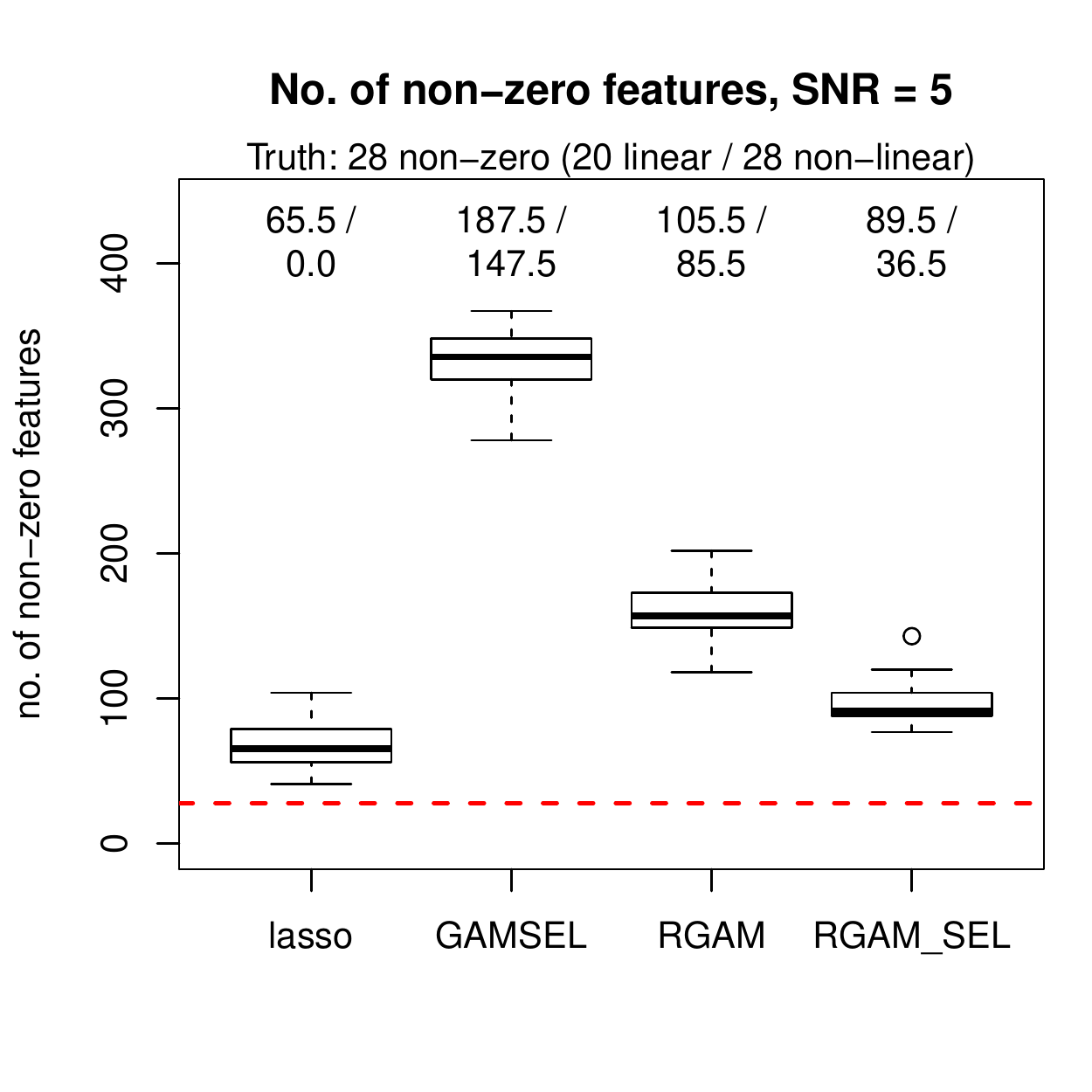}\includegraphics[width=2.0in]{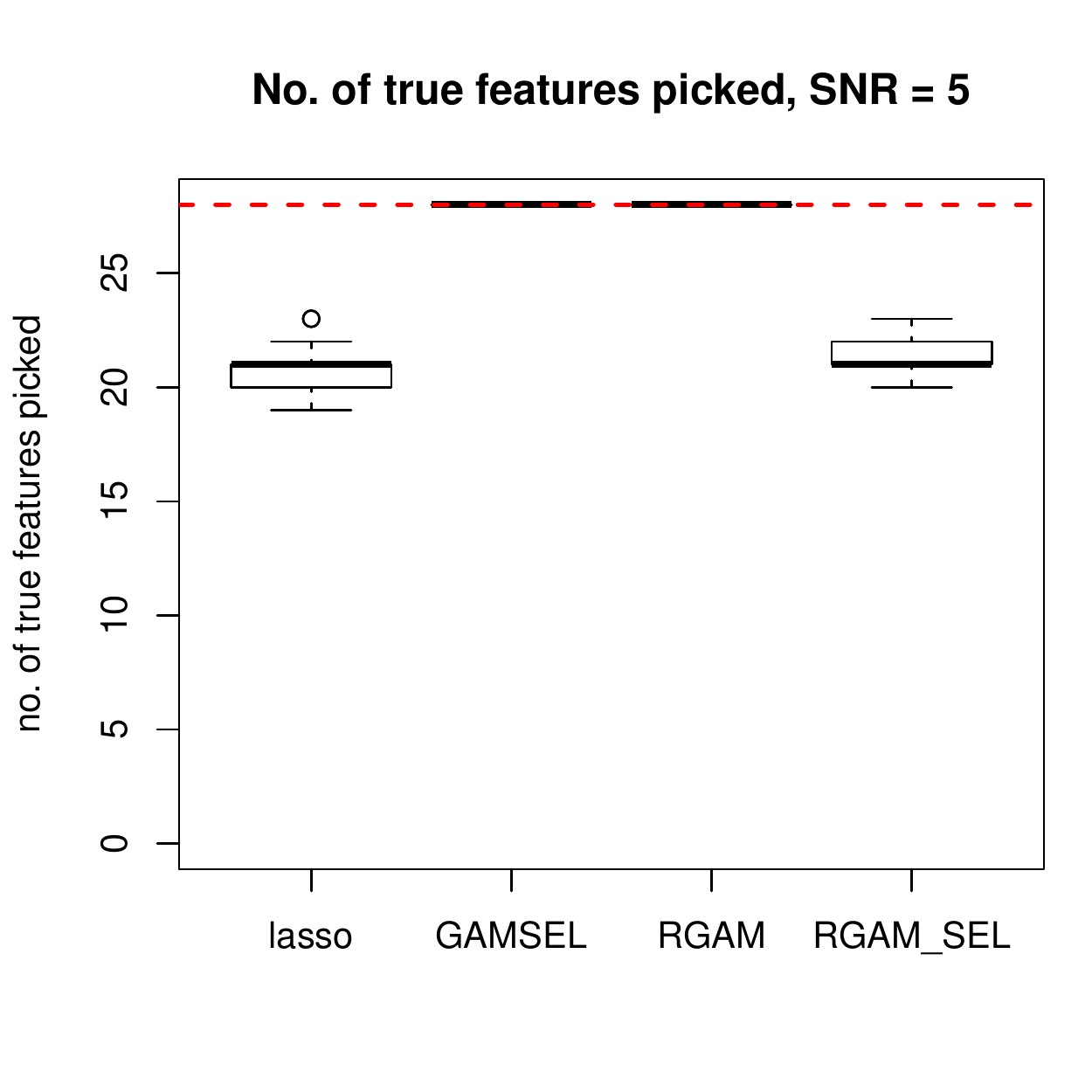}
\end{figure}

\bibliographystyle{agsm}
\bibliography{rgam}

\end{document}